\documentclass{article}

\usepackage{varioref}
\usepackage{amsmath,amsfonts,amsthm,mathrsfs}
\usepackage[normalem]{ulem}
\usepackage{multirow,color,graphics}
\usepackage{amsmath}
\usepackage{amsfonts}
\usepackage{amssymb}
\usepackage{jheppub}
\usepackage{enumerate}
\usepackage{fancyvrb}
\usepackage{verbatim}
\usepackage{wrapfig}
\usepackage{appendix}
\usepackage{amstext}
\usepackage{amssymb}
\usepackage{graphicx}
\usepackage{color}
\usepackage{varioref}
\usepackage{multirow,graphics}
\usepackage{epstopdf}
\usepackage{bbm}
\usepackage{slashed}
\usepackage{relsize}

\newcommand{\nn}{\nonumber}

\numberwithin{equation}{section}

\usepackage{tikz}
\usetikzlibrary{plotmarks,calc,decorations, decorations.pathmorphing, patterns}
\usetikzlibrary{arrows}
\tikzstyle dynkin node=[very thick,shape=circle,draw,inner sep=0pt,minimum size=5mm]
\tikzstyle dynkin line=[very thick]
\tikzstyle inverse line=[gray,line width=1.46pt,line cap=round, dash pattern=on 0pt off 2\pgflinewidth]
\tikzstyle red phase=[red,decoration={snake,amplitude=0.1mm,segment length=1.6mm},decorate]
\tikzstyle blue phase=[blue,decoration={snake,amplitude=0.1mm,segment length=0.9mm},decorate]
\tikzstyle green phase=[green,decoration={snake,amplitude=0.1mm,segment length=0.9mm},decorate]
\tikzstyle brown phase=[brown,decoration={snake,amplitude=0.1mm,segment length=0.9mm},decorate]
\newcommand{\boundellipse}[3]
{(#1) ellipse (#2 and #3)
}
\usetikzlibrary{decorations.pathmorphing}
\tikzstyle arrow=[thick,rounded corners=18pt,-latex]
\tikzstyle box=[draw,rounded corners,outer sep=4pt]
\tikzstyle B node=[outer sep=0pt]
\tikzstyle Q node=[inner sep=1pt,outer sep=0pt]
\definecolor{purple_nice}{rgb}{0.4,0.2,0.7}
\definecolor{fuel_blue}{RGB}{42,162,185}

\def\<{\langle}
\def\>{\rangle}

\newcommand{\chiFT}{{$\chi \textbf{FT}_4$}}
\newcommand{\sym}{${\cal N}=4$ SYM}

\newcommand{\p}{\partial}
\newcommand{\bQ}{{\bf Q}}
\newcommand{\bP}{{\bf P}}
\newcommand{\bp}{{\bf p}}

\newcommand{\tr}{{\text{tr}}}

\newcommand{\la}[1]{\label{#1}}
\newcommand{\eq}[1]{(\ref{#1})}

\def\CN{{\cal N}}
\def\CO{{\cal O}}

\def\bQ{{\bf Q}}
\def\bP{{\bf P}}

\def\su{{\mathsf u}}

\def\su{v}

\makeatletter
\@ifundefined{usebibtex}{} {}
\makeatother

\newcommand{\beq}{\begin{equation}}
\newcommand{\eeq}{\end{equation}}
\newcommand{\beqq}{\begin{equation*}}
\newcommand{\eeqq}{\end{equation*}}
\newcommand\beqa{\begin{eqnarray}}
\newcommand\eeqa{\end{eqnarray}}
\newcommand\beqaa{\begin{eqnarray*}}
\newcommand\eeqaa{\end{eqnarray*}}
\newcommand\bea{\begin{array}}
\newcommand\eea{\end{array}}
\newcommand\beaa{\begin{array}}
\newcommand\eeaa{\end{array}}

\def\bQ{{\bf Q}}

\def\Tr{\text{Tr}}

\def\L{J}
\def\mathfrakM{\varphi}

\title{
\Large   Integrability of Conformal Fishnet Theory}

\author[a,b]{Nikolay Gromov}
\author[c,d]{~Vladimir Kazakov}
\author[e]{~Gregory Korchemsky}
\author[c]{~Stefano Negro}
\author[c]{~Grigory Sizov}

\affiliation[a]{Mathematics Department, King's College London,
The Strand, London WC2R 2LS, UK.}
\affiliation[b]{St.Petersburg INP, Gatchina, 188 300, St.Petersburg,
Russia}
\affiliation[c]{Laboratoire de Physique Th\'eorique, D\'epartement de Physique de l'ENS, \'Ecole Normale Sup\'erieure, rue Lhomond 75005
Paris, France}
\affiliation[d]{Universit\'e Paris-VI, PSL Research University, Sorbonne Universit\'es, UPMC Univ. Paris 06, CNRS, 75005 Paris, France}
\affiliation[e]{Institut de Physique Th\'eorique\footnote{Unit\'e Mixte de Recherche 3681 du CNRS}, Universit\'e Paris Saclay, CNRS, CEA, F-91191 Gif-sur-Yvette }

\abstract{ 
We study integrability of fishnet-type Feynman graphs arising in planar four-dimensional bi-scalar chiral 
theory recently proposed in arXiv:1512.06704 as a special double scaling limit of gamma-deformed \({\cal N} = 4\) SYM theory. 
We show that the transfer matrix ``building" the fishnet graphs emerges from the $R-$matrix of non-compact conformal \(SU(2,2)\) Heisenberg spin chain with spins belonging to principal series representations of the four-dimensional conformal group. We demonstrate explicitly a  relationship between this integrable spin chain and the Quantum Spectral Curve (QSC) of \({\cal N} = 4\) SYM.
Using QSC and spin chain methods, we construct Baxter equation for $Q-$functions of the conformal spin
chain needed for computation of  the anomalous dimensions of operators of the type \(\tr(\phi_1^J)\) where  \(\phi_1\) is one of the 
two scalars of the theory. For  \(J = 3\) we derive from QSC a quantization condition that fixes the relevant solution of Baxter equation. The scaling dimensions of the operators only receive contributions from wheel-like graphs. 
We develop integrability techniques to compute the divergent part of these graphs and use it to present the weak coupling expansion of dimensions to  very high orders. Then we  apply our exact equations to calculate the anomalous dimensions with \(J=3\) to practically unlimited precision  at any coupling.  These  equations also describe an infinite tower of local conformal operators  all carrying the same charge \(J=3\). 
  The method should be applicable for any \(J\) and, in principle, to any  local operators of bi-scalar theory.
We show that at strong coupling the scaling dimensions can be derived from semiclassical quantization of finite gap solutions describing an integrable system of 
 noncompact $SU(2,2)$ spins. 
This bears similarities with the classical strings arising in the strongly coupled limit of \({\cal N} = 4\) SYM. 
}

\begin{document}



\maketitle
\newpage
{\addtocounter{page}2}



\section{Introduction: bi-scalar theory and results for ``wheel" graphs}

The \(\gamma\)-deformed planar \(\CN=4\) SYM theory has an interesting double scaling  limit ~\cite{Gurdogan:2015csr} combining a vanishing 't~Hooft coupling constant \(g^2=N_cg_{_{YM}}^2\to0\) and the increasing twist parameters \(q_j=e^{-i\gamma_j/2}\to\infty\)\footnote{Here \(\gamma_j\) are the angles of twist parameters for three Cartan \(U(1)\) subgroups of \(SU(1,5)\) R-symmetry of the model.} in such a way that \(\xi_j=gq_i\) are kept fixed\footnote{A very similar limit was proposed previously in the context of the cusped Wilson Lines in \cite{Correa:2012nk}.}.
In the particular case $\xi_1=\xi_2=0$ and $\xi\equiv\xi_3\ne0$,  the limiting theory describes two complex scalar fields  and is
dubbed the bi-scalar \chiFT\,~\cite{Gurdogan:2015csr}. The Lagrangian of this theory is given by
\begin{align}
\label{bi-scalarL}
{\cal L}_{\phi}= \frac{N_c}{2}\Tr\,\,
\left(\p^\mu\phi^\dagger_1 \p_\mu\phi_1+\p^\mu\phi^\dagger_2 \p_\mu\phi_2+2\xi^2\,\phi_1^\dagger \phi_2^\dagger \phi_1\phi_2\right)\,,
\end{align}
where $\phi_i = \phi_i^a T^a$ are complex scalar fields and $T^a$ are the generators of the $SU(N_c)$ gauge group in the fundamental representation,
normalized as $\tr(T^a T^b)=\delta^{ab}$.
Notice that the quartic scalar interaction term in (\ref{bi-scalarL}) is complex making the theory nonunitary. As we show below, this leads to a number
of unusual properties of the bi-scalar \chiFT\,.


The \(\gamma\)-deformed version of   \(\CN=4\) SYM breaks the global \(PSU(2,2|4)\) superconformal symmetry down to  \(SU(2,2)\times U(1)^3\).  In the bi-scalar theory (\ref{bi-scalarL}) this symmetry is further reduced to \(SU(2,2)\times U(1)^2\).  For the large majority of physical quantities, such as the multi-point correlators\footnote{As was argued in \cite{Sieg:2016vap,Gurdogan:2015csr} only the correlators containing the operators of length \(\L=2\), such as \(\Tr(\phi^2)\) in their initial or final states, appear to break the conformal invariance due to the appearance of double-trace couplings in the effective lagrangian and generate a scale.} or amplitudes, this   theory shows a CFT behavior in the planar limit.
This bi-scalar \chiFT\ is shown to have a very limited set of planar Feynman graphs in the perturbative expansion of any physical correlator. The number of planar graphs   at each loop order  is not growing with the order. Moreover, at sufficiently large loop orders, these graphs have in the bulk the ``fishnet" structure of a regular square lattice pointing on the  explicit integrability of the model  (\ref{bi-scalarL})  in the planar limit        \cite{Gurdogan:2015csr}. This integrability is due to the fact noticed in     \cite{Zamolodchikov:1980mb} that these fishnet graphs define an integrable lattice model. 

In this paper, we show that
the bi-scalar model \eqref{bi-scalarL} represents a  four-dimensional field-theoretical realization of the planar fishnet lattice model and identify the underlying
integrable model as being a noncompact Heisenberg spin chain with spins belonging to infinite dimensional representation of the four-dimensional conformal group
\(SU(2,2)\).
This fact sheds a certain light on the origins of, still mysterious, integrability of the planar  \(\CN=4\) SYM~\cite{Beisert:2010jr}. Similar noncompact Heisenberg
spin chains have been previously encountered in the study of high-energy asymptotics in QCD \cite{Lipatov:1993yb,Faddeev:1994zg}. We shall apply the technique developed in \cite{Derkachov:2001yn,Derkachov:2002wz}
to clarify the basic features of integrability of the fishnet graphs using the method of the Baxter $Q-$operator. In particular, we describe the Lax operators,
transfer-matrices and construct the Baxter $TQ$ equation that we later use to compute the anomalous dimensions of certain operators in the bi-scalar model
(\ref{bi-scalarL}). However, in order to extract the anomalous dimensions from the $TQ$ relations we had to use an additional insight coming from the side of ${\cal N}=4$ SYM, where the problem of finding the spectrum is solved by the Quantum Spectral Curve (QSC) method
\cite{Gromov2014,Gromov:2014caa,Gromov:2015dfa,Kazakov:2015efa}.

The integrability of the bi-scalar \chiFT\,, together with a specific, limited Feynman graph content of this theory provides us with a powerful method of computing new multi-loop massless four-dimensional Feynman integrals.  In the paper \cite{Caetano:2016ydc} a large variety of such graphs, relevant for the computations of anomalous dimensions, is described and some of these graphs are computed by the help of integrability.  Similar double scaling limit has been observed in
\cite{Caetano:2016ydc} for the ABJ(M) model, for which the QSC is also known \cite{Cavaglia:2014exa,Bombardelli:2017vhk}. In the  ABJ(M) theory the Feynman graphs are dominated in the bulk by the regular rectangular ``fishnet" lattice structure.
 In the paper \cite{Chicherin:2017cns}, a similar ``fishnet" structure for the bi-scalar amplitudes and single-trace correlators in the theory \eqref{bi-scalarL} has been observed and their integrable structure in terms of an explicit Yangian symmetry has been established. The authors of ~\cite{Mamroud:2017uyz}  introduce a similar tri-scalar  \(\chi\){\bf FT}\(_6\) with  \(\phi^3-\)type chiral interactions, realizing the hexagonal finsnet graphs of~\cite{Zamolodchikov:1980mb} and compute the simplest all-loop 2-point correlation function.  This theory  seems to define a genuine CFT\(_6\) in planar limit for all local operators. In the paper \cite{Basso:2017jwq} a single trace 4-point correlation function
 given by a rectangular  fishnet graph has been explicitly computed using the  bootstrap methods for AdS/CFT integrability.  

\begin{figure}
\center{\includegraphics[scale=0.4]{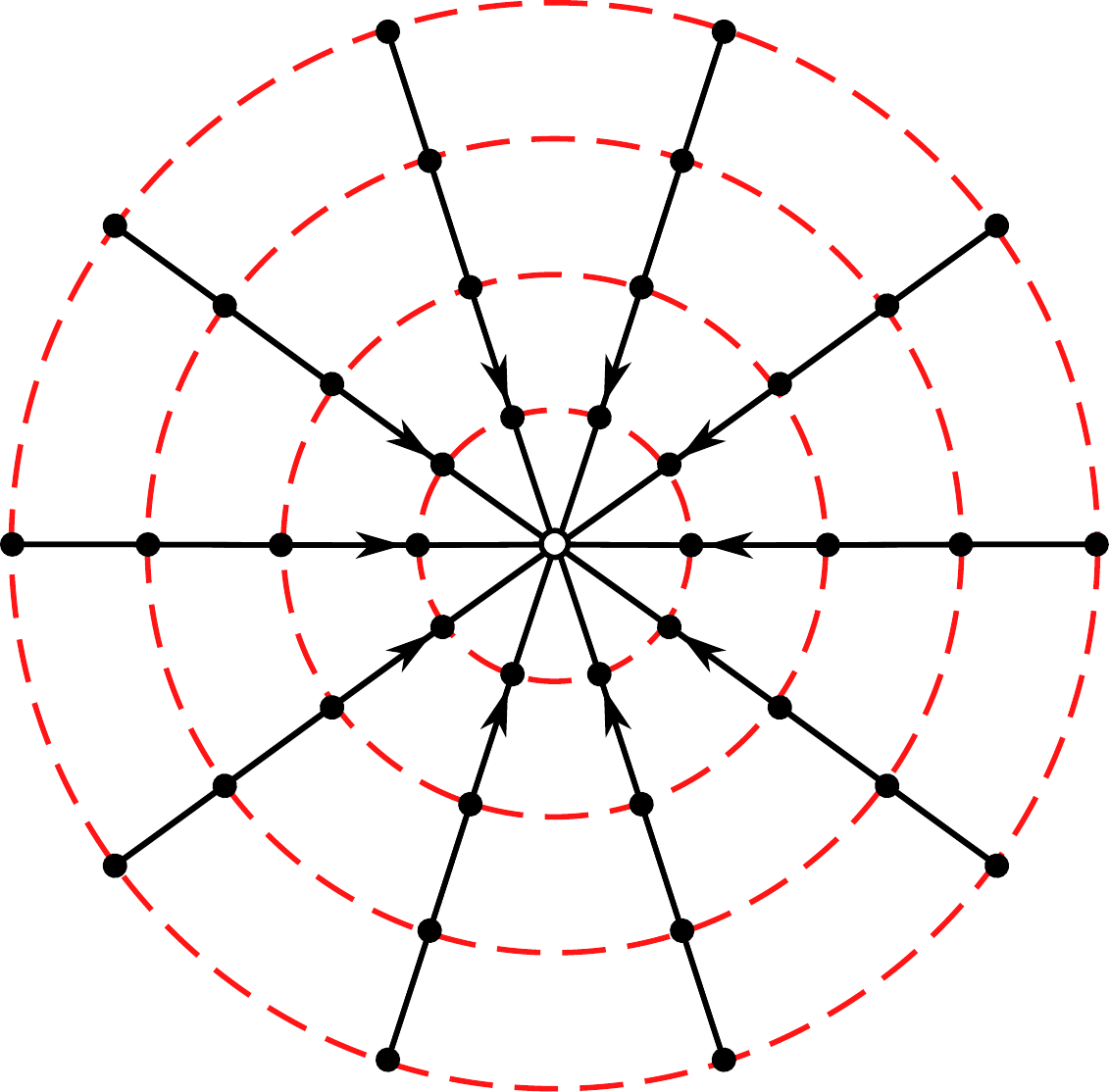}\label{fig:multiwheel}}
\caption{The ``wheel" Feynman graph with \(M\) frames defining the order \(\xi^{2M{\L}}\) of the perturbative expansion of anomalous dimension of the BMN vacuum operator \(\Tr\phi_1^{\L}\) (\(M=4\,,\,\, {\L}=10\) on this picture).   }
\end{figure}

The operators in the theory \eqref{bi-scalarL} can be classified with respect to the irreducible representations of  its global symmetry   \(SU(2,2)\times U(1)^2\), characterized by the values of Cartan generators \((\Delta,S,\dot S| J_1,J_2)\), where $\Delta$ is the scaling dimension of the operator, the pair $(S, \dot S)$
defines its Lorentz spin and the $U(1)$ charges $J_i$ count the difference between the total number of $\phi_i$ and $\phi_i^\dagger$.
In this paper, we study the scaling dimensions of a family of scalar single-trace operators of the following schematic form
\begin{align}\label{O-def}
\mathcal O_{\L,n,\ell} = P_{2\ell}(\partial) \tr[ \phi_1^\L \phi_2^n (\phi_2^\dagger)^n] +\dots \,,
\end{align}
where $P_{2\ell}(\partial)$ denotes  $2\ell$ derivatives acting on scalar fields inside the trace with all Lorentz indices contracted. The dots
stand for similar operators with the scalar fields $\phi_1$, $\phi_2$ and $\phi_2^\dagger$ exchanged inside the trace. The operators (\ref{O-def})
belong to the representation \((\Delta,0,0| \L,0)\) with $\Delta=\L+2(n+\ell)$ for zero coupling $\xi=0$. For nonzero coupling,
the operators (\ref{O-def}) with the same $\Delta$ and $\L$ mix with each other and their scaling dimensions can be found by diagonalizing the corresponding mixing matrix.
In the special case $n=\ell=0$ the operator (\ref{O-def}) takes the form
\begin{equation}\label{L-wheel}
\CO_\L=\Tr(\phi_1^\L)\,.
\end{equation}
Similar operator can be also defined in  \(\CN=4\) SYM in which case it is protected from quantum corrections and is known as the BMN vacuum operator.
In the \(\gamma\)-deformed \(\CN=4\) SYM, the operator (\ref{L-wheel}) is not protected and its scaling dimension starts to depend on the
coupling. Furthermore, in the bi-scalar theory (\ref{bi-scalarL})  the coupling dependent corrections to the scaling dimension $\Delta$
only come from the wheel-type Feynman graphs ~\cite{Gurdogan:2015csr}   shown in Fig.~\ref{fig:multiwheel}.   Each wheel contains $\L$ interaction vertices and the contribution
to the scaling dimension of the wheel graph with $M$ frames scales as $\xi^{2{\L}M}$. For $M=2$ and arbitrary ${\L}$, the contribution of the double-wheel (with two frames) graph to the
scaling dimension of the operator (\ref{L-wheel}) in \(\gamma\)-deformed \(\CN=4\) SYM was found in  \cite{Ahn2011}
using the TBA-type computations and represented in ~\cite{Gurdogan:2015csr}  in terms of explicit multiple zeta values (MZV).

In this paper, we use integrability technique to compute the scaling dimensions of the operators (\ref{O-def}) with ${\L}=3$ for an arbitrary coupling $\xi$. At zero
coupling, the scaling dimensions of such operators take odd values $\Delta(0)=3,5,\dots$. For $\Delta(0)=3$ there is only one operator $\tr(\phi_1^3)$. For $\Delta(0)=5$
we can separate all operators into those involving derivatives, $\tr(\phi_1^2\Box\phi_1)$ and $\tr(\phi_1 \partial^\mu \phi_1\partial_\mu \phi_1 )$,  and those built from five scalar fields.
Making use of the equations of motion in the theory (\ref{bi-scalarL}) and discarding operators with total derivatives, we can express the former operators
in terms of the latter. The relevant operators are of the following form
$\tr(\phi_1^3\phi_2\phi_2^\dagger)$, $\tr(\phi_1^2\phi_2\phi_1\phi_2^\dagger)$,
$\tr(\phi_1\phi_2\phi_1^2\phi_2^\dagger)$ and $\tr(\phi_2\phi_1^3\phi_2^\dagger)$. Examining their mixing matrix, we find that,
due to a particular form of the interaction term in (\ref{bi-scalarL}), these operators form the so called logarithmic multiplet, typical for a non-unitary theory \cite{Gainutdinov2013,Caetano:TBP}. As
a result, for $\Delta(0)=5$ there are only two unprotected conformal operators. We determine their scaling dimension for an arbitrary coupling $\xi$. The two other operators form as logarithmic multiplet leading to a logarithmic factor in their correlation functions.
For $\Delta(0)\ge 7$, the basis of the operators (\ref{O-def}) contains operators with derivatives and the spectrum of their scaling dimensions has a more complicated
structure. In particular, in virtue of integrability, the spectrum contains pairs of operators with coinciding scaling dimensions.
We argue that for each odd $\Delta(0)\ge 7$ there are two operators that are not degenerate. We determine the scaling dimensions of such operators for an arbitrary coupling.


To compute the scaling dimensions of the operators (\ref{O-def}), we employ the method of  quantum spectral curve (QSC). This method has been
previously used to compute the  anomalous dimensions of  local single-trace operators in  planar  \(\CN=4\) SYM theory \cite{Gromov2014,Gromov:2014caa}
and it can be generalized to twisted version of this theory \cite{Gromov:2015dfa,Kazakov:2015efa}. The twisted QSC method (tQSC) should be adopted for the
twisted \(\CN=4\) SYM theory in the double scaling limit, which is the principal technical problem we are confronted to. A similar problem 
has been recently solved in \cite{Gromov:2016rrp}  for the case of the cusped Wilson line in a very similar limit to the one leading to the bi-scalar model \eqref{bi-scalarL}. 
We will adopt the most important elements of the method of  \cite{Gromov:2016rrp}  for our current, more complicated case.

The tQSC method consists of two ingredients: i)~derivation of the 
Baxter equation for particular set of the operators (\ref{O-def}), ii)~derivation of the quantization condition  which fixes the dependence of the parameters of this
equation on the coupling constant $\xi$. This information is sufficient to determine the scaling dimension of the operators $\Delta=\Delta(\xi)$.
We solve  the problem i)  by applying the double expansion with respect to the parameters \(g\) and \(1/q_3\), of the coefficients of the most general Baxter equation derived within the QSC method  
in \cite{Alfimov:2014bwa}. 
To check this result, we use the fact that the same Baxter equation describes the conformal, \(SO(1,5)\sim SU(2,2)\) spin chain in principal series representation, for  a given set of Cartan charges \((\Delta,0,0)\) corresponding to the scalar operators (\ref{O-def}).
It is known to have a universal form of a fourth-order finite difference equation with the coefficients depending on the integrals of motion including $\Delta$. For the sake of consistency, we also re-derive  the Baxter equation directly from the equivalent conformal spin chain describing the dynamics of ``fishnet" diagrams. 
Moreover, we show that the same equation can be defived
by imposing a certain symmetry among its coefficients, inspired by algebraic properties of the conformal Lax operator~\cite{Derkachov2009,Derkachov2011} along with the \(u\to - u \) symmetry and large \(u\) asymptotics of its 4 solutions.  The problem ii), allowing us to fix the dependence of $\Delta $ and the remaining coefficients of Baxter equation on the
coupling constant, is solved by exploiting the explicit analyticity properties of the QSC equations \cite{Gromov2014,Gromov:2014caa,Kazakov:2015efa,Gromov:2015dfa}, including the Riemann-Hilbert relations completing the general QSC Baxter equation, in the form proposed in \cite{Gromov:2015wca}. The direct derivation of the same quantization conditions from the conformal $SU(2,2)$
spin chain is not yet available.


Let us summarize the main results of this paper.

We show that for nondegenerate operators mentioned above the Baxter equation possesses the additional symmetry under
the exchange of the spectral parameter $u\to -u$. This symmetry fixes the values of all but two integrals of motion of the spin chain
and leads to a remarkable factorization of the 4th order Baxter equation into two 2nd order finite-difference equations
of the form
\begin{equation}\label{Baxter24}
\left(\frac{(\Delta-1)(\Delta-3)}{4u^2}-\frac{\,m}{u^3}-2\right)q(u)+q(u+i)+q(u-i)=0\,.
\end{equation}
This equation depends on two integrals of motion, $\Delta$ and $m$, and coincides with the Baxter equation for the $SL(2)$
spin chain of length ${\L}=3$ and spin $0$ representation at each site.
The quantization condition which fixes the dependence of $\Delta$ and $m$ on the coupling constant
is
\begin{equation}
m^2=-\xi^6\,,\qquad\qquad
q_2(0,m)\,q_4(0,-m)\,+\,q_2(0,-m)\,q_4(0,m)=0 \,,
\label{quant24}
\end{equation}
where $q_2(u,m)$ and $q_4(u,m)$ denote special solutions to (\ref{Baxter24})
which have the following large \(u\) asymptotic expansions  (the reason to label them as  \(2\) and \(4\)
will be clear below)
\begin{align}\notag\label{pure-def}
{}& q_2(u,m)\sim u^{\Delta/2-1/2}\left(1+O({1}/{u})\right),\qquad
\\[2mm]
{}& q_4(u,m)\sim u^{-\Delta/2+3/2}\left(1+O({1}/{u})\right).
\end{align}
It is important here that the expansion on the right-hand side runs in powers of $1/u$ only and there is no admixture of
$O(u^{-\Delta/2+3/2})$ and  $O(u^{\Delta/2-1/2})$ terms in the first and in the second relations, respectively. That is the reason why
we refer to (\ref{pure-def}) as ``pure" solutions.

\begin{figure}
\includegraphics[scale=0.6]{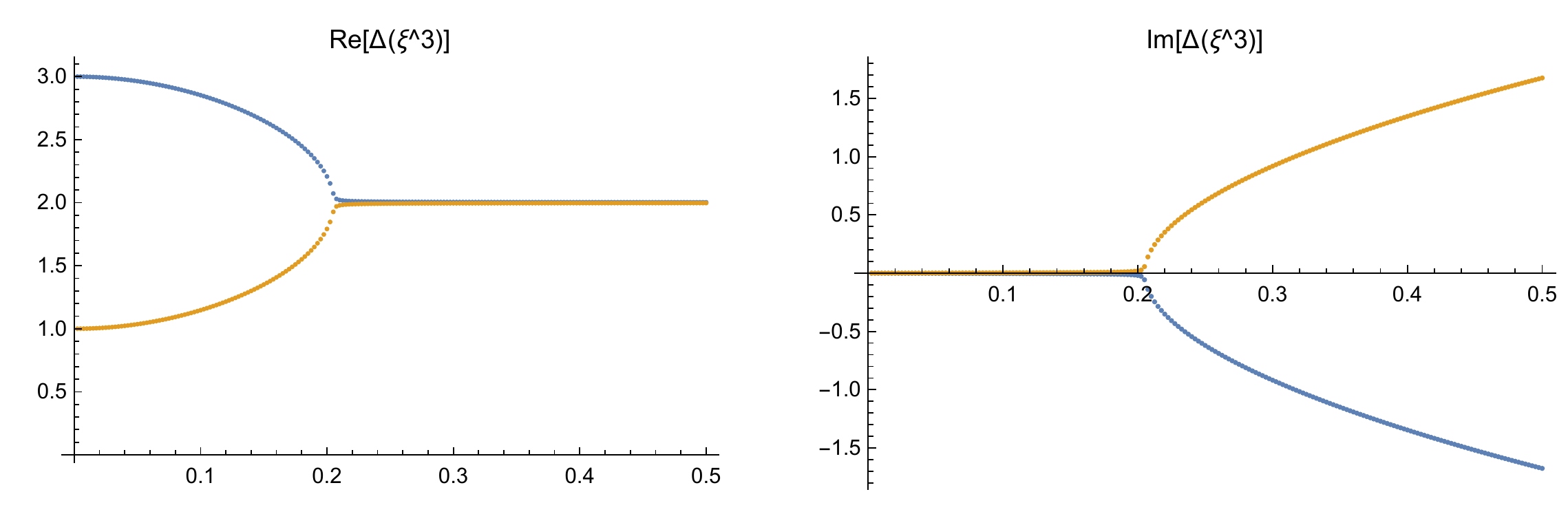}
\caption{Numerical results for  the scaling dimension of the operator  \(\tr(\phi_1^3)\) as a function of the coupling \(\xi^3\). We observe a "phase transition" at \(\xi^3\simeq 0.21\) where the scaling dimension takes the value $\Delta=2$ and becomes imaginary. This point defines the radius of convergency of the weak coupling expansion.  The second branch, starting from \(\Delta(0)=1\), arises due to the  symmetry of the Baxter equation (\ref{Baxter24}) under \(\Delta\to 4-\Delta\).
\label{fig:numericsL3}
}
\end{figure}

\begin{figure}
\center{\includegraphics[scale=0.4]{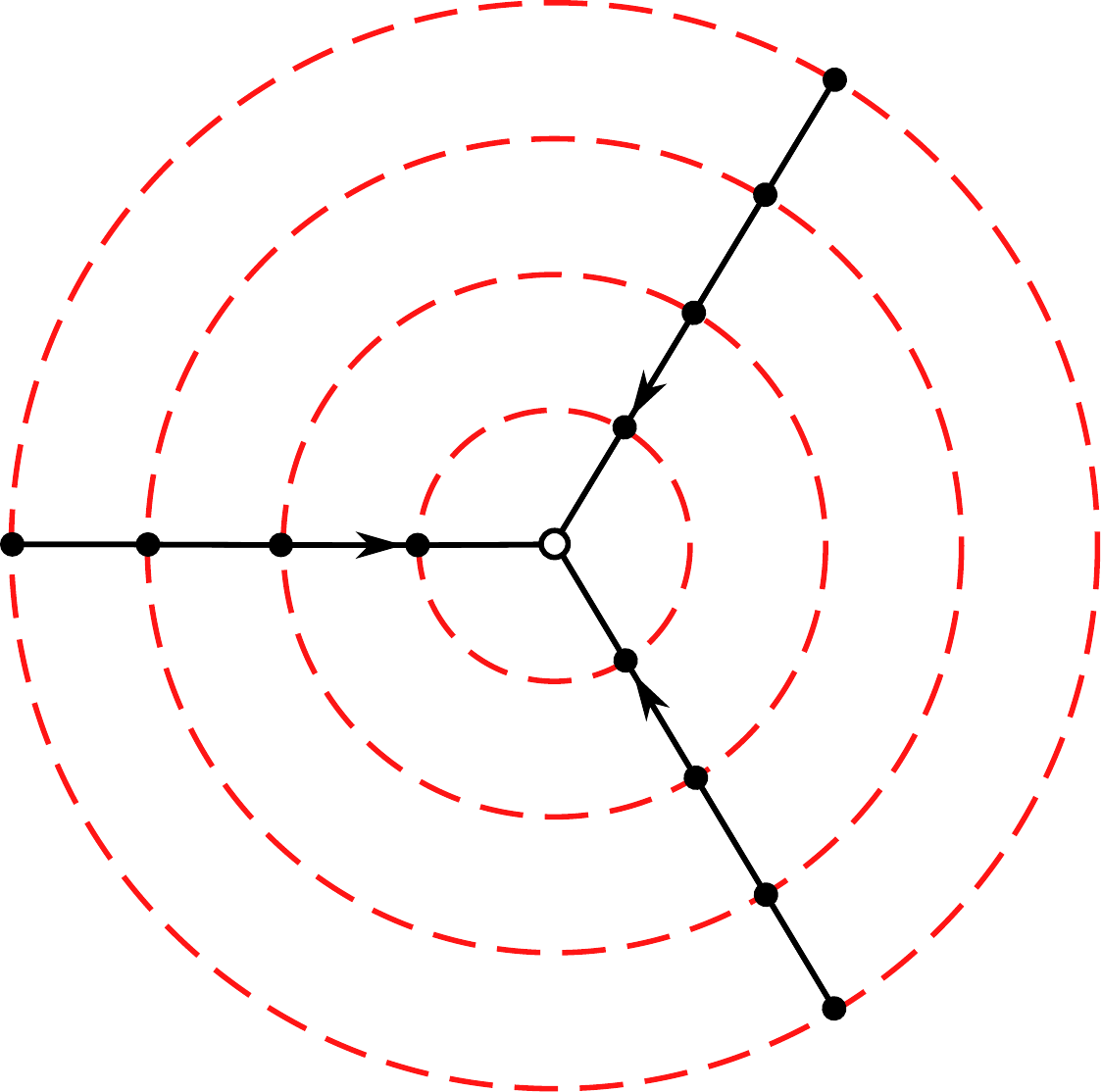}}
\caption{The ``wheel" Feynman graph corresponding to $O(\xi^{24})$ term in the weak coupling expansion \eqref{D3exp}
of anomalous dimension of the operator \(\Tr( \phi_1^3)\). \label{fig:L3graph}  }
\end{figure}

Having constructed the functions $q_2(u,m)$ and $q_4(u,m)$, satisfying (\ref{Baxter24}) and (\ref{pure-def}),  we can use (\ref{quant24}) to compute the scaling dimension $\Delta$ for any value of the coupling constant $\xi$.
This procedure can be carried out numerically, with practically unlimited precision following the general method \cite{Gromov:2015wca,Gromov:2016rrp}. As an example, we show in the
Fig.~\ref{fig:numericsL3} our results for the scaling dimension $\Delta_3(\xi)$ of the simplest operator (\ref{L-wheel}) for ${\L}=3$. At weak coupling, $\Delta_3(\xi)$ receives corrections from the wheel diagrams shown in Fig.~\ref{fig:L3graph}.
As a function of the coupling constant, $\Delta_3-2$ decreases with $\xi$ and turns from real to purely imaginary values at
$\xi=\xi_\star$ with
\(\xi_\star^3\simeq 0.2\), indicating that the weak coupling expansion has a finite radius of convergency. We find that 
$(\Delta_3(\xi)-2)^2$ is a smooth real  function of the coupling and, therefore, $\Delta_3(\xi)$ has a square-root singularity at $\xi=\xi_\star$.
For $\xi>\xi_\star$, the scaling dimension
takes the form $\Delta_3=2 + i d(\xi)$ where real valued function $d(\xi)$ monotonically increases with the coupling and scales as
$d \sim \xi^{3/2}$ for $\xi \to\infty$.

The relations (\ref{Baxter24}) -- (\ref{pure-def}) are invariant under $\Delta\to 4-\Delta$. Therefore, for any solution $\Delta(\xi)$
to the quantizaton condition  there should exist another one $4-\Delta(\xi)$. Then, the appearance of the imaginary part of $\Delta(\xi)$
can be interpreted as a result of the collision of two `energy levels' $\Delta(\xi)$ and $4-\Delta(\xi)$ at $\xi=\xi_\star$  (see Fig.~\ref{fig:numericsL3}).
Although the level crossing cannot happen in a unitary conformal field theory, it occurs in the theory (\ref{bi-scalarL}) since it is not unitary.

At weak coupling the quantization conditions (\ref{quant24}) can be solved analytically order-by-order in $\xi^6$.
The approach is algorithmic and is only limited by the computer power available. The first four terms of
the weak coupling expansion of $\Delta_3-3$  are
\footnote{Strictly speaking, expansion of the scaling dimension in the theory \eq{bi-scalarL} runs in powers of
the fine structure constant $\alpha = \xi^2/(4\pi^2)$. To simplify formulae we do not display $1/(4\pi^2)$ factor in
what follows. }
\begin{eqnarray}
&&\Delta_3-3=-12 \zeta _3\xi^6+ \xi ^{12}\left(189 \zeta _7-144 \zeta _3{}^2\right)
 \notag\\
\nn
&&+\xi ^{18} \left(-1944 \zeta _{8,2,1}-3024 \zeta
_3{}^3-3024 \zeta _5 \zeta _3{}^2+6804 \zeta _7 \zeta
_3+\frac{198 \pi ^8 \zeta _3}{175}+\frac{612 \pi ^6 \zeta
_5}{35}+270 \pi ^4 \zeta _7+5994 \pi ^2 \zeta _9-\frac{925911
\zeta _{11}}{8}\right)\\
\nn
&&+\xi ^{24} \left(-93312 \zeta _3 \zeta
_{8,2,1}+\frac{10368}{5} \pi ^4 \zeta _{8,2,1}+5184 \pi ^2
\zeta _{9,3,1}+51840 \pi ^2 \zeta _{10,2,1}-148716 \zeta
_{11,3,1}-1061910 \zeta _{12,2,1}\right.\\ \nn&&\left.+62208 \zeta
_{10,2,1,1,1}-77760 \zeta _3{}^4-145152 \zeta _5 \zeta
_3{}^3-\frac{576}{7} \pi ^6 \zeta _3{}^3-864 \pi ^4 \zeta _5
\zeta _3{}^2-2592 \pi ^2 \zeta _7 \zeta _3{}^2+244944 \zeta _7
\zeta _3{}^2\right.\\ \nn&&\left.+186588 \zeta _9 \zeta _3{}^2+\frac{9504}{175} \pi
^8 \zeta _3{}^2-2592 \pi ^2 \zeta _5{}^2 \zeta
_3+\frac{29376}{35} \pi ^6 \zeta _5 \zeta _3+298404 \zeta _5
\zeta _7 \zeta _3+12960 \pi ^4 \zeta _7 \zeta _3+287712 \pi ^2
\zeta _9 \zeta _3\right.\\ \nn&&\left.-5555466 \zeta _{11} \zeta _3+\frac{2910394
\pi ^{12} \zeta _3}{2627625}+57672 \zeta _5{}^3-71442 \zeta
_7{}^2+\frac{13953 \pi ^{10} \zeta _5}{1925}+\frac{7293 \pi ^8
\zeta _7}{175}-\frac{19959 \pi ^6 \zeta _9}{5}\right.\\ &&\left.+\frac{119979
\pi ^4 \zeta _{11}}{2}+\frac{10738413 \pi ^2 \zeta
_{13}}{2}-\frac{4607294013 \zeta _{15}}{80}\right)+O\left(\xi
^{30}\right)\,,
\label{D3exp}
\end{eqnarray}
where $\zeta _{i_1,\dots, i_k}=\sum_{n_1> \dots>n_k >0} 1/(n_1^{i_1}\dots  n_k^{i_k} )$ are  multiple zeta functions. 
Here the coefficients in front of \(\xi^{6M}\) give the residues
at simple pole in dimensionally regularized Feynman integrals corresponding to  the \({\L}=3\) wheel graphs with \(M=1,2,3,4\) frames (see Fig. \ref{fig:L3graph}).
The first two terms of (\ref{D3exp}) (up to double wrapping) coincide with  the known results \cite{Broadhurst:1985vq,Panzer:2013cha}.

At strong coupling, for $\xi\gg 1$, the scaling dimension $\Delta_3-2$ takes pure imaginary values that scale as $\xi^{3/2}$. In this limit, the quantization conditions
(\ref{quant24}) can be solved using semiclassical methods. We find that the strong coupling expansion runs in powers of $1/\xi^3$
and the first few terms are given by
\begin{eqnarray}\notag
\Delta_3-2=&&i\,2\sqrt{2}\,\xi ^{3/2} \left[1+\frac{1}{(16\xi^3)}+\frac{55}{2}\frac{1}{(16\xi^3)^2}+\frac{2537}{2}\frac{1}{(16\xi^3)^3}+\right.\\
&&\left. +\frac{830731 }{8}\frac{1}{(16\xi^3)^4}+\frac{98920663}{8}\frac{1}{(16\xi^3)^5}+\frac{31690179795}{16}\frac{1}{(16\xi^3)^6} +{\cal O}(\xi^{-21})\right]\;.
\end{eqnarray}
Notice that the expansion coefficients grow factorially indicating that the series is not Borel summable. In a close analogy with
the strong coupling expansion of the cusp anomalous dimension in planar $\mathcal N=4$ SYM \cite{Basso:2009gh}, this suggests that $\Delta_3$
receives at strong coupling exponentially small, nonperturbative corrections of the form $e^{-c\xi^3}$.

The same technique can be applied to compute the scaling dimensions of the operators (\ref{O-def}) with the same $R-$charge
${\L}=3$ and $\Delta(0)=5,7,\dots$ in a free theory. As explained above, for $\Delta(0)=5$ we encounter two unprotected operators.
We present the numerical and analytic study of quantum corrections to their scaling dimensions both at weak and at strong
coupling. We show that the resulting expressions have a different behaviour with the coupling constant as compared with
$\Delta_3$ (see Fig.~\ref{fig:numericsL3}). Namely, they acquire an imaginary part  at weak coupling and increase monotonically
with the coupling. For $\Delta(0)\ge 7$ we compute the scaling dimensions of a pair of nondegenerate operators, one for each $\Delta(0)$. We demonstrate
that for $\Delta(0)=7, 11, 15,\dots$ and $\Delta(0)=9,13,17,\dots$ the dependence of the scaling dimensions of the coupling follows the same
pattern as $\Delta_3(\xi)$ and  $\Delta_5(\xi)$, respectively (see Fig.~\ref{Fig:res} below).
 The limit of large length  \(\xi\)
for such operators will be also treated and the explicit formula for the dimension  will be found in the form of a Bohr-Sommerfeld type quantisation condition.


\section{$SU(2,2)$ picture for fishnet graphs}
\label{sec:conf_spin_chain_picture}

To prepare the formalism for computing 
the wheel-graphs  shown in Fig.~\ref{fig:multiwheel}, we demonstrate in this section that these graphs  can be identified as
transfer matrices of an integrable conformal \(SU(2,2)\) spin chain.

\begin{figure}[t]
\center{\includegraphics[scale=1]{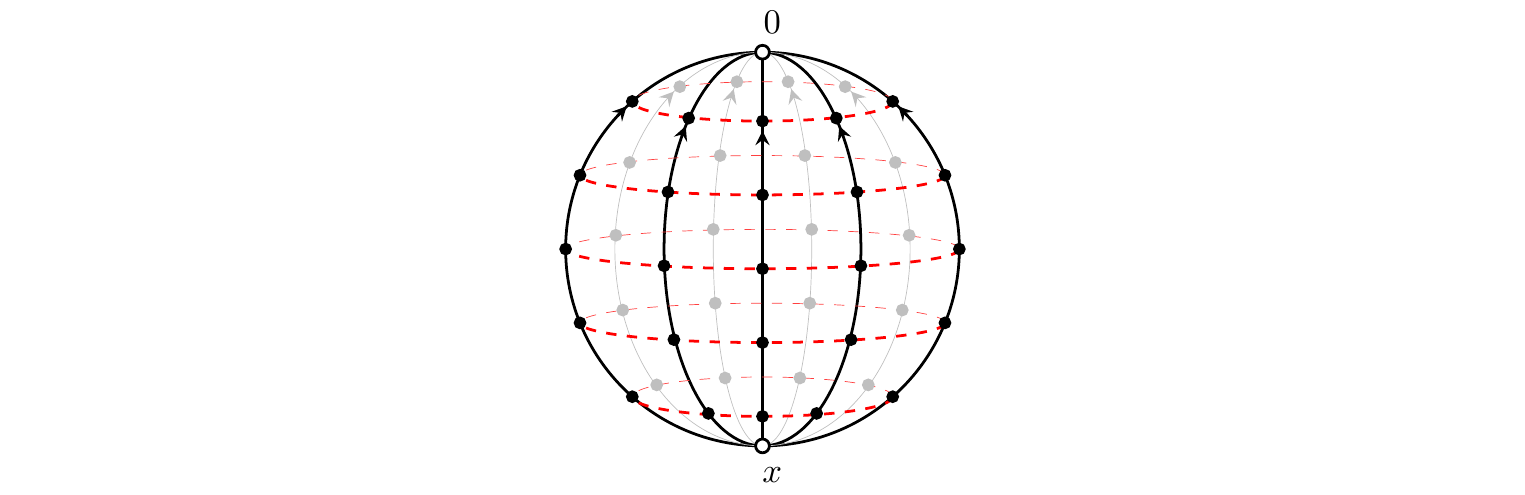}}
\caption{A ``wheel" Feynman graph with $M$ frames and ${\L}$ spokes whose external legs have been joined at the point $x$ ($M=5$ and ${\L}=9$ in this picture).}
\label{fig:globe_graph}
\end{figure}

To begin with, we consider the two-point correlation function of  the operators (\ref{L-wheel})
\begin{align}\label{D2}
D(x)=\langle \mathcal O_{\L}(x) \bar{\mathcal O}_{\L}(0) \rangle =  {d_{\L}(\xi)\over (x^2)^{{\L}+\gamma_{\L}(\xi)}}\,,
\end{align}
where the normalization constant $d_{\L}(\xi)$ and the anomalous dimension $\gamma_{\L}(\xi)$  depend on the coupling
constant and the length \(L=J\) of the operator. As was shown  \cite{Gurdogan:2015csr}, in the planar limit this correlation function
receives contribution from globe-like graphs shown in Fig.~\ref{fig:globe_graph}.
The same graphs can be viewed as fishnet graphs with periodic boundary conditions along horizontal (longitudinal) direction and all lines in the vertical (latitudinal)
direction joined at the points $x$ and $0$, respectively, as depicted in Fig.~\ref{fig:globe_graph}. The contribution of such graphs to (\ref{D2}) can be written as
\begin{align}\label{D-repr}
D(x) = \sum_{M=0}^\infty  \xi^{2{\L}M}\int   \prod_{i=1}^{\L} {d^4 y_i \over y_i^2} \mathcal T_{{\L},M} (x_1,\dots,x_{\L}|y_1,\dots,y_{\L})\bigg|_{x_1=\dots=x_{\L}=x} \,,
\end{align}
where $\mathcal T_{{\L},M} (x_1,\dots,x_{\L}|y_1,\dots,y_{\L})$ describes a cylinder-like fishnet graph with $M$ wheels and ${\L}$ external
points. For $M=0$ we have
\begin{align}
\mathcal T_{{\L},0}(x_1,\dots,x_{\L}|y_1,\dots,y_{\L}) = \prod_i \delta^{(4)}(x_i-y_i)\,,
\end{align}
so that the corresponding contribution to (\ref{D-repr}) is given by the product of ${\L}$ scalar propagators. For $M\ge 1$,
due to the structure of wheel graphs, $\mathcal T_{{\L},M}$ admits the following recursive representation
\begin{align}\notag\label{TN}
\mathcal T_{{\L},M} (x_1,\dots,x_{\L}|y_1,\dots,y_{\L}) {}&= \int \prod_{i=1}^{\L} d^4 z_i \,\mathcal H_{\L}(x_1,\dots,x_{\L}|z_1,\dots,z_{\L})
\mathcal T_{{\L},M-1} (z_1,\dots,z_{\L}|y_1,\dots,y_{\L})
\\
{}& \equiv  \mathcal H_{\L}\circ \mathcal T_{{\L},M-1}(x_1,\dots,x_{\L}|y_1,\dots,y_{\L}) \,,
\end{align}
where the kernel $\mathcal H_{\L}$ represents one wheel of the diagram in Fig.~\ref{fig:multiwheel}. It is given by the
product of scalar propagators
\begin{align}\notag\label{H-L}
\mathcal H_{\L}(x_1,\dots,x_{\L}|y_1,\dots,y_{\L}) {}&=
\underset{\hskip0.5cm y_1\, y_2\, y_3\,y_4\,\dots\,\dots\dots y_{\L}\,y_{{\L}+1}}{\overset{x_1\, x_2\, x_3\,x_4\,\dots\dots\dots\,x_{\L}}{\mathlarger{\mathlarger{\perp\!\perp\!\perp\!\perp\cdots\perp}}}},
\\
{}& = {1\over (2\pi)^{4{\L}}} \prod_{i=1}^{\L} {1\over (x_i-y_i)^2(y_i-y_{i+1})^2 }\,,
\end{align}
with periodic boundary condition $y_1=y_{{\L}+1}$.   Here each horizontal link is a free scalar propagator   \((2\pi)^{-2}(y_j-y_{j+1})^{-2}\). Each vertical line  produces similar propagator connecting points $x_i$ and $y_i$. Replacing
the scalar propagator with inverse d'Alembert operator,   \( \Box_i^{-1}(x_i,y_i) = (2\pi)^{-2}/(x_i-y_i)^{2}\), we find
that (\ref{H-L}) can be identified as a kernel of the following ``graph building" operator defined in~\cite{Gurdogan:2015csr}
\begin{align}\label{HL-op}
\mathcal H_{\L} ={1\over (2\pi)^{2{\L}}}  \prod_{i=1}^{\L} \Box_i^{-1} \prod_{i=1}^{\L} {1\over (x_i-x_{i+1})^2}\,,
\end{align}
with $x_{\L+1}=x_1$.
Applying (\ref{TN}) we obtain the following concise operator representation for fishnet with $M$ rows
\begin{align}\label{T-H-rel}
\mathcal T_{{\L},M} = \mathcal H_{\L} \circ  \mathcal H_{\L} \circ \dots \circ  \mathcal H_{\L} \equiv (\mathcal H_{\L})^M \,.
\end{align}
Here the operator $\mathcal H_{\L}$ adds an extra row to fishnet graph and `$\circ$' denotes the convolution of the kernel (\ref{H-L}). As we show in the next subsection,  $\mathcal H_{\L}$ can
be identified as a transfer matrix of the non-compact Heisenberg spin chain of length ${\L}$ with spins being the generators of
the conformal group \(SU(2,2)\).

We expect that the anomalous dimension in (\ref{D2}) is different from zero. This means that the correlation function (\ref{D-repr})
has to develop ultraviolet divergences and, therefore, requires a regularization. If we introduced dimensional regularization with
$D=4-2\epsilon$,
these divergences would appear as poles of (\ref{D-repr}) in $1/\epsilon$.
Notice that for arbitrary $x_i$ and $y_i$, the function $\mathcal T_{{\L},M} (x_1,\dots,x_{\L}|y_1,\dots,y_{\L}) $ is well-defined in $D=4$ dimensions and does not require any regularization. The divergences appear
in (\ref{D-repr}) when we identify the points $y_1=\dots=y_{\L}$ and/or integrate over $x_i\to x$. In both cases they come from integration
in (\ref{TN}) in the vicinity of the two points $z_i=0$ and $z_i=x$ and have a clear UV origin.
To lowest order $O(\xi^{2{\L}})$, the dimensionally regularized integral in (\ref{D-repr}) has simple pole $1/\epsilon$ whose residue
gives the anomalous dimension. At high orders, the integrals entering the two-point correlation function (\ref{D-repr}) have overlapping
divergences and, as a consequence, they produce higher power of $1/\epsilon$. The divergent part of (\ref{D2})
has the following form
\begin{align}\label{D-pole}
D(x) = e^{-\gamma_{\L}(\xi)/\epsilon + O(\epsilon^0)}\,.
\end{align}
The question arises how to use a four-dimensional representation  (\ref{D-repr}) to extract correctly the divergent
part of the correlation function (\ref{D-pole}). To this end we notice that, in distinction from $D(x)$, its logarithm
has at a most simple pole $\ln D(x) \sim -\gamma_{\L}(\xi)/\epsilon$ and, therefore, does not contain overlapping divergences\footnote{Except the case \(J=2\) which  has overlapping singularities and leads to  the leading asymptotics \(\ln D(x) \sim -1/\epsilon^2\)~ \cite{Sieg:2016vap,Gurdogan:2015csr}.}.
The simple pole of  $\ln D(x)$ originate from the two integration regions in which all internal vertices approach one of the
external points, $z_i=x$ or $z_i=0$. Then, applying the dilatation operator to $\ln D(x)$, we can remove its divergent
part and evaluated the anomalous dimension
$\gamma_{\L}(\xi) = - [(x\partial_x)/2 + {\L}] \ln D(x)$ using the four-dimensional representation (\ref{D-repr}).

\subsection{``Graph-building" transfer matrix and its integrable $SO(1,5)$ spin chain realization}
\label{subsec:graph_build_transf_mat}

Let us show that the integral operator $\mathcal H_{\L}$ defined in (\ref{TN}) and (\ref{H-L}) commutes with the generators of the conformal $SO(1,5)$ group
\begin{align}\label{conf}
[ G, \mathcal H_{\L} ] =0\,,\qquad G  = \{ P_\mu, \,  M_{\mu\nu}, \,  D, \,  K_\mu \}\,,
\end{align}
where $G=\sum_{i=1}^{\L} G_{i}$ are given by the sum of differential operators acting on points $x_1,\dots,x_{\L}$
\begin{align}\label{gen}\notag
{}& P_\mu = \partial_\mu \,, && M_{\mu\nu} = x_\mu \partial_\nu - x_\nu \partial_\mu \,,\qquad
\\[2mm]
{}& D= (x\partial) + 1\,, && K_\mu = 2x_\mu (x\partial) - x^2 \partial_\mu + 2 x_\mu\,.
\end{align}
We recall that general expressions for the generators of the conformal group depend on the parameters $(\Delta,S,\dot S)$
defining the scaling dimension and the Lorentz spin.
The relations (\ref{gen}) correspond to a  free scalar field  representation of the conformal group
 with $\Delta=1$ and $S=\dot S=0$.

To verify (\ref{conf}) we apply its both sides to a test function $\Phi(\boldsymbol{x})$ depending on the set of coordinates
$\boldsymbol{x}=(x_1,\dots,x_{\L})$
\begin{align}\label{conf1}
\int d^4 \boldsymbol{y} \, \left[ G_x\mathcal H_{\L} (\boldsymbol{x}|\boldsymbol{y}) \Phi(\boldsymbol{y}) -   \mathcal H_{\L} (\boldsymbol{x}|\boldsymbol{y}) G_y \Phi(\boldsymbol{y})\right]
=   \int d^4 \boldsymbol{y} \, \left[ G_x\mathcal H_{\L} (\boldsymbol{x}|\boldsymbol{y}) - G_y^\dagger   \mathcal H_{\L} (\boldsymbol{x}|\boldsymbol{y})\right]   \Phi(\boldsymbol{y}) = 0\,,
\end{align}
where we used a shorthand notation for $d\boldsymbol{y}=dy_1 \dots dy_{\L}$ and introduced subscript to indicate that $G_x$ acts on $x$ variables.
Here in the second relation we integrated by parts and introduced notation for the conjugated generators $G_y^\dagger$. For instance,
$P_\mu^\dagger = - P_\mu$ and $D^\dagger = - (x\partial)-3$. As follows from (\ref{conf1}), the conformal symmetry implies that
the kernel of the operator $\mathcal H_{\L}$ has to satisfy the following relation
\begin{align}\label{diff}
\left(G_x  - G_y^\dagger \right)  \mathcal H_{\L} (\boldsymbol{x}|\boldsymbol{y}) = 0\,.
\end{align}
It is then straightforward to check that the kernel (\ref{H-L}) verifies this relation indeed.

\begin{figure}[t!bp]
\centerline{
\includegraphics[width = 0.37\textwidth]{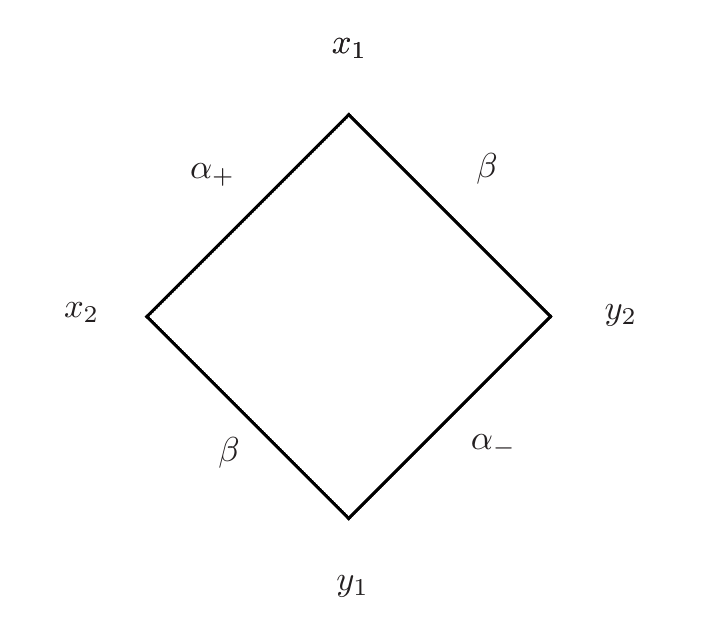}}
\caption{Diagrammatic representation of the $R-$operator (\ref{R}). Solid line with index $\alpha$ stands for $1/(x^2)^\alpha$.
The values of indices depend on the spectral parameter and are given by $\alpha_+=-1-u$, $\alpha_-=1-u$ and $\beta=2+u$. }
\label{fig:R}
\end{figure}

Having identified the representation of the conformal group (\ref{gen}), we can now construct the Heisenberg spin chain with the spin generators given by (\ref{gen}) and establish the relation between $\mathcal H_{\L}$, Eq.~(\ref{HL-op}), and transfer matrices of this model.
As a first step, we define the $R-$operator acting on the tensor product of two representations
(\ref{gen}). 
Similar to (\ref{H-L}) it can be realized as an integral operator \cite{Derkachov:2001yn}
\begin{align}\label{R12}
R_{12}(u) \, \Phi(x_1,x_2) = \int d^4 y_1 d^4 y_2\, R_u(x_1,x_2|y_1,y_2) \Phi(y_1,y_2) \,,
\end{align}
with $\Phi(x_1,x_2)$ being a test function. The requirement for the operator $R_{12}(u)$ to satisfy the Yang-Baxter equation  leads to a differential
equation for  the kernel $R_u(x_1,x_2|y_1,y_2)$. Its solution is given by \cite{Chicherin:2012yn}
\begin{align}\label{R}
R_u(x_1,x_2|y_1,y_2) =    {c(u) \over [(x_1-x_2)^2]^{-u-1} [(x_1-y_2)^2 (x_2-y_1)^2]^{u+2} [(y_1-y_2)^2]^{-u+1} }\,,
\end{align}
where  $c(u)$ is a  normalization factor
\begin{align}\label{c-cons}
c(u) = {2^{4u}\over \pi^4} {\Gamma^2(u+2)\over \Gamma^2(-u)} \,.
\end{align}
Its value was fixed by imposing the so-called $T-$inversion relation
$
R_{12}(u) R_{12}(-u) = \mathbf{1}
$, or equivalently
\begin{align}
\int d^4 y_1 d^4 y_2\, R_u(x_1,x_2|y_1,y_2)R_{-u}(y_1,y_2|z_1,z_2) = \delta^{(4)}(x_1-z_1)\delta^{(4)}(x_2-z_2)\,.
\end{align}
It proves convenient to use a diagrammatic representation for the $R-$operator (\ref{R12}) and (\ref{R}).
Representing each factor of the form $1/[(x-y)^2]^\alpha$  as a solid line connecting points $x$ and $y$ with index $\alpha$ attached to it, we can
depict (\ref{R}) as a rectangular graph shown in Fig.~\ref{fig:R}. Notice that the indices of all four lines depend on the spectral parameter.

Let us examine (\ref{R12}) for $u\to 0$. Putting $u=-\epsilon$ in (\ref{R}) and (\ref{R12}) we find for $\epsilon\to 0$
\begin{align} \label{R(0)}
R_{12}(\epsilon)   \Phi(x_1,x_2) {}&= {\epsilon^2\over \pi^4}  \int {(x_1-x_2)^2\, d^4 y_1 d^4 y_2 \,\Phi(y_1,y_2)  \over  [(x_1-y_2)^2]^{2-\epsilon}[(x_2-y_1)^2]^{2-\epsilon} (y_1-y_2)^2 } \,,
\end{align}
where we replaced $c(\epsilon)$ by its leading asymptotic behaviour. For the expression on the right-hand side to be different from zero,
the integral should produce a double pole $1/\epsilon^2$. Indeed, making use of the identity
\begin{align}\label{id}
\lim_{\epsilon\to 0} {\epsilon \over \pi^2 [(x-y)^2]^{2-\epsilon} } =  \delta^{(4)}(x-y)\,,
\end{align}
we find that integration in (\ref{R(0)}) yields $R_{12}(0) \Phi(x_1,x_2) = \Phi(x_2,x_1)$, so that $R_{12}(0)$ coincides with the permutation operator,
\begin{align}\label{P-12}
R_{12}(0)=P_{12}\,.
\end{align}
The same property can be easily found from the diagrammatic representation of the $R-$operator (see Fig.~\ref{fig:R}). The operator $R_{12}(0)$
is described by the diagram in which two lines carry index $\beta=2$.
According to (\ref{id}), such lines collapse into two points, $x_1=y_2$ and $x_2=y_1$, leading to (\ref{P-12}).

\begin{figure}[t!bp]
\centerline{
\includegraphics[width = \textwidth]{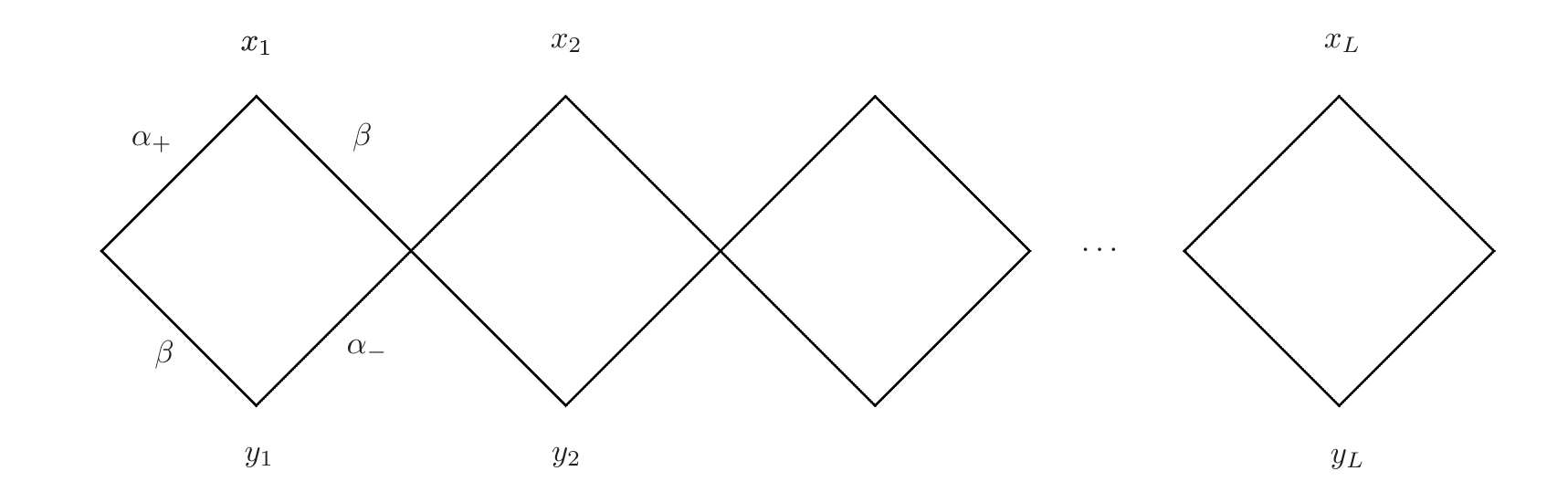}}
\caption{Diagrammatic representation of the transfer matrix (\ref{T-kernel1}).
It is obtained by gluing together ${\L}$ rectangles, each representing the $R-$operator. The leftmost and rightmost vertices
are identified. The values of indices $\alpha_\pm$ and $\beta$ are the same as in Fig.~\ref{fig:R}.
}
\label{fig:T}
\end{figure}

We can now use the $R-$operator (\ref{R12}) to construct the so-called fundamental transfer matrix
\begin{align}\label{T}
T_{\L}(u) = \tr_0[ R_{10}(u) R_{20}(u) \dots R_{{\L}0}(u)]\,.
\end{align}
It acts on the tensor product of ${\L}$ copies of the conformal group representation (\ref{gen}) and the trace is taken over the same
(infinite-dimensional) auxiliary space representation. Replacing each $R-$operator in (\ref{T}) by its integral representation (\ref{R12})
we can realize $T_{\L}(u)$ as an integral operator
\begin{align}\label{T-kernel}
T_{\L}(u)  \Phi(x_1,\dots,x_{\L}) = \int d^4 y_1 \dots d^4 y_{\L}\, T_{{\L},u}(\boldsymbol{x}|\boldsymbol{y})\Phi(y_1,\dots,y_{\L})\,,
\end{align}
where $\Phi$ is a test function and  the kernel $T_{{\L},u}(\boldsymbol{x}|\boldsymbol{y})$ is given by a ${\L}-$fold integral
over the auxiliary space
\begin{align}\label{T-kernel1}
T_{{\L},u}(\boldsymbol{x}|\boldsymbol{y}) = \int d^4 x_0 d^4 y_0   d^4 z_0  \dots d^4 w_0\, R_u(x_1,x_0|y_1,y_0) R_u(x_2,y_0|y_2,z_0)
\dots R_u(x_{\L},w_0|y_{\L},x_0)\,.
\end{align}
Using diagrammatic form of the $R-$operator, we can represent this integral in the form of ${\L}$ rectangles glued together through common vertices
in a pairwise manner as shown in Fig.~\ref{fig:T}. In the standard manner, in virtue of Yang-Baxter equation, the transfer matrices (\ref{T}) commute among themselves
\begin{align}
[T_{\L}(u), T_{\L}(v)]=0\,,
\end{align}
as well as with the local integrals of motion of the spin chain.

Let us examine (\ref{T-kernel}) and (\ref{T-kernel1}) for two special values of the spectral parameter, $u\to 0$ and $u\to -1$. In both
cases, some of the propagators in the integral representation of the transfer matrix have index $2$ which allows us to apply
the identity (\ref{id}). For $u=-\epsilon$ and $\epsilon\to 0$ we find
\begin{align}
T_{{\L},u=0}(\boldsymbol{x}|\boldsymbol{y}) = \prod_{i=1}^{\L} \delta^{(4)}(x_i - y_{i+1})\,,
\end{align}
or equivalently $T_{\L}(0)$ coincides with the operator of the cyclic shift.
The same relation can be also obtained using (\ref{P-12}).
For $u=-1+\epsilon$ and $\epsilon\to 0$ we find in the similar manner
\begin{align}
T_{{\L},u=-1+\epsilon}(\boldsymbol{x}|\boldsymbol{y}) =  &{}  {1\over (16\pi^2\epsilon)^{\L}} \prod_{i=1}^{\L}
{1\over  (x_i-y_i)^2 (y_{i}-y_{i+1})^2  }\,,
\end{align}
where $y_{\L+1}=y_1$ and the additional power of $1/\epsilon$ appears due to the fact that $c(-1+\epsilon)$ defined in (\ref{c-cons}) is regular for $\epsilon\to 0$.
Substituting this relation into (\ref{T-kernel}) we obtain the following result for the transfer matrix
\begin{align}
T_{{\L}}(-1+\epsilon) =  {1\over (4\epsilon)^{\L}}    {1\over \Box_1 \dots \Box_{\L}} \prod_{i=1}^{\L}
{1 \over (x_{i}-x_{i+1})^2  }\sim {1\over \epsilon^{\L}} \mathcal H_{\L} \,.
\end{align}
Comparing this relation with (\ref{HL-op}) we conclude that the operator $\mathcal H_{\L}$ defines the leading
asymptotic behaviour of the transfer matrix $T_{{\L}}(-1+\epsilon)$ for
$\epsilon\to 0$. Since the transfer matrix commutes with the local integrals of motion for an arbitrary spectral parameter,
the same is true for the operator $\mathcal H_{\L}$.

Having established integrability of the operator $\mathcal H_{\L}$, we can now turn to computing the correlation function
(\ref{D-repr}) and (\ref{T-H-rel}). A direct calculation of (\ref{D-repr}) would require diagonalizing the operator $\mathcal H_{\L}$
with subsequent summation over the whole spectrum of its eigenstates.
Since the underlying Heisenberg spin chain is defined for an infinite-dimensional representation of the conformal group,
this proves to be an extremely nontrivial task. In the next section, we present another approach based on the Baxter
$TQ$ equation, combined with the QSC formalism, that allows us to avoid these
difficulties and obtain a compact representation for the anomalous dimension $\gamma_{\L}(\xi)$.


\section{The Baxter $TQ$ equation}\label{sec:BaxterBootstrap}

We have shown in the previous section that the corrections to the scaling dimension of the scalar operators (\ref{L-wheel}) can be
described in terms of the Heisenberg spin chain with the spin operators being the generators of the conformal group. In this section,
we derive the Baxter $TQ$ equation in this integrable model for the states  corresponding
to a special class of scalar operators defined in (\ref{O-def}).



The Baxter equation is a fourth-order finite difference equation for the function $q(u)$ with the coefficients
given by transfer matrices of \(SU(2,2)\) spin chain in specific, antisymmetric representations. We argue below that, for some states with  the  charges \((\Delta,0,0|J,0)\), the
general form of this equation can be determined, up to a few constants (to be fixed by additional quantization conditions, see section~\ref{sec:quantization}),  from the known asymptotic behaviour of four solutions for the Baxter function $q(u)$ at infinity
combined with the symmetry under parity transformation $u\to -u$.   We partially justify these
assumptions in  Appendix~\ref{subsec:Baxter_from_lax} by making use of the Lax operator formalism.


It was shown  in the paper \cite{Derkachov2011} (see eq.(5.5) there) that the Baxter TQ-equation for \(SL(4)\) takes a standard form of a linear 4th order finite difference equation \eq{eq:Baxter_EQ_gen_form} on Q-functions, with coefficients being the fundamental transfer-matrices of the \(SL(4)\) spin chain with spins in principal series representations.
  In Appendix~\ref{subsec:Baxter_from_lax} we use this formula and  \(u\leftrightarrow -u\) symmetry arguments for the states with the  charges \((\Delta,0,0|J,0)\) to bring the Baxter equation to the
following symmetric  form
\begin{equation}
A\left(u+i\right)q\left(u+2i\right)-B\left(u+\frac{i}{2}\right)q\left(u+i\right)+C\left(u\right)q\left(u\right)-B\left(u-\frac{i}{2}\right)q\left(u-i\right)+A\left(u-i\right)q\left(u-2i\right)=0\;,
\label{eq:symmetricform}
\end{equation}
where
\(A(u)=u^{\L}\) is completely fixed by the choice of the representation (\ref{gen}), whereas
\(B(u) \)  and \(u^{\L}C(u)\) are polynomials in $u$ of degree \({\L}\) and \(2{\L}\), respectively. The coefficients \(A(u)\) \(B(u)\) and \(C(u)\) are explicitly calculated in Appendix~\ref{subsec:Baxter_from_lax}  in terms of global charges of the state  and they contain a certain number  of state-dependent constants, to be defined later by quantization conditions proper to our problem.   To fix the form of polynomials \(A(u),B(u),C(u)\) it appears to be enough to impose two
additional conditions:  if  \(q(u)\) is a real solution to \eq{eq:symmetricform} then  \(q(-u)\) should also be a solution. This gives the parity condition on the coefficient functions:
\begin{equation}\label{Csymmetry}
B(u)=(-1)^{\L} B(-u)\,,\qquad  C(u)=(-1)^{\L}C(-u).
\end{equation}
Secondly, for large $u$, the solution to \eq{eq:symmetricform} should have  asymptotic behavior $q\sim u^\delta$ with
the exponent $\delta$ taking the following values~(see next section or the twisted QSC formalism of~ \cite{Kazakov:2015efa})
\begin{equation}
\delta=\left\{\frac{\Delta-J }{2},\,\frac{\Delta-J }{2} +1,\, 2-\frac{\Delta +J}{2} ,\, 3-\frac{\Delta +J}{2}\right\}. \label{eq:q_asympt}
\end{equation}
These conditions fix the functions $B(u)$ and $C(u)$ to be
\begin{align}\notag\label{BC}
{}& B(u) = 4u^J-\frac{1}{2}(\alpha+3 J-4)u^{J-2}+bu^{J-4}+\sum_{k=3}^{[J/2]} d_ku^{J-2k}\,,
\\
& C(u) =  6u^J-(\alpha+3 J-4)u^{J-2}+\frac{(\alpha -4)^2+32 b+3 J^2+2 (\alpha -7) J}{16}u^{J-4}+\sum_{k=3}^{J}c_ku^{J-2k}\,,
\end{align}
where  \(\alpha=(\Delta-2)^2\). These expressions depend on \(1+(J-2)+([J/2]-2)=J+[J/2]-3 \)  arbitrary constants
\(b, c_k,d_k\),  to be fixed  by the additional quantization conditions.


Let us examine the relations (\ref{eq:symmetricform}) and (\ref{BC}) for $J=2,3,4$.

\paragraph{Baxter equation for \({\L}=2\)}

\begin{align}
&\left(u+i\right)^{2}q\left(u+2i\right)+\left(u-i\right)^{2}q\left(u-2i\right) +\left[6u^{2}-\alpha-2+ \frac{\alpha(\alpha-4)}{16\,u^{2}}\right]q\left(u\right) = \nonumber \\
&=\left[4\left(u+\frac{i}{2}\right)^{2}-\frac{\alpha+2}{2}\right]q\left(u+i\right)+\left[4\left(u-\frac{i}{2}\right)^{2}-\frac{\alpha+2}{2}\right]q\left(u-i\right)\,,
\label{L2BaxterLax}
\end{align}
where all constants are fixed in terms of \(\alpha=(\Delta-2)^2\).
We  notice that this equation factorizes as
\beq\label{L2BaxterLax-fac}
\left[\left(\frac{(\Delta-2)(\Delta-4)}{4u^2}-2\right)+D+D^{-1}\right]u^2\left[\left(\frac{\Delta(\Delta-2)}{4u^2}-2\right)+D+D^{-1}\right]q(u)=0\,,
\eeq
where \(D=e^{i\p_u}\) is the shift operator.

\paragraph{Baxter equation for \({\L}=3\)}

\begin{align}
&\left(u+i\right)^{3}q\left(u+2i\right)+\left(u-i\right)^{3}q\left(u-2i\right) +\left[6u^{3}-\left(\alpha+5\right)u+\frac{\left(\alpha-1\right)^{2}}{16u}+\frac{m^2}{u^{3}}\right]q\left(u\right) = \nonumber\\
&=\left(u+\frac{i}{2}\right)\left[4\left(u+\frac{i}{2}\right)^{2}-\frac{\alpha+5}{2}\right]q\left(u+i\right)+\left(u-\frac{i}{2}\right)\left[4\left(u-\frac{i}{2}\right)^{2}-\frac{\alpha+5}{2}\right]q\left(u-i\right) \;.
\label{L3BaxterLax}\end{align}
where \(\alpha=(\Delta-2)^2\) and $m$ is arbitrary. Remarkably, this equation also factorizes 
\begin{equation}
\left[\left(\frac{(\Delta-1)(\Delta-3)}{4u^2}+\frac{m}{u^3}-2\right)+D+D^{-1}\right]u^3\left[\left(\frac{(\Delta-1)(\Delta-3)}{4u^2}-\frac{m}{u^3}-2\right)+D+D^{-1}\right]q(u)=0\,.
\label{factorL3}\end{equation}
The values of the parameters \(m\)  and  \(\alpha=(\Delta-2)^2\) as functions of the coupling \(\xi\)
will be fixed  in section~\ref{sec:quantization} from an additional quantization condition.

\paragraph{Baxter equation for \({\L}=4\)}
\begin{align}\nonumber
&\left(u+i\right)^{4}q\left(u+2i\right)+\left(u-i\right)^{4}q\left(u-2i\right) +
\left[6u^{4}-\left(\alpha+8\right)u^{2}+ \left(2b+\frac{\alpha^2+8}{16}\right)+\frac{c_1}{u^2}+\frac{c_2}{u^4}\right]q\left(u\right) = \nonumber\\
&=\left[4\left(u+\frac{i}{2}\right)^4-\frac{\alpha+8}{2}\left(u+\frac{i}{2}\right)^{2}+b\right]q\left(u+i\right)
+\left[4\left(u-\frac{i}{2}\right)^4-\frac{\alpha+8}{2}\left(u-\frac{i}{2}\right)^{2}+b\right]q\left(u-i\right).
\label{L4BaxterLax}
\end{align}
It depends on 3 extra constants, \(b,c_1,c_2\), apart from \(\alpha=(\Delta-2)^2\), to be fixed from quantization conditions, yet to be derived.

The Baxter equations (\ref{L2BaxterLax}), (\ref{L3BaxterLax}) and (\ref{L4BaxterLax})  match perfectly the equations derived in detail for particular cases \({\L}=2,3,4\) in Appendix~\ref{subsec:Baxter_from_lax} from the Lax operator formalism and they will be confirmed in section~\ref{sec:BaxterfromQSC} from the QSC formalism.


\section{Baxter equations for  bi-scalar \chiFT  ${ }$ from the double scaling limit of  QSC}
\label{sec:BaxterfromQSC}

In this section we use an alternative integrability based approach which originates from ${\cal N}=4$ SYM. The integrability structure for the spectrum in this theory is very well studied and is given in terms of so-called Quantum Spectral Curve (QSC) \cite{Gromov2014,Gromov:2014caa}. 
We will use that fact that \chiFT{} can be obtained as a limit of a twisted version of ${\cal N}=4$ SYM in order to get further clues about the spectrum of anomalous dimensions and integrability structure of \chiFT. In particular in this section we demonstrate how the Baxter equations obtained from the integrability of the Feynman graphs can be obtained directly from QSC establishing an important link between these two seemingly different integrability based approaches. We will see that QSC in addition to the Baxter equation also provides an essential missing ingredient -- the quantization condition for the spectrum, which we were not able to obtain from the Feynman graphs.
The quantization condition is discussed in Section \ref{sec:quantization}. In this section we describe how the Baxter equation arises from QSC approach in the double scaling limit where the twisted SYM reduces to the scalar \chiFT. 


\subsection{QSC generalities}
The description of QSC and the notations closely follow \cite{Gromov:2014caa,Gromov:2015dfa,Kazakov:2015efa}.
In formal terms, the QSC is represented by a Grassmannian consisting of  $2^8$ so-called Q-functions of spectral parameter \(u\) related to each other by Pl\"ucker QQ-relations, in such a way that  8 of them are enough to parameterize all other.  General Q-functions can have quite complicated analytical structure as functions of the spectral parameter. In practice it is sufficient to restrict ourselves to a  subset of functions with simplest analytic properties, namely
\beqa \notag
&& \bQ_i(u), \ \bQ^i(u),\; \quad (i=1, \dots, 4) \\
&& \bP_a(u), \ \bP^a(u),\; \quad(a= 1,\dots, 4)
\label{QPfunctions}
\\ \notag
&& Q_{a|i}(u),\;\quad (i=1, \dots, 4, \ a= 1,\dots, 4)\;.
\eeqa
They are related to each other by the following set of equations:
\beqa\notag
&& Q_{a|i}^+(u) -Q_{a|i}^-(u)=\bP_a(u) \bQ_i(u)\,,
\\ \label{QSCequations}
&& \bQ_i(u)=-Q_{a|i}^{\pm}(u)\bP^a(u),\;\; \qquad \bP_a(u)= -Q_{a|i}^{\pm}(u)\bQ^i(u)\,,
\\\notag
&& \bP_a(u) \bP^a(u)=0,\;\;\qquad\hspace*{12mm}  \bQ_i(u) \bQ^i(u)=0\,,
\eeqa
and similar relations with all upper/lower  indices lowered/raised. Here and below we use a common notation
\beq
f^{\pm}(u) = f(u\pm i/2),\qquad  f^{[\pm k]} (u)= f(u\pm i k/2 )\,.
\eeq


Apart from these algebraic relations, there are also analytic constraints on the structure of cuts of Q-functions in the complex plane, which, in particular, express their monodromies around the branch points through other Q-functions. 

\begin{figure}[h]
\begin{center}
\includegraphics[scale=0.3]{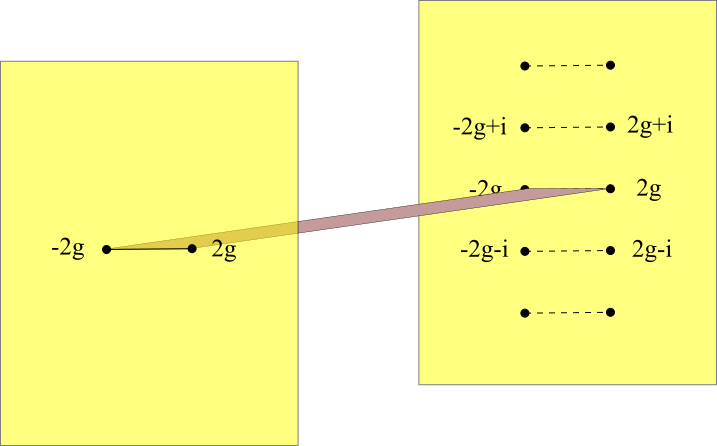}\hspace{20mm}
\includegraphics[scale=0.3]{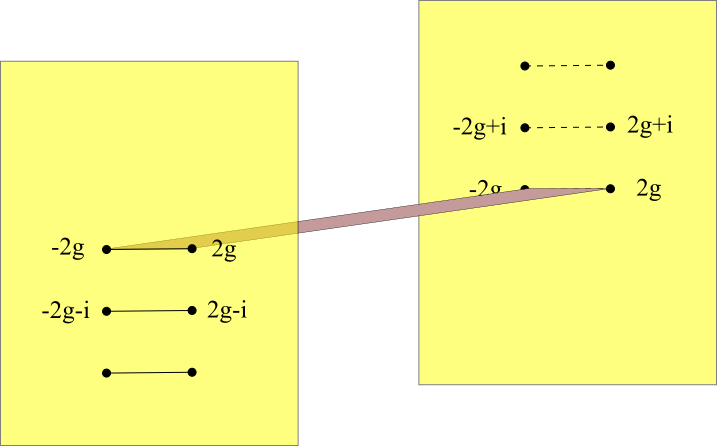}
\end{center}
\caption{{\bf Left}: Analytic structure of $\bP_a(u)$ and $\bP^a(u)$: $\bP(u)$ has one cut on the first sheet (left). Analytical continuation through this cut  leads to the second sheet corresponding to $\tilde \bP(u)$ (right), which has an infinite ladder of cuts
{\bf Right}: Analytic structure  of $\bQ_i(u)$ and $\bQ^i(u)$: $\bQ(u)$ has an infinite ladder of cuts in the lower half-plane (left). Analytical continuation through the cut on the real axis leads to a sheet corresponding to $\tilde\bQ(u)$, which has an infinite ladder of cuts in the upper half-plane (right).
\label{Fig:cutsP}\label{Fig:cutsQ}
}
\end{figure}

The basic analytic structure is the following: $\bP_a$ and $\bP^a$ have one cut on the real axis, from $-2g$ to $2g$, as shown in Fig. \ref{Fig:cutsP}. This cut can be resolved by a transfromation to Zhukovsky variable $x(u)$, defined as $x(u)+1/x(u)=u/g$, $|x(u)|>1$. Thus it will be convenient to parametrize $\bP$ in  $x$ instead of $u$.

Functions $\bQ_i$ and $\bQ^i$ have infinite ladder of Zhukovsky cuts: the original one, from $-2g$ to $2g$ and its copies, shifted by an integer number of $i$ into the lower half plane, as shown in Fig.~\ref{Fig:cutsQ}. 
For any function of spectral parameter $Q(u)$ we denote by $\tilde Q(u)$  the analytic continuation of $Q(u)$ under the Zhukovsky cut on the real axis. It is sufficient to impose the gluing condition on $\bQ_i$
at the cut $[-2g,2g]$ \cite{Gromov:2015vua}\footnote{We used here the left-right symmetry relations \eqref{LRsymmetry} and already put at this stage \(\kappa=\hat\kappa\).}
\begin{align}\notag
{}& \tilde  \bQ_1(-u)=\beta_1\bQ_3(u)\,, &&
\tilde\bQ_2(-u)=\beta_2\bQ_4(u)\,,
\\[2mm]
{}& \tilde\bQ_3(-u)=1/\beta_1\bQ_1(u)\,, &&
\tilde\bQ_4(-u)=1/\beta_2\bQ_2(u)\,,
\label{RiemannHilbert}
\end{align}
which schematically can be written as $\tilde{\bQ}_i(-u)={H_{i}}^j\bQ_j(u)$\footnote{In the unitary theory like $N=4$ SYM one can use the complex conjugation instead, which is applicable for all operators and not only for the parity symmetric ones. Also, for this gluing condition to be of this simple form that is crucial to choose $\bQ_i$ to be of a ``pure" form, meaning that their large $u$ asymptotic are of the form $u^{-\hat{\nu_i}}\sum_{i=0}^\infty \frac{a_i}{u^i}$ to all orders in $1/u$. In general this condition is not sufficient to uniquely determine $\bQ_i$ as $\nu_i$ for different $i$ 
could differ by an integer, allowing for $\bQ_i$ to mix. There are several ways to deal with this problem, one possibility is to introduce a twist in $AdS^5$. Another possibility, which we use in this paper, is to make use of the $u\to-u$ symmetry, applicable for some states, and keep only even powers $u^{-\hat{\nu_i}}\sum_{i=0}^\infty \frac{a_{2i}}{u^{2i}}$ which also leads to the correct non-ambiguous gluing. Finally, it was noticed in
\cite{Gromov:2015vua} that the conditions \eq{RiemannHilbert} are over-defined and it is usually sufficient to impose only one of them.},
to close the system of equations. This means that imposing the above relation will give us a discrete set of isolated solutions each corresponding to a certain state of the theory. 

In order to identify
a particular state we have first to know its quantum numbers, which are hidden in the large $u$ asymptotic of $\bQ_i$ and $\bP_a$ as we discuss below.

\paragraph{Asymptotics and quantum numbers.}
Asymptotic behavoir of Q-functions at large $u$ is determined by the twists and by the quantum numbers of the particular state of \sym\  we are studying. Here we will consider a particular kind of state: BMN vacuum $\tr (\phi_1^{\L})$, where  \(\phi_1\) is one of  three complex scalars of \sym\ theory\footnote{We use here notation  $\phi_i$ for complex scalars, instead of the familiar \(\{X,Y,Z\}\) since two of them appear with a different notation in bi-scalar model \eqref{bi-scalarL}, say \(Z=\phi_1,\,\,Y=\phi_2\).    }. In the notation of  \cite{Kazakov:2015efa} its quantum numbers are
\beq \label{charges}
\{{\L},0,0 | \Delta,0,0\}\,.
\eeq
 In the untwisted theory $\tr (\phi_1^{\L})$ would have been protected BPS state, but in the presence of twists it has a nontrivial scaling dimenstion $\Delta_{\L}(g)$ which we will be computing. This scaling dimension was computed at weak coupling up to $g^{4{\L}-2}$ (two wrappings) in \cite{Ahn2011} in terms of infinite double sums and integrals and it was brought to the standard explicit MZV form in \cite{Gurdogan:2015csr}.   For  the case of our current interest, the bi-scalar model, the single wrapped graph at any \({\L}\) was computed in  \cite{Broadhurst:1985vq} and the double-wrapped graph at  \({\L}=3\)  was computed in \cite{Panzer:2013cha}. We use these results to verify our   computation.

In the general case of twisted QSC, the asymptotics of one-indexed functions are given by \cite{Kazakov:2015efa}
\beqa \nn 
&&\bP_a\sim  A_a x_a^{i u} u^{-\hat \lambda_a}\,,\qquad   \bP^a\sim  A^a x_a^{-i u} u^{\hat \lambda_a^*},
\\[2mm]
\nn
&&\bQ_i\sim B_i u^{-\hat \nu_i}\left(1+\frac{b_{i,1}}{u}+\frac{b_{i,2}}{u^2}+\dots\right)\,, \\
\label{asymptoticsQi}
&&\bQ^i\sim B^i u^{\hat \nu_i^*}\left(1+\frac{b^{i,1}}{u}+\frac{b^{i,2}}{u^2}+\dots\right)\;.
\eeqa
In a particular case of BMN vacuum with charges \eqref{charges}
 the powers $\hat \lambda_a,\hat \lambda_a^*,\hat \nu_i,\hat \nu_i^*$ are determined by the quantum numbers of the state
\beqa\label{twists} \nn
&&\hat \lambda=\hat \lambda^*=\left\{\frac{\L}{2},\frac{\L}{2},-\frac{\L}{2},-\frac{\L}{2}\right\},\\ \nn
&&\hat \nu_i=\left\{-\frac{\Delta}{2},-1-\frac{\Delta}{2},-2+\frac{\Delta}{2},-3+\frac{\Delta}{2}\right\},\\
&&\hat \nu_i^*=\left\{-\frac{\Delta}{2}+3,-\frac{\Delta}{2}+2,\frac{\Delta}{2}+1,\frac{\Delta}{2}\right\}\,,
\eeqa
and the twists are chosen to be $x_a=\left\{\kappa^\L,\kappa^{-\L},\hat\kappa^\L,\hat\kappa^{-\L}\right\}$.   They are related to the twist parameters introduced at the beginning of introduction as
\begin{align}\label{kap}
\kappa=q_3q_2\,,\qquad \qquad \hat\kappa=q_3/q_2\,,
\end{align}
whereas the dependence on \(q_1\) is absent for BMN vacuum state.

We can plug the asymptotics \eq{asymptoticsQi}  into the equations \eqref{QSCequations} and obtain a set of constraints for combinations  $A_a A^a$ and $B_i B^i$ (no summation). After the fixing the rescaling symmetry the solution for $A_a$ can be chosen to be \cite{Kazakov:2015efa}
\beqa\label{AA}\nn
&&A_1=-A_2=\frac{\hat \kappa^\L (\kappa^\L-1)^3}{(1+\kappa^\L)(\kappa^\L-\hat \kappa^\L)((\kappa \hat \kappa)^\L-1)}\\
&&A_3=-A_4=-\frac{\kappa^\L (\hat\kappa^\L-1)^3}{(1+\hat\kappa^\L)(\kappa^\L-\hat\kappa^\L)((\kappa \hat \kappa)^\L-1)}\;.
\eeqa
In a similar way for $B_i$:
\beqa
&&B_1B^1=-B_4 B^4=\frac{i(\kappa^\L-1)^2(\hat \kappa^\L-1)^2}{(\kappa \hat \kappa)^\L(\Delta-2)(\Delta-3)}\notag\\
&&B_2B^2=-B_3 B^3=-\frac{i(\kappa^\L-1)^2(\hat \kappa^\L-1)^2}{(\kappa \hat \kappa)^\L(\Delta-1)(\Delta-2)}
\label{BB}\;.
\eeqa

Since, as we discussed, the $\bP_a$ and $\bP^a$ functions has only one cut, they admit the following Laurent expansion
\beqa
\bP_a(u)=x_a^{i u} (g x(u))^{-\hat \lambda_a}\bp_a(u)\;\;,\;\;
\bP^a(u)=x_a^{-i u} (g x(u))^{\hat \lambda_a^*}\bp^a(u),
\eeqa
where 
\beqa\la{LS}
\bp_a=A_a\sum_{n=0}^\infty \frac{b_{a,n}}{x^n}\;\;,\;\;\bp^a=A^a\sum_{n=0}^\infty \frac{b^{a,n}}{x^n}\;.
\eeqa
 The constants $b_{a,n}$ and $b^{a,n}$ contain all the information about the state. In order to constrain them we have to find $\bQ_i$ and $\bQ^i$ for the given $\bP_a$
and $\bP^a$ and impose the analyticity condition \eq{RiemannHilbert} on the cut $[-2g,2g]$. The Fourth-order Baxter equation is an efficient way of doing this.

\paragraph{Fourth-order Baxter equation.}
Instead of using the chain of QQ-relations \eqref{QSCequations}
one can equivalently deduce the $\bQ_i$ using the given $\bP_a$ and $\bP^a$ functions
from the ``Fourth-order Baxter equation". More precisely,
four one-indexed functions \(\bQ_j\) are the $4$ linear independent solutions of the following 4th order finite-difference equation~\cite{Alfimov:2014bwa}:
\begin{align}\label{Baxter}\nn
\bQ_i^{[+4]}D_0-\bQ_i^{[+2]} \left[D_1-\bP_a^{[+2]}\bP^{a[+4]}D_0 \right]+\bQ_i \left[D_2-\bP_a \bP^{a[+2]}D_1+\bP_a \bP^{a[+4]}D_0\right]
\\
 -\bQ_i^{[-2]}
\left[\bar D_1+\bP_a^{[-2]}\bP^{a[-4]}\bar D_0 \right]+\bQ_i^{[-4]}
\bar D_0=0\,,
\end{align}
where $D_i$ are determinants of matrices with elements of form $\bP_a^{[k]}$ given in the appendix \ref{sec:Coefficients}.

\subsection{Double-scaling limit}
The limit we have to take to obtain the \chiFT{} is $g\to 0$
and $s\equiv\sqrt{\kappa\hat{\kappa}}\to\infty$ with $\xi=g s$ fixed.
First obvious thing which happens in this limit the branch points of the Zhukovsky cuts collapse into a point producing poles. Furthermore,
the coefficients $b_{a,n}$ and $b^{a,n}$ in \eq{LS} typically scale as $g^{2n}$ 
and thus the infinite series truncates. In Appendix~\ref{sec:appQSC} we argue that for the BMN state the following scaling in $g$ has to be imposed:
\beqa\label{Pa1}\nn
\label{Pa2}\nn
&&\bp_a=\left\{A_1 f_1(u),A_2 f_1(-u),A_3 g_1(u),A_4 g_1(-u)\right\}\\[2mm]
&&\bp^a=\left\{ f_2(u), f_2(-u), g_2(u), g_2(-u)\right\}.
\eeqa
where
$f_i,g_i$ as series in $x(u)$:
\beqa\label{fg_ansatz}\nn
&&f_1=1+g^{2\L} \sum\limits_{n=1}^{\infty}\frac{g^{2n-2}c_{1,n}}{(g x)^n}\\\nn
&&g_1=(g x)^{-\L}\left(u^\L+  \sum_{k=0}^{\L-1} c_{2,-k}u^k+ \sum\limits_{n=1}^{\infty}\frac{g^{2n}c_{2,n}}{(g x)^n}\right)\\\nn
&&f_2=(g x)^{-\L}\left(u^\L+  \sum_{k=0}^{\L-1} c_{3,-k}u^k+ \sum\limits_{n=1}^{\infty}\frac{g^{2n}c_{3,n}}{(g x)^n}\right)\\
&&g_2=1+g^{2\L} \sum\limits_{n=1}^{\infty}\frac{g^{2n-2}c_{4,n}}{(g x)^n}\;.
\eeqa
This follows the argument similar to \cite{Marboe:2014gma}.

We see that the terms with heigher dergree of $1/x^n$ get more and more suppressed. However, we still have to keep a few first term simply because the Baxter equation \eq{Baxter}
contains $\bP_a$ with shifts $u\to u+  i n$, $n=-2,\dots,2$.
Due to the twists tending to infinity along the complex axes the factors $x_a^{iu}$ in $\bP_a$ we could get an enhancement of these suppressed with $g$ terms. At the end we have to plug the truncated expressions for $\bP_a$, which become rational functions with poles at $u=0$ in the $g\to 0$ limit,
into the equation \eq{Baxter} we obtain a finite difference equation with rational coefficients. We worked out explicitly the $J=2,3,4$ cases in the Appendix~\ref{sec:appQSC}. Below we present the results.

\paragraph{Baxter equation for twist 2.}\label{factorL2}

For $\L=2$  the only unfixed coefficient left is $\Delta$ and the Baxter equation \eqref{Baxter} coincides in the double scaling limit with \eq{L2BaxterLax-fac}. The expression on the left-hand side of  \eq{L2BaxterLax-fac} is given by the product of  two
finite difference operators separated by the factor of $u^2$.
It is easy to verify that \eq{L2BaxterLax-fac} stays invariant under the exchange of these two operators.
 As a consequence, the four solutions to \eq{L2BaxterLax-fac} have to satisfy 2nd order finite difference equations, e.g.
\begin{eqnarray} \label{bax2-fact}
\left(\frac{\Delta(\Delta-2)}{4u^2}-2\right)q(u)+q(u+i)+q(u-i)=0\,,
\end{eqnarray}
and the second equation is obtained by replacing $\Delta\to 4-\Delta$. By matching the asymptotics \eqref{asymptoticsQi},
we find that  $\bQ_2$ and $\bQ_3$ satisfy \eq{bax2-fact} whereas $\bQ_1$ and $\bQ_4$ satisfy the second equation.
The 2nd order Baxter equation \eq{bax2-fact} has been previously studied in \cite{Derkachov:2002wz,Alfimov:2014bwa}
and its explicit solutions have been found in terms of hypergeometric ${}_3F_2-$functions
\beq\label{q2}
q_0(u)=2iu _3F_2\left(i u+1,\frac{\Delta}{2},1-\frac{\Delta}{2};1,2;1\right)\,,
\eeq
and the second solution is given by $q_0(-u)$. Using these results, we construct their linear combinations
\beqa\notag
&& \bQ_2(u)=
-u B_2
\frac{\tan \frac{\pi  \Delta }{2}\;
        \Gamma \left(\frac{\Delta }{2}+1\right) \Gamma \left(\frac{\Delta
        }{2}\right)^2}{4e^{\frac{\pi  \Delta }{4i}}  \Gamma (\Delta -1)}
\left[\frac{q_0(-u)}{\sin\frac{\pi \Delta}{2} }+\left(-i \coth(\pi u)+\cot\frac{\pi \Delta}{2}\right)q_0(u)\right]
\,,
\\
&& \bQ_3(u)=u B_3
\frac{i  \tan \frac{\pi  \Delta
        }{2}\; \Gamma \left(1-\frac{\Delta }{2}\right)^2 \Gamma
        \left(2-\frac{\Delta }{2}\right)}{4e^{\frac{i \pi  \Delta}{4}  } \Gamma (1-\Delta )}
\left[\frac{q_0(-u)}{\sin\frac{\pi \Delta}{2} }-\left(i \coth(\pi u)+\cot\frac{\pi \Delta}{2}\right)q_0(u)\right] 
\,.
\eeqa
These expressions are solutions to the Baxter equation \eq{Baxter} with the correct pole structure. Moreover, they are ``pure" solutions in the sense that their expansion at infinity has a form $u^{\alpha}\left(1+c_1/u^2+\dots\right)$.
The two remaining solutions $\bQ_4$ and $\bQ_1$ are obtained from $\bQ_2$ and $\bQ_3$ by replacing $\Delta$ with $4-\Delta$ (and changing $B_i$ accordingly).

\paragraph{Baxter equation for twist 3.}
For $\L=3$ the equation is a bit more complicated. 
After using the identification
\beq\label{qQ0}
q_i(u)=\bQ_i(u)u^{-\L/2}
\eeq
the equation \eqref{Baxter} reduces in the double scaling limit to the Baxter equation \eq{L3BaxterLax}
where \(\alpha=(\Delta-2)^2\) and \(m\) are two yet unfixed parameters. Its factorized form is given by \eq{factorL3}.
In other words again we reproduced precisely the equation obtained from the integrability of the Feynman graphs! In the next section where we will derive the quantization conditions for fixing \(m\) and, finally, \(\Delta(\xi).\) Thus we will extract the all-loop anomalous dimension of \(\Tr\phi_1^3\) operator, and of some related operators.
It can be also easily checked that the equation \eq{L3BaxterLax} is invariant w.r.t. the change \(m\to -m\). That is why we can find all four solutions from a much simple 2nd order Baxter equations
\begin{equation}
\left(\frac{(\Delta-1)(\Delta-3)}{4u^2}-\frac{m}{u^3}-2\right)q(u)+q(u+i)+q(u-i)=0
\label{baxter3}
\end{equation} 
as another pair of solutions can be obtained by replacing \(m\to -m\).


\paragraph{Baxter equation for twist 4.} For $J=4$ with the identification \eq{qQ0} we obtained again the same Baxter equation as in \eq{L4BaxterLax}, for which we have not been able to show the factorization property.
We postpone further investigation of this state for a future publication.

In conclusion, the derivation in this section of Baxter equations from QSC formalism in the double scaling  for \(\L=2,3,4\)  confirms the Baxter equation for arbitrary \(\L\) given at the end of the previous section, obtained from the group-theoretical considerations, in usual assumption of universality of Baxter equations (independence on auxiliary state representation). Now we have to impose additional quantization conditions on possible solutions of Baxter equation, which we do in the next section, so far only for \(\L=3\) case, from the double scaling limit of QSC formalism.


\section{Quantization condition for the Baxter equation}
\label{sec:quantization}

In this section we  focus on the operators (\ref{O-def}) with the $R-$charge $\L=3$ whose scaling dimensions $\Delta$ are described by the Baxter equation (\ref{baxter3}). The method presented here should also apply for general \(\L\) and even to general operators in the bi-scalar theory \eqref{bi-scalarL}.

The Baxter equation (\ref{baxter3}) does not single out a particular value of the parameters $\Delta$ and $m$. Furthermore, it does not depend
on the coupling constant $\xi$.  The goal of this section is to use the underlying QSC description of the full ${\cal N}=4$
theory to derive the quantization conditions which fix the dependence $\Delta(\xi)$ and $m(\xi)$. A priori, these conditions can be derived from the
first principles using integrability of the fish-net diagrams \cite{Zamolodchikov:1980mb,Gurdogan:2015csr}. However, this is beyond the scope of the current paper.


Let us first  discuss analytic properties of the solutions of the Baxter equation \eq{baxter3}.
To build a solution
we start from $u$ with large positive imaginary part, for which
the finite difference equation \eq{baxter3} can be replaced by an ordinary differential equation. For arbitrary $\Delta$ and $m$, it has
two linear independent power-like solutions that we can choose to be ``pure solutions''
\beqa \notag\label{as-inf}
&&q_2(u,m)= u^{\Delta/2-1/2}\left(1+\frac{a_1}{u}+O\left(\frac{1}{u^2}\right)\right),
\\\la{bax3}
&& q_4(u,m)= u^{-\Delta/2+3/2}\left(1+\frac{b_1}{u}+O\left(\frac{1}{u^2}\right)\right).
\eeqa
For arbitrary non-integer $\Delta$ these solutions are well defined from the requirement that $q_2(u,m)$ does not contain $O(u^{-\Delta/2+3/2})$ terms and
 $q_4(u,m)$ does not contain \(O(u^{\Delta/2-1/2})\) terms. The expansion coefficients $a_i$ and $b_i$ can be fixed from  \eq{baxter3}, e.g.
$$
a_1=\frac{m}{\Delta-3}\,,\qquad \qquad b_1=\frac{m}{1-\Delta}\,.
$$
Using \eq{bax3} we can apply the finite difference equation \eq{baxter3} to recursively decrease the imaginary part of $u$ in integer steps. Since the coefficients of the Baxter equation \eq{baxter3} are analytic for ${\rm Im} \, u>0$ and have a 3rd order pole at $u=0$,
the solutions constructed in this way are also analytical in the upper half-plane and have generically a 3rd order pole
at the points $u=-i n$ with $n=1,2,3,\dots$. We will discuss in more details how to build these solutions in the section~\ref{sec:numerics}.

Having constructed  solutions to the Baxter equation \eq{baxter3}, we can now identify four $\bQ-$functions defined in \eq{QPfunctions}.
In the double scaling limit we have
\begin{align}
& {\bf Q}_4(u)=\frac{u^{3/2}}{2}[q_4(u,m)+q_4(u,-m)]\,,
&& \quad {\bf Q}_1(u)=
\frac{-is^6}
{2m(\Delta-2)}
u^{3/2}[q_2(u,m)-q_2(u,-m)]\,,
\notag\\
& {\bf Q}_2(u)=\frac{u^{3/2}}{2}[q_2(u,m)+q_2(u,-m)]\,,
&& \quad {\bf Q}_3(u)=
\frac{is^6}
{2m(\Delta-2)}
u^{3/2}[q_4(u,m)-q_4(u,-m)]\,.
\label{Qq}
\end{align}
It is straightforward to check that these expressions verify the defining relations \eq{asymptoticsQi} and their normalization is fixed by \eq{BB} (we recall that $s=\sqrt{\kappa\hat{\kappa}}$ and $\kappa, \hat{\kappa}\to\infty$ in the double scaling limit). To simplify the calculation, we have assumed $\bQ-$functions to be even functions of $m$. 
We can relax this condition and carry out the calculation in the general case. This will not affect the final result for $\Delta(\xi)$ but will significantly complicate  some intermediate expressions. At the same time that is clear that the $m\to-m$ symmetry of the system has to be reflected in the $\bQ_i$ so instead of deriving the expressions for $\bQ_2$ and $\bQ_4$ we fix the proportion of $q_4(u,m)$ and $q_4(u,-m)$ from the symmetry requirement from the beginning.

The quantization condition for $\Delta(\xi)$ arises from the requirement for the functions \eq{Qq} to have correct analytic properties in twisted \({\cal N}=4\) SYM presented at the beginning of section~\ref{sec:BaxterfromQSC}. We recall that for a finite value of the twist the functions $\bQ_i$ have the cut $[-2g,2g]$.
Going under the cut, these functions have to satisfy the gluing conditions  \cite{Gromov:2015vua}. In the double-scaling limit, the cut $[-2g,2g]$ shrinks into a point and the relations
 \eq{RiemannHilbert} lead to nontrivial constraints for the functions $\bQ_i(u)$ near the origin \cite{Gromov:2015vua}.

Another condition comes from the  symmetry of the Baxter equation \eq{baxter3} under parity transformation, $u\to -u$ and $m\to -m$. Indeed, the functions
$q_i(u,m)$ and $q_i(-u,-m)$ satisfy \eq{baxter3}, implying that the functions $\bQ_i(-u)$ can be expanded over the basis of the solutions \eq{Qq}
with some periodic coefficients:
\begin{equation}\la{defOmega}
\bQ_i(-u)={\Omega_{i}}^j(u)\bQ_j(u)\,,
\end{equation}
where ${\Omega_{i}}^j(u)$ has the following i-periodicity property\footnote{i-periodic functions play the same role  in finite difference equations of Baxter type as constants in linear differential equations.}
\begin{eqnarray}
{\Omega_{i}}^j(u+i)={\Omega_{i}}^j(u)\,.
\end{eqnarray}
Similar relation should also hold   under the cut $\tilde \bQ_i(-u)={\tilde\Omega_{i}}^j(u)\tilde\bQ_j(u)$.

As we show in the Appendix \ref{app:Omega}, the discontinuity of ${\Omega_{i}}^j(u)$ across the cut,
 \(\Delta{\Omega}_i^{\;j}\equiv \tilde{\Omega}_i^{\;j}-{\Omega}_i^{\;j}\), satisfies the following exact Riemann-Hilbert equation (valid for any value of the twist parameter)
\begin{equation}\label{dOmega}
\Delta{\Omega}_i^{\;j}(u)\equiv
\tilde{\Omega}_i^{\;j}(u)-{\Omega}_i^{\;j}(u)
=-\tilde\bQ_i(-u)\tilde{\bQ}^j(u)+
\bQ_i(-u){\bQ}^j(u)\;.
\end{equation}
In the double-scaling limit, this relation allows us to compute $\Delta{\Omega}_i^{\;j}(u)$ for $u\to 0$ in terms of solutions to the Baxter equation,
$q_2(u,\pm m)$ and $q_4(u,\pm m)$, evaluated at the origin.
Indeed, for $u\to 0$, substituting   \eq{Qq}  and \eq{RiemannHilbert}  into \eq{dOmega} we find after some algebra
\footnote{in order to simplify the expression we imposed already $\beta_2=-\beta_1$. This condition can be deduced alongside with the quantization condition in the way it is described in the next section.}
\begin{eqnarray}
&&\Delta{\Omega}_i^{\;j}/u^3= \left(
\begin{array}{cccc}
 z  &
0 &  w  &    v \\
0 & - z  & t
& - w  \\
 { w}/{ \beta_1^2 } & -  v/\beta_1^2
  &
 z
& 0 \\ -t/\beta_1^2 &
- { w}/{ \beta_1^2 } &
0 & - z \\
\end{array}
\right) + O(u)\,,
\label{DeltaO}
\end{eqnarray}
where $s=\sqrt{\kappa\hat{\kappa}}$ and the notation was introduced for
\begin{align}\notag
{}& z=s^6 \frac{ q_2 \dot{q}_4-q_4 \dot{q}_2 }{2 m (\Delta -2)}\,,
&&
w=s^6\frac{  \left(\dot{q}_4^2-q_4^2\right) \beta_1^2+q_2^2-\dot{q}_2^2 }{4 m (\Delta -2)} \,,\qquad
\\
{}& v=s^{12} \frac{
 \beta_1^2
\left(q_4-\dot{q}_4\right){}^2-\left(q_2-\dot{q}_2\right){}^2 }
{4i m^2 (\Delta -2)^2}\,,
&&
t= \frac{\beta_1^2
\left(q_4+\dot{q}_4\right){}^2-\left(q_2+\dot{q}_2\right){}^2}{4i} \,,
\end{align}
with compact notations \begin{equation}q_j\equiv q_j(0,m)\,,\qquad \dot q_j\equiv q_j(0,-m)\,,\qquad (j=2,4).\end{equation}

We conclude from \eq{DeltaO}  that $\Delta{\Omega}_i^{\;j}$ vanishes  near the origin in the double scaling limit as  $\Delta{\Omega}_i^{\;j}\sim u^3$. In the next section, we will find ${\Omega}_i^{\;j}(u)$ independently for an arbitrary coupling from the Baxter equation. Comparing the results of these two calculations we will be able to determine the quantization condition for $\Delta$ and $m$.

\subsection{Extracting $\Omega_i^j$ from the  Baxter equation}

As was mentioned in the beginning of this section, the Baxter equation \eq{baxter3} is invariant under simultaneous
change $u\to -u$ and $m\to -m$. Following the same logic that led \eq{defOmega}, its solutions have to satisfy
the relation
\begin{equation}\la{defom}
q_j(-u,-m)={\sigma_j}^k(u) q_k(u,m)\,, \qquad(j, k=2,4)\,,
\end{equation}
where ${\sigma_l}^k(u)$ are some nontrivial i-periodic functions.
We can use the Baxter equation to find the leading behavior of ${\sigma_j}^k(u)$ for \(u\to 0\).

We recall that solutions to
the Baxter equatin $q_j(-u,-m)$ have 3rd order poles at $u=in$ (with $n=1,2,\dots$) whereas $q_k(u,m)$ are regular at these
points. Together with periodicity condition ${\sigma_l}^k(u+i)={\sigma_l}^k(u)$, this implies that ${\sigma_l}^k(u)$
must have $3$rd order pole at the origin. Since the functions $q_j(-u,-m)$ and $q_k(u,m)$ are regular for $u\to 0$, it
follows from \eq{defom} that the
residue at this pole have to satisfy
\begin{equation}\la{lin2}
\lim_{u\to 0} u^3{\sigma_j}^k(u) \,q_k(m,0)=0\,.
\end{equation}

Using the Baxter equation \eq{bax3} we get for $u\to 0$
\begin{eqnarray}
q_j(u-i,-m)= -m\frac{q_j(0,-m)}{u^3}+{\cal O}\left(1/u^2\right)\;.
\end{eqnarray}
Replacing $u\to - u+i$ in \eq{defom} and matching it into the last relation we obtain
\begin{equation}\la{lin1}
m q_j(0,-m)\simeq -u^3{\sigma_j}^k(-u) q_k(i,m)+ O(u)\,.
\end{equation}
The relations \eq{lin2} and \eq{lin1} viewed as a linear system of equations on $u^3{\sigma_l}^k(u)$ for $u\to 0$
lead to
\begin{equation}
u^3{\sigma_j}^k(u)= \frac{m
}{q_2(i,m)q_4(0,m)-q_4(i,m)q_2(0,m)}
\left(
\begin{array}{cc}
\dot q_2 q_4\;\; & -\dot q_2 q_2 \\
\dot q_4 q_4\;\; & -q_2 \dot q_4 \\
\end{array}
\right)_{jk} + O(u)\,.
\end{equation}
 We notice that the expression in the denominator coincides
with the Wronskian of the finite difference equation \eq{baxter3}. As such, it should not depend on $u$ and can be computed
using asymptotic behavior of functions at infinity \eq{as-inf}
\begin{align}\label{Wr}
q_2(u+i,m)q_4(u,m)-q_4(u+i,m)q_2(u,m)=i(\Delta-2)\,.
\end{align}
In this way, we finally obtain
\begin{align}\label{ome}
{\sigma_j}^k(u)=\frac{m}{i u^3(\Delta-2)}
\left(
\begin{array}{cc}
\dot q_2 q_4\;\; & -\dot q_2 q_2 \\
\dot q_4 q_4\;\; & -q_2 \dot q_4 \\
\end{array}
\right)_{jk} + O(1/u^2) \,.
\end{align}
We can now combine the relations \eq{defom} and \eq{Qq} to establish linear relations between the functions
$\bQ_i(-u)$ to $\bQ_j(u)$.
Using the definition \eq{defOmega} we obtain from \eq{ome} the leading asymptotic behavior of ${\Omega_{i}}^j(u)$
at the origin
\begin{equation}
{\Omega_{i}}^j = \frac{1}{u^3}\left(
\begin{array}{cccc}
\frac{m \left(q_4 \dot{q}_2-q_2 \dot{q}_4\right)}{2 (\Delta -2)} & -\frac{i
s^6 \left(q_4 \dot{q}_2+q_2 \dot{q}_4\right)}{2 (\Delta -2)^2} & 0 &
\frac{i s^6 q_2 \dot{q}_2}{(\Delta -2)^2} \\
-\frac{i m^2 \left(q_4 \dot{q}_2+q_2 \dot{q}_4\right)}{2 s^6} & \frac{m
\left(q_2 \dot{q}_4-q_4 \dot{q}_2\right)}{2 (\Delta -2)} & -\frac{i m^2 q_2
\dot{q}_2}{s^6} & 0 \\
0 & \frac{i s^6 q_4 \dot{q}_4}{(\Delta -2)^2} & \frac{m \left(q_4
\dot{q}_2-q_2 \dot{q}_4\right)}{2 (\Delta -2)} & -\frac{i s^6 \left(q_4
\dot{q}_2+q_2 \dot{q}_4\right)}{2 (\Delta -2)^2} \\
-\frac{i m^2 q_4 \dot{q}_4}{s^6} & 0 & -\frac{i m^2 \left(q_4 \dot{q}_2+q_2
\dot{q}_4\right)}{2 s^6} & \frac{m \left(q_2 \dot{q}_4-q_4
\dot{q}_2\right)}{2 (\Delta -2)} \\
\end{array}
\right)+ O(1/u^2) \,.
\label{Ou3}
\end{equation}

In the next section we compare this relation
with the expression for the discontinuity of ${\Omega_{i}}^j$ given by \eq{DeltaO}
to obtain the constraints on the parameters of the Baxter equation $m$ and $\Delta$.

\subsection{Quantization condition from gluing}

We demonstrated in the previous subsection that the matrix ${\Omega_i}^j(u)$ has $3^{\rm rd}$ order pole at $u=0$.
We would like to stress that the relation \eq{Ou3} holds in the double scaling limit, for $\xi=g s$ fixed with $s\to\infty$ and
$g\to 0$.
Due to our conventions different $\bQ_i$ scale differently with $g$
and as a consequence different components of ${\Omega_i}^j$ scale differently in our limit. This is totally due to the choice of our normalization. We introduce $\gamma_{ij}$ so that ${\Omega_i}^j\sim s^{\gamma_{ij}}$.
 As explained in section~\ref{sec:BaxterfromQSC},  at finite coupling $g$ the only singularities ${\Omega_i}^j(u)$
could have at finite \(u\) are due to the branch
cuts of the Zhukovsky variables $x(u)$ locate at \(u=\pm 2g\). As a consequence, it can be represented in the form
\begin{align}\label{sqrt}
 s^{-\gamma_{ij}}\Omega_{i}^{\;j}(u)=\sqrt{u^2-4g^2} \;f_{ij}(g,u)+h_{ij}(g,u)\,,
\end{align}
where $4\times 4$ matrices $f(g,u)$ and $h(g,u)$ are regular around the origin and each term in their small $g$ expansion should be regular as well. 

The poles of ${\Omega_i}^j(u)$ at the origin can only appear as an effect of expansion of the Zhukovsky
cut
\begin{align}\label{sqrt-exp}
\sqrt{u^2-4g^2}=u-\frac{2 g^2}{u}-\frac{2 g^4}{u^3}-\frac{4 g^6}{u^5}+O\left(g^8\right)\;.
\end{align}
At the same time, computing discontinuity of ${\Omega_i}^j(u)$ across the cut $[-2g,2g]$ we find from \eq{sqrt} for $g\to 0$
\begin{align}\label{f(g)}
f_{ij}(g,u)= s^{-\gamma_{ij}} \frac{ \Omega_i^{\; j}(u)-{\tilde\Omega}_i^{\; j}(u)}{2\sqrt{u^2-4g^2}}=-
s^{-\gamma_{ij}}\frac{\Delta{\Omega}_i^{\; j}(u)}{2u}\left(1+ O(g^2)\right)\,.
\end{align}
According to \eq{DeltaO},  \(\Delta{\Omega_i}^j\sim u^3\) for $u\to 0$ in the double scaling limit. Then, it follows from the
last relation that
the series expansion of $f(g,u)$  in $g$ and $u$  must be of the form
\begin{align}\label{f-f}
f(g,u)=g^{-n}(f_{0,2}u^2+f_{0,3}u^3+\dots)+g^{-n+2}(f_{1,0}+f_{1,1}u+\dots)+g^{-n+4}(f_{2,0}+f_{2,1}u+\dots)\,,
\end{align}
where we assume that $f(g,u)$ scales as $g^{-n}$. The expansion coefficients depend on $\xi$ and $\Delta$ (but not of $g$). Substituting the last relation into \eq{sqrt} and taking into account \eq{sqrt-exp} we get
\begin{align}\label{Omega-f}
s^{-\gamma}{\Omega}(u)=-2g^{4-n}\frac{f_{0,2}+f_{1,0}}{u}
-2g^{6-n}\left[\frac{2f_{0,2}+f_{1,0}}{u^3}+
\frac{2f_{0,3}+f_{1,1}}{u^2}+\frac{2f_{0,4}+f_{1,2}+f_{2,0}}{u}\right]+\dots\,,
\end{align}
where we use a shorthand notation for a matrix $s^{-\gamma}\,{\Omega}\equiv s^{-\gamma_{ij}}{\Omega_i^{\; j}}(u)$ and
dots denote terms regular for $u\to 0$ and/or suppressed by powers of $g^2$.
 Notice that regular function \(h(g,u)\) does not contribute here to the leading, singular terms.

We expect that in the double scaling limit the matrix \eq{Omega-f} should take the expected form \eq{Ou3}. In particular,
it should scale as $1/u^3$ for $u\to 0$.
Note that l.h.s. of \eq{Omega-f} scales as $g^0$. 
We have to deduce the value of $n$. If we assume that $n=4$
the second term in \eq{Omega-f} become negligible and we will not be able to reproduce the $1/u^3$ singularity. If we take $n=6$ we get a contradiction as the l.h.s. scales as $g^0$ whereas in the r.h.s. we get a term $1/g^2$. The only way to avoid this problem is to set
 $f_{0,2}+f_{1,0}=0$. Using this relation
and setting $n=6$ we get for the residue of ${\Omega}(u)$ at 3rd order pole at $u=0$
\beq\la{onpole}
\lim_{u\to 0} u^3\,\Omega =
-2g^6s^{\gamma} f_{0,2} =\lim_{u\to 0} \Delta \Omega/u^3\,,
\eeq
where in the second relation we took into account \eq{f(g)} and \eq{f-f}. This equation provides us with  a set of nontrivial relations
between the entries of matrices \eq{DeltaO} and \eq{Ou3}.
In particular, we notice that the matrix element $\Delta{\Omega_2}^1$ is zero implying that ${\Omega_2}^1$ should vanish.
This leads to the {\it quantization condition}
\beq\label{quant}
q_4(0,m)q_2(0,-m)+q_2(0,m)q_4(0,-m)=0\;.
\eeq
In the similar manner, the vanishing of ${\Omega_{3}}^1$ implies $\Delta{\Omega_{3}}^1=0$ leading to
$  \left(\dot{q}_4^2-q_4^2\right) \beta_1^2+q_2^2-\dot{q}_2^2=0$. Together with \eq{quant} this gives
\beq
\beta_1^2=\frac{q^2_2}{q_4^2}=\frac{\dot q^2_2}{\dot q_4^2}\;.
\eeq
Finally imposing the relation \eq{onpole} for $\Omega_1{}^1$ and $\Delta\Omega_1{}^1$ we find that the parameter \(m^2\) is
related to the coupling constant in the bi-scalar theory \eqref{bi-scalarL}
\beq
m^2=-g^6s^6\;=-\xi^6.
\label{m_ident}\eeq
It is straightforward to verify that all remaining conditions \eq{onpole} are satisfied automatically!

The quantization condition \eq{quant} fixes the dependence of $\Delta$ on $m$. Together with \eq{m_ident} this allows us
to find
 the spectrum of anomalous dimensions of the states/operators \eq{O-def}  with charges \((\Delta(\xi),0,0|\L,0)\), obeying the parity invariance.
\footnote{We remind that the full superconformal symmetry of \({\cal N}=4\) SYM is broken by \(\gamma\)-deformation to \(PSU(2,2|4)\to SU(2,2)\times U(1)^3\) and an arbitrary state is still characterized by Cartan charges \((\Delta,S_1,S_2|J_1,J_2,J_3)\). For the bi-scalar theory under consideration the remaining symmetry is \(SU(2,2)\times U(1)^2\) (only two complex scalars out of three left) we label the states/operators with Cartan charges \((\Delta,S_1,S_2|J_1,J_2)\). For our particular BMN-state and its analogs considered further we take \((\Delta,0,0|\L,0)\).}

In the next sections we will explore these quantization conditions along with the Baxter equation \eq{baxter3} to study the dimensions of the underlying operators with charge \(J=3\) numerically, as well as in strong and weak coupling approximations.

\section{Numerical solution}
\label{sec:numerics}

In this section, we describe the numerical solution to Baxter equation \eq{baxter3} supplemented
with the quantization conditions \eq{quant} and \eq{m_ident}.
The numerical method for solving the whole QSC was developed in \cite{Gromov:2015wca} and it can be adopted for our
much simpler case (see also \cite{Gromov:2016rrp}). One first constructs analytically the solution to the Baxter equation
\eq{baxter3}
for large values of the spectral parameter $u$ where one can simply take the asymptotic series \eq{bax3} and find the coefficients of the expansion by plugging it into the equation and expanding for large $u$.
We kept around $12$ first orders which allowed us to gave extremely accurate approximation for $q_2(u,m)$
and $q_4(u,m)$ for ${\rm Im}\;u>100$. After that we used the exact Baxter equation \eq{baxter3} recursively to decrease the imaginary part of $u$ until we reach $u=0$. After that we define the function
\begin{align}
F(\Delta,\xi)=q_4(0,m) q_2(0,-m)+q_2(0,m) q_4(0,-m)\,,
\end{align}
which must be zero for the physical values of $\Delta$ and $m^2=-\xi^6$. One can simply use the \verb"FindRoot"
function in {\it Mathematica} to find $\Delta(\xi)$ determined by the condition $F(\Delta(\xi),\xi)=0$.

In this way we obtained the results for the scaling dimension of
the operator $\tr(\phi_1^3)$ shown in Fig.~\ref{fig:numericsL3}. We see that the dimension approaches $\Delta=2$,
where it collides  at $\xi^3\simeq 0.21$ with the ``mirror" solution obtained by $\Delta\to 4-\Delta$.
After that the two dimensions stay in the plane ${\rm Re}\,\Delta=2$, while their imaginary parts increase at large $\xi$ as $\Delta\sim\xi^{3/2}$. In the next section we describe this strong coupling asymptotics analytically.

\begin{figure}[h]
\begin{center}
\includegraphics[scale=0.55]{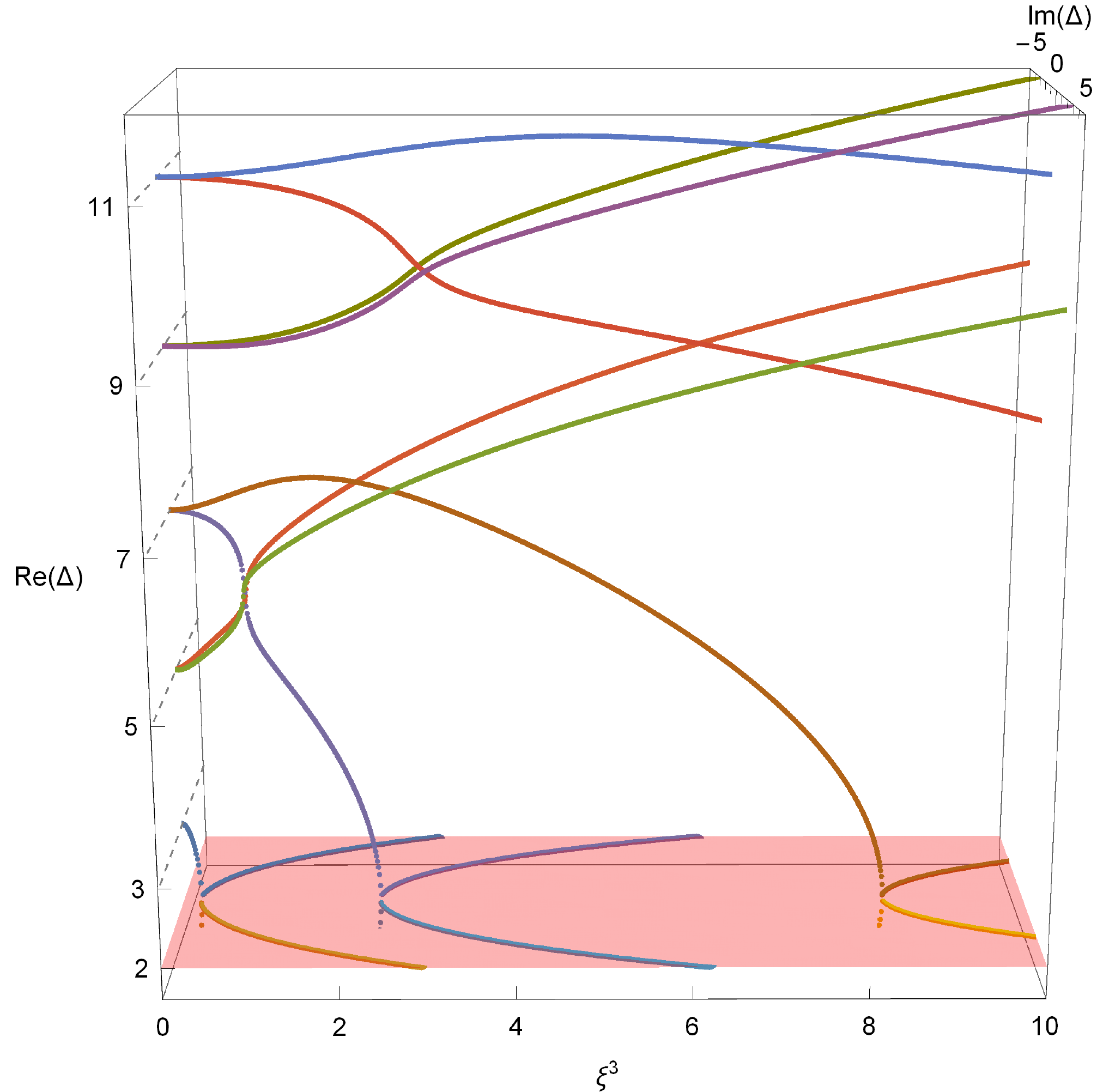}
\end{center}
\caption{Real and imaginary part of the scaling dimension of the nine lowest lying states with $\L=3$. The curve that starts at $\Delta(0)=3$ corresponds to the operator $\tr(\phi_1^3)$. The pair of states that start at $\Delta(0)=3+2k$ with $k=1,2,3,4$  correspond to the operators of the form \eq{O-def} (or rather to their linear combinations diagonalizing the dilatation operator). }
\label{Fig:res}
\end{figure}

We also found that the quantization condition \eq{quant} also describes other states with Cartan charges \((\Delta,0,0|L,0)\). As we first found  numerically (and then confirmed analytically, see the next section) the quantization condition $F(\Delta,\xi)=0$ is satisfied for several values of $\Delta$! At zero coupling,
these extra solutions all start at odd integer values of $\Delta$. For $\Delta{(0)}=5+4n\;,\;n=0,1,2,\dots$ the scaling dimension
$\Delta(\xi)$ become complex for arbitrary small $\xi>0$ whereas solutions with $\Delta{(0)}=3+4n\;,\;n=1,2,\dots$ are real for small $\xi$. Our numerical analysis suggests that, similar to the state with $\Delta{(0)}=3$, all solutions with $\Delta{(0)}=3+4n$
reach the value $\Delta=2$ and then acquire an imaginary part.

Finally, using the high precision numerics we extracted the expansion coefficients of the weak coupling expansion of $\Delta(\xi)$ for
the two lowest lying states
\begin{align}\label{Delta3}
\Delta_3&= 3-14.4246828379151314247968579381 \xi
^6-17.4934615492599154108489144266 \xi
^{12}\\
\nn&-1198.90916684527343375296340880 \xi
^{18}-4689.74599336134323308194176857 \xi
^{24}\\
\nn&-280246.105267046718780737267525 \xi
^{30}-1.76612732373253221270019811001\times 10^6\; \xi
^{36}\\
\nn&-8.77012836297716360442838113467\times 10^7\; \xi
^{42}+O\left(\xi ^{49}\right)\,,
\\[2mm]
\Delta_5&=
5-2.00000000000000000000000000000 i \xi
^3+3.00000000000000000000000000000 \xi
^6\\ \nn
&+7.75000000000000000000000000000 i \xi
^9-20.6438292905212171438007855155 \xi
^{12}\\ \nn
&-67.5066068073454771471035348196 i \xi
^{15}+233.347926388436938426879094509 \xi
^{18}\\ \nn
&+845.865771390416192186499791683 i \xi
^{21}-3168.44499021745756976755618573 \xi
^{24}+O\left(\xi
^{30}\right)\,.
\end{align}
We use these results in the next section to verify our analytic expressions.

\section{Weak coupling solution}

In this section, we describe the method for finding analytical solutions to the  Baxter equation \eq{baxter3}  with quantization condition \eq{quant} at weak coupling.
We keep the presentation short since it goes along the same lines as in \cite{Gromov:2016rrp,Gromov:2015vua}.

\subsection{Perturbative solution of the Baxter equation}

At the first step we have to find two linearly independent solutions to the Baxter equation \eq{baxter3} with the parameter $m$ satisfying \eq{m_ident}. At weak coupling, the solutions to \eq{baxter3} can be constructed perturbatively in powers of $m$.
To lowest order, for $m=0$ and $L\equiv \Delta(0)$ an odd integer,  the Baxter equation \eq{baxter3} reduces to that for the
$SL(2)$ spin chain of length $2$. As such, it has a solution \eq{q2} which for odd $L$ reduces to a polynomial.
The second solution then can be deduced from the Wronskian relation \eq{Wr}. It is a meromorphic function of $u$ with the second-order poles located in the lower half-plane
\begin{equation}
q_{I}={\cal P}_{(L-1)/2}(u)\,,\qquad\qquad q_{II}={\cal P}_{(L-1)/2}(u)\eta_2(u)+{\cal Q}_{(L-3)/2}(u)\,,
\end{equation}
where ${\cal P}_n$ and ${\cal Q}_n$ are polynomials of degree $n$ and the notation was introduced for the special
function $\eta_{s_1,\dots,s_k}(u)$  with an appropriate pole structure\footnote{The sum is divergent when $s_1=1$. The
divergent part does not depend on $u$ and can be regularized so that
$\eta_1(u)=i \psi(-i u)$ and $\eta_{1,s}(u)=\eta_1(u)\eta_s(u) -\eta_{s+1}(u)-\eta_{s,1}(u) $. More complicated $\eta_{1s_2,\dots,s_3}$
can be obtained recursively as explained in \cite{Leurent:2013mr}.
} \cite{Leurent:2013mr}
\begin{equation}\label{eta}
\eta_{s_1,\dots,s_k}(u)=\sum_{n_1>n_2>\dots>n_k\geq 0}\frac{1}{(u+in_1)^{s_1}\dots (u+in_k)^{s_k}}\;.
\end{equation}
In this way we find that the solutions to  \eq{baxter3} satisfying \eq{as-inf} are
given to the leading order in \(m=i\xi^3\) by
\begin{eqnarray}
\nn L=3&&\;\;:\;\;q_I=u\;\;,\;\;q_{II}=1\,,
\\[1.5mm]
L=5&&\;\;:\;\;q_I=u^2\;\;,\;\;q_{II}=u^2 \eta_2(u)+i u-\frac12\,,
\label{q-LO}
\\
\nn L=7&&\;\;:\;\;q_I=u^3\;\;,\;\;q_{II}=u^3 \eta_2(u)+i u^2-\frac{u}{2}-\frac{i}{6}\,.
\end{eqnarray}
To incorporate corrections in $m$, we use the ansatz $\Delta=L + \sum_k m^k \Delta^{(k)}  $
and look for solutions to \eq{baxter3} in the form $q_I(u)+\delta q_I(u)$ with $\delta q_I(u)=c_1(u) q_I(u)+c_2(u) q_{II}(u)$
and similar for $q_{II}(u)+\delta q_{II}(u)$. This leads to the system of first-order finite difference equations for the
coefficient functions $c_1(u)$ and $c_2(u)$ which can be solved order-by-order in $m$ in terms of the functions \eq{eta}
with polynomial coefficients. 

Let us consider the state with $L=3$. Numerical solution \eq{Delta3} suggests that corrections to $\Delta$ run in
powers of $m^2=-\xi^6$. Using the ansatz
\begin{align}\label{delta}
\Delta=3-m^2\delta + O(m^4)\,,
\end{align}
we compute corrections to the solutions \eq{q-LO}
\begin{eqnarray}
q_I&=&u-i m \left(\eta _1-\eta _2 u\right)+m^2 \left(-\eta _{1,2}+\eta _{2,1}+u \eta
_{1,3}-u \eta _{2,2}-\frac{i \delta }{2}+\frac{1}{2} i   \eta _1
u \delta+\frac{u \delta}{2}\right)\\
\nn&-&\frac{1}{2} i m^3 \left(-2 \eta _{1,2,2}+2 \eta
_{1,3,1}+2 \eta _{2,1,2}-2 \eta _{2,2,1}-i u \eta _{2,1} \delta+2 u \eta
_{1,2,3}-2 u \eta _{1,3,2}-2 u \eta _{2,1,3}\right.\\
\nn&&+\left.2 u \eta _{2,2,2}+2   \eta
_1\delta -2   \eta _2 u\delta\right)+{\cal O}\left(m^4\right)\,,\\
q_{II}&=&1-i m \left(\eta _2-\eta _3 u\right)+m^2 \left(-\eta _{2,2}+\eta _{3,1}+u \eta
_{2,3}-u \eta _{3,2}-\frac{1}{2} i\eta _1\delta +\frac{1}{2} i \eta _2
u\delta\right)\\
\nn&-&\frac{1}{2} i m^3 \left(-i \delta  \eta _{1,2}-2 \eta _{2,2,2}+2 \eta
_{2,3,1}+2 \eta _{3,1,2}-2 \eta _{3,2,1}+i \delta  u \eta _{1,3}-i \delta  u
\eta _{3,1}+2 u \eta _{2,2,3}\right.\\
\nn&&\left.-2 u \eta _{2,3,2}-2 u \eta _{3,1,3}+2 u \eta
_{3,2,2}+\delta  \eta _2-\delta  \eta _3 u\right)+{\cal O}\left(m^4\right)\;.
\end{eqnarray}
At the next stage we have to find particular linear combinations of these functions, $q_2(u)$ and $q_4(u)$, which have the correct
asymptotic behavior \eq{bax3} at infinity.

As follows from  \eq{bax3}, the leading correction to $q_2(u)$ and $q_4(u)$ at large $u$ should scale as $\delta q_2 = -\frac12 (m^2 \delta) u \ln u $ and $\delta q_4 = \frac12(m^2 \delta)\ln u  $, respectively.
Expanding $q_I(u)$ and $q_{II}(u)$ at large $u$ and matching the coefficients we find that solutions to the Baxter equation
satisfying \eq{bax3} are given by
\begin{eqnarray}\notag
q_2(u,m)&=&\left(
1-\frac{1}{4} i (\pi -2 i) m^2 \delta
\right)q_I(u)+\left(-\frac{1}{m\delta}+\frac{1}{4} i (\pi +2 i) m+\frac{1}{4} i m^2\delta\right)q_{II}(u)\,,
\\
q_4(u,m)&=&\left(1+\frac{1}{4} i (\pi +2 i) m^2\delta\right)q_{II}(u)\;.
\end{eqnarray}
To analyze the quantization condition \eq{quant},
we have to evaluate these expressions at the origin. Making use of the identity
$$
\eta_{s_1,s_2,\dots,s_k}(i)=(-i)^{s_1+s_2+\dots+s_k}\zeta_{s_1,s_2,\dots,s_k}\,,
$$
as well as  relations between the multiple zeta values, e.g. $\zeta_{2,2,2}={\pi ^6}/{5040}$,
we arrive at
\begin{eqnarray}
q_2(0)&=&-\frac{1}{m \delta}-\frac{i \zeta _2}{\delta }+m \left(\frac{\zeta _{2,2}}{\delta
}-\frac{\zeta _{3,1}}{\delta }-\frac{\tilde\zeta _1}{2}+\frac{i \pi
}{4}-\frac{1}{2}\right)+\frac{i m^2}{4 \delta } \bigg(4 \zeta _{2,2,2}-4 \zeta _{2,3,1}
\\
&-&4 \zeta _{3,1,2}  +4 \zeta
_{3,2,1}+\delta ^2+ (i \pi    \zeta _2-4   \zeta _2+2   \zeta
_3-2   \tilde\zeta _{1,2}+4
\zeta _{2,1})\delta \bigg)+{\cal O}\left(m^3\right)\,,
\nn \\
q_4(0)&=&1+i \zeta _2 m+m^2 \left(-\zeta _{2,2}+\zeta _{3,1}-\frac{\tilde\zeta
_1 \delta}{2}+\frac{i \pi  \delta }{4}-\frac{\delta }{2}\right)\\
&+&\frac{1}{4} m^3
\left(-4 i \zeta _{2,2,2}+4 i \zeta _{2,3,1}+4 i \zeta
_{3,1,2}-4 i \zeta _{3,2,1}-(\pi \zeta _2 +2 i\zeta
_3+2 i \tilde \zeta _{1,2}) \delta \right)+{\cal O}\left(m^4\right)\,,
\nn
\end{eqnarray}
where in our conventions $\tilde\zeta_1=i\eta_1(i)=\gamma$ is the Euler constant and
$\tilde \zeta_{1,2} =i^3\eta_{1,2}(i)=\frac{\gamma  \pi ^2}{6}-2 \zeta_3$.
We use these relations to find
\begin{equation}
q_2(0,m)q_4(0,-m)+q_4(0,m)q_2(0,-m)=\frac{1}{2} i m^2 (\delta +12 \zeta_3)+O\left(m^3\right)\,.
\end{equation}
The quantization condition \eq{quant} yields $\delta=-12\zeta_3$ and, together with \eq{delta},  fixes
the dependence of the scaling dimension on the coupling constant.

With a help of {\it Mathematica} we pushed the calculation of \eq{delta} up to the order $m^{12}$
and arrived at the weak coupling expansion of $\Delta(\xi)$ given by \eq{D3exp} in the Introduction.

The above analysis can be repeated for the states with $L =5,7,9$. We present below the results for
weak coupling expansion of scaling dimensions of these states.

\paragraph{States with $L=5$:} Solving the quantization conditions \eq{quant}, we  found two states with $\Delta=5$
at zero coupling. In distinction from \eq{delta}, their scaling dimensions at weak coupling run in powers of $m$, or equivalently
in powers of $i \xi^3$, and are complex conjugated to each other:
\begin{eqnarray}
\nn\Delta_{5,A}=5-2 i \xi ^3+3 \xi ^6+\frac{31 i \xi ^9}{4}+\xi ^{12} \left(3 \zeta_3-\frac{97}{4}\right)+i\xi ^{15} \left(\frac{27 \zeta_3}{2}-\frac{5359 }{64}\right)+\xi ^{18} \left(-\frac{219 \zeta_3}{4}-\frac{15 \zeta_5}{2}+\frac{4911}{16}\right)\\
\Delta_{5,B}=5+2 i \xi ^3+3 \xi ^6-\frac{31 i \xi ^9}{4}+\xi ^{12} \left(3 \zeta_3-\frac{97}{4}\right)-i\xi ^{15} \left(\frac{27 \zeta_3}{2}-\frac{5359}{64}\right)+\xi ^{18} \left(-\frac{219 \zeta_3}{4}-\frac{15 \zeta_5}{2}+\frac{4911}{16}\right)\notag\\
\label{DeltaL5}\end{eqnarray}

\paragraph{States with $L=7$:} As in the previous case,  we found two states. Similar to the state with
$L=3$, their scaling dimensions
are real at weak coupling and have an expansion in powers of $\xi^6$
\begin{eqnarray}\notag
\Delta_{7,A}&=&7-\frac{\xi ^6}{2}-\frac{17 \xi ^{12}}{64}-\frac{891 \xi ^{18}}{4096}-\frac{27465 \xi ^{24}}{131072}\,, \\
\Delta_{7,B}&=&7+\frac{\xi ^6}{2}-\frac{23 \xi ^{12}}{64}+\xi
^{18} \left(\frac{15283}{36864}-\frac{\zeta_3}{12}\right)+\xi ^{24} \left(\frac{65 \zeta_3}{384}-\frac{678575}{1179648}\right)\,.
\end{eqnarray}
\paragraph{State with $L=9$:} We found two states and their properties are similar to those of $L=5$ states
\begin{eqnarray}\notag
\Delta_{9,A}&=&9-\frac{i \xi ^3}{3}+\frac{7 \xi ^6}{216}-\frac{223 i \xi ^9}{10368}+\xi ^{12} \left(\frac{\zeta_3}{432}+\frac{17029}{1119744}\right)+i\xi ^{15} \left(\frac{1424867
}{214990848}-\frac{31\zeta_3}{31104}\right),\\
\Delta_{9,B}&=&9+\frac{i \xi ^3}{3}+\frac{7 \xi ^6}{216}+\frac{223 i \xi ^9}{10368}+\xi ^{12} \left(\frac{\zeta_3}{432}+\frac{17029}{1119744}\right)-i\xi ^{15} \left(\frac{1424867
}{214990848}-\frac{31\zeta_3}{31104}\right).
\end{eqnarray}
We verified that these relations are in perfect agreement with the numerical consideration at weak coupling.
The strong coupling expansion of the scaling dimensions is discussed in  section \ref{sec:strong}.


\subsection{Logarithmic multiplet}
\label{sec:log_multiplet}

As was mentioned above, the scaling dimensions of operators with bare dimension $\Delta(0)=5,9,\dots$ have expansion
in powers of $\xi^3$ and not in powers of $\xi^6$ as it happens for operators with bare dimension $\Delta(0)=3,7,\dots$. In this section we show that this property reflects a very unusual feature of bi-scalar \chiFT\  theory
first noticed by Joao Caetano~\cite{Caetano:TBP}: the operators with $\Delta(0)=5,9,\dots$ behave as conformal primary operators in a logarithmic four-dimensional conformal field theory.

To simplify the consideration, we examine the simplest case of operators with bare dimension $\Delta(0)=5$ and the $R-$charge
$\L=3$. As was mentioned in the Introduction,
such operators can be obtained from $\tr(\phi_1^3)$ operator by inserting a pair of scalar fields $\phi_2$ and
$\phi_2^\dagger$ inside the trace and by dressing scalar fields with derivatives, $\tr(\phi_1^2\Box\phi_1)$ and
$\tr(\phi_1\partial^\mu \phi_1\partial_\mu \phi_1)$. In the latter case, we can use the equations of motion in the theory
(\ref{bi-scalarL})  to show that the operators with derivatives can be expressed in terms
of the former operators as well as the conformal descendant operator $\Box\tr(\phi_1^3)$. This allows us to define the
basis  of dimension$-5$ operators
\begin{align}\label{O's}
O_1=\tr(\phi_1^3 \phi_2 \phi_2^\dagger)\,,\qquad
O_2=\tr(\phi_1^2 \phi_2\phi_1 \phi_2^\dagger)\,,\qquad
O_3=\tr(\phi_1 \phi_2\phi_1^2 \phi_2^\dagger)\,,\qquad
O_4=\tr(\phi_2\phi_1^3 \phi_2^\dagger)\,.
\end{align}
At quantum level, these operators mix with each other and their anomalous dimensions can be found by diagonalizing
the corresponding mixing matrix $V_{ij}$
\begin{align}\label{RG}
\mu {d\over d\mu} O_i(x) = -V_{ij} O_j(x)\,,
\end{align}
where $\mu$ is an ultraviolet cut-off and the matrix elements $V_{ij}$ describe the mixing $O_i\to O_j$.

\begin{figure}
\center{\includegraphics[scale=0.8]{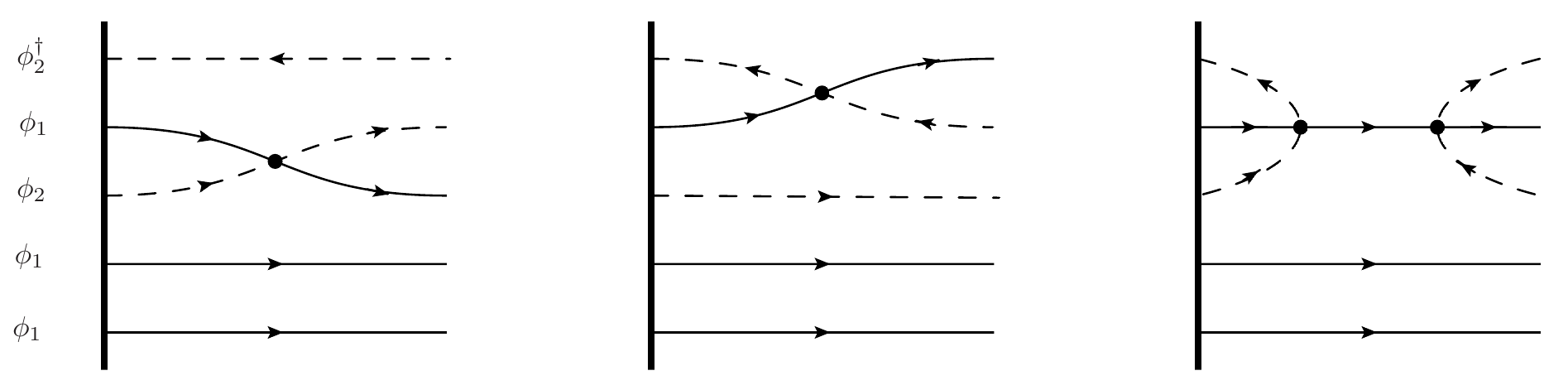}}
\caption{\label{fig:trans}Mixing of $O_3=\tr(\phi_2^\dagger\phi_1 \phi_2\phi_1^2 )$ with the operators $O_4$ and $O_2$ to the lowest order in the coupling. Three diagrams correspond to three transitions defined in (\ref{trans}). Solid line denotes field $\phi_1$,
dashed line represents $\phi_2$ and $\phi_2^\dagger$ depending on the direction of the arrow.}
\end{figure}

To the lowest order in the coupling, the quartic scalar interaction vertex in (\ref{bi-scalarL}) induces the following transitions
\begin{align}\label{trans}
\phi_1\phi_2\to \phi_2\phi_1\,,\qquad\quad \phi_2^\dagger\phi_1\to \phi_1\phi_2^\dagger
\,,\qquad\quad \phi_2^\dagger\phi_1\phi_2 \to \phi_1 \to \phi_2\phi_1\phi_2^\dagger\,.
\end{align}
The corresponding Feynman diagrams are shown in Fig.~\ref{fig:trans}. Applying these rules we find that the operators
(\ref{O's}) mix as follows
\begin{align}\label{O-mix}
O_1 \to 2 O_2\,,\qquad\quad O_2 \to 2 O_3\,,\qquad\quad O_3 \to 2 O_4 + O_2\,,
\end{align}
where the factor of $2$ is due to the fact that the first two transitions in (\ref{trans}) yield the same operator. The last transition in
(\ref{trans}) produces the additional mixing between the operators $O_3$ and $O_2$. Notice that the operator $O_4$ is
not affected by the transitions (\ref{trans}).

As follows from (\ref{O-mix}), the mixing matrix for the operators (\ref{O's}) takes the following form to leading order in $\xi^2$
\begin{align}\label{mix-mat}
V = \left[\begin{array}{cccc}0 & 2\xi^2 \gamma_1 & {\cal O}(\xi^4) & {\cal O}(\xi^6) \\0 & 0 & 2\xi^2 \gamma_1 & {\cal O}(\xi^4) \\0 & \xi^4 \gamma_2 & 0 & 2\xi^2 \gamma_1 \\0 & 0 & 0 & 0\end{array}\right]\,.
\end{align}
Here $\xi^2\gamma_1$ and $\xi^4\gamma_2$ describe the first two and the last transitions  in (\ref{trans}), respectively. The terms appearing as \({\cal O}(\xi^4)\) 
and  \({\cal O}(\xi^6)\)  in the matrix do not contribute to anomalous dimensions 
so we ignore them in what follows. Using the dimensional regularization with $D=4-2\epsilon$, we can find them as the residue at a simple
pole $1/\epsilon$ of diagrams shown in Fig.~\ref{fig:trans}
\begin{align} \label{one-loop}
\gamma_1 = -2 \,, 
\qquad\qquad
\gamma_2 =  1 \,. 
\end{align}
Notice that $\gamma_1$ and $\gamma_2$ have an opposite sign.

The mixing matrix (\ref{mix-mat}) has a number of unusual properties.
In unitarity conformal field theory this matrix has to be Hermitian. Since \chiFT\ theory is not unitary, the matrix
(\ref{mix-mat}) does not have this property. This implies that its eigenvalues are, in general, complex valued functions of the
coupling $\xi$. Indeed, the scaling dimensions \eq{DeltaL5} develop
imaginary part at weak coupling.

Secondly, the matrix (\ref{mix-mat}) has rank $3$ and it can be brought to the Jordan canonical form by a similarity transformation
\begin{align}
V=  g^{-1}  J   g\,,\qquad \quad
J =\left(
\begin{array}{cccc}
 0 & -1 & 0 & 0 \\
 0 & 0 & 0 & 0 \\
 0 & 0 & 2 i \xi ^3 & 0 \\
 0 & 0 & 0 & -2 i \xi ^3 \\
\end{array}
\right)
,\qquad\quad   g=\left(
\begin{array}{cccc}
 \xi ^2 & 0 & 4 & 0 \\
 0 & 0 & 0 & 16 \xi ^2 \\
 0 & \xi ^2 & 2 i \xi  & -4 \\
 0 & \xi ^2 & -2 i \xi  & -4 \\
\end{array}
\right).
\end{align}
If the matrix $J$ were diagonalizable, its eigenspectrum would define four different conformal operators. Since $J$ contains $2\times 2$ Jordan
block, the situation is more complex. Namely, we can use the lower diagonal $2\times 2$ block of the matrix $J$ to define two  conformal operators
$\xi^2 O_2\mp 2i\xi O_3-4 O_4$
with the anomalous dimension
\begin{align} \label{gpm}
\gamma_\pm 
= \pm 2i\, \xi^3 
+{\cal O}(\xi^6)\,.
\end{align}
Notice that this expression scales as $O(\xi^3)$, in agreement with (\ref{DeltaL5}).

The Jordan block of $J$ describes the mixing matrix for the pair of the operators $16 O_4$ and $O_1+4O_3/\xi^2$ that we denote as $A$ and $B$.
The form of this block is fixed by the interaction term in \eq{bi-scalarL} and is protected from quantum corrections. The pair of the
operators $A$ and $B$ belongs to the same conformal multiplet with the conformal weight $\Delta=5$, a phenomenon typical for logarithmic conformal field theories \cite{Gurarie:1993xq}. To show this, we define the set of operators conjugated to \eq{O's}
\begin{align}\notag
{}& \bar O_1= \tr((\phi_1^\dagger)^3 \phi_2^\dagger \phi_2)\,,&&
\bar O_2=\tr((\phi_1^\dagger)^2 \phi_2^\dagger \phi_1^\dagger \phi_2)\,,
\\
{}& \bar O_3=\tr(\phi_1^\dagger  \phi_2^\dagger (\phi_1^\dagger)^2\phi_2)\,,&&
\bar O_4=\tr(\phi_2^\dagger (\phi_1^\dagger)^3\phi_2)\,,
\end{align}
so that $\bar O_i(x) = O_{5-i}^\dagger(x)$ with $i=1,\dots,4$. The reason why we label these operators in such a way is that they
satisfy the same evolution equation \eq{RG} as operators $O_i$ with the mixing matrix given by \eq{mix-mat}. As a consequence, two of the operators 
have the anomalous dimension \eq{gpm} and the two remaining ones $\bar A=16 \bar O_4$ and $\bar B= \bar O_1+4\bar O_3/\xi^2$ have a mixing matrix given by $2\times 2$ Jordan cell. Computing the correlation functions of the operators $A(x), B(x)$ and
$\bar A(0),\bar B(0)$ we find  
\begin{align}\notag\label{log}
{}& \langle {A(x) \bar A(0)}\rangle = 0\,, &&\langle {B(x) \bar B(0)}\rangle = c{\ln (x^2\mu^2)\over (x^2)^5}\,,
\\
{}& \langle {A(x) \bar B(0)}\rangle = {c\over (x^2)^5}\,, &&
\langle {B(x) \bar A(0)}\rangle = {c\over (x^2)^5}\,,
\end{align}
where $c$ is the normalization factor. Here $ \langle {A(x) \bar B(0)}\rangle\sim  \langle {O_4(x) \bar O_1(0)}\rangle=\langle {O_4(x) O_4^\dagger(0)}\rangle$ is given by the product of five scalar propagators. At the same time,
the correlation function
$\langle {A(x) \bar A(0)}\rangle \sim \langle O_4(x) \bar O_4(0)\rangle$ vanishes since, due to different ordering of scalar fields
in the operators $O_4$ and $\bar O_4$,  the same product of scalar propagators
is accompanied by a nonplanar color factor. The correlation function $\langle {B(x) \bar B(0)}\rangle$ receives a logarithmically
enhanced correction coming from the transition $O_3\to O_4$ (see the first two diagrams in Fig.~\ref{fig:trans}) and from similar
transition $\bar O_3\to \bar O_4$.
It is easy to verify that the relations \eq{log} are in agreement with the evolution equation
\eq{RG}. They coincide with analogous expressions for two-point correlation functions in a logarithmic conformal field theory \cite{Gurarie:1993xq}.


It is straightforward to extend the above analysis to higher orders in the coupling. We can use the transitions (\ref{trans}) to generate
higher order Feynman diagrams. Starting from order $O(\xi^6)$ a new transition appears. To see this we notice that the rightmost diagram
in Fig.~\ref{fig:trans} has an intermediate state of three scalar fields $\phi_1$. In a close analogy with $\tr(\phi_1^3)$ operator,
we can dress this state by  an arbitrary number of wheel graphs. Such graphs provide higher order contribution to the transition
$O_3\to O_2$ and modify the eigenvalues of the mixing matrix.

\section{Strong coupling expansion}
\label{sec:strong}

In this section, we study the properties of scaling dimensions of the operators in bi-scalar \chiFT\ theory \eq{bi-scalarL} at strong coupling $\xi \gg 1$. The numerical results shown in Fig.~\ref{Fig:res} suggest that the scaling dimensions exhibit remarkable
regularity at strong coupling and
fall into two
different groups. The first group consists of functions $\Delta(\xi)$ that  start at zero coupling at $\Delta(0) =3,7,11, \dots$ and behave at strong coupling as \begin{equation}\label{Delta-as}\Delta(\xi)=2+ i d(\xi)\,,\qquad \text{with}\quad d(\xi)\sim \xi^{3/2}.\end{equation} The second group consists of functions that start at $\Delta(0)=5,9,13,\dots$,
take complex values for $\xi\neq 0$ and scale at strong coupling as $\Delta(\xi) \sim \xi$. To explain these properties, we solve
the Baxter equation \eq{baxter3} with quantization conditions \eq{quant} at strong coupling using semiclassical methods.

We remind that in planar $\mathcal N=4$ SYM theory the scaling dimensions of operators are identified through the
AdS/CFT correspondence with energies of classical strings on the $AdS^5\times S_5$ background. The analysis in this section
suggests that asymptotic behavior of the scaling dimensions in strongly coupled bi-scalar \chiFT\ theory is described by
a classical integrable conformal spin chain with a finite number of non-compact spins. We postpone its
detailed exploration to a future publication.


\subsection{Baxter equation at strong coupling} \label{8.1}

Our strategy in this section is to solve the Baxter equation \eq{baxter3} and, then, use the quantization conditions
\eq{quant} and \eq{m_ident} to find $\Delta(\xi)$ at large $\xi$.

It is convenient to change variables $u=iv$ and
introduce notation for
\begin{align}
q(v) \equiv  {q_2(iv)}/{i^{\frac{\Delta-1}2}}\,, \qquad\qquad \bar q(v)\equiv  {q_4(iv)}/{i^{\frac{3-\Delta}2}}\,.
\end{align}
By definition, these functions have third-order poles at negative integer $v$ and for large positive $v$ they behave as
\begin{align}\label{q(v)}
q(v) = v^{\frac{\Delta-1}2} (1+ O(1/v))\,,\qquad\qquad \bar q(v) = v^{\frac{3-\Delta}2} (1+ O(1/v))\,.
\end{align}
In addition, they satisfy the Baxter equation
\begin{align}\label{bax-t}
q(v+1)+q(v-1) = t(v) \, q(v)\,,
\end{align}
where the notation is introduced for
\begin{align}\label{t-d}
t(v) = 2 - { \xi^2 \mathfrak d \over v^2} + {\xi^3\over v^3}\,,\qquad\qquad \mathfrak d=-{(\Delta-2)^2-1\over 4\xi^2 }\,.
\end{align}
Here we used \eq{m_ident} to replace $m=i\xi^3$. As we show below, the scaling dimensions are real functions of $m$ and,
therefore, the second solution  $m=-i\xi^3$ leads to a complex conjugated expression for $\Delta$. To find the function
$\Delta(\xi)$ for $\xi\gg 1$ we  have to solve the quantization condition \eq{quant} that takes the form
\begin{align}\label{Phi}
F(v)= q(v,\xi) \bar q(v,-\xi)+ \bar q(v,\xi)q(v,-\xi) \to 0 \,,\qquad\qquad \text{for $v\to 0$} \,,
\end{align}
where $q(v,\xi)$ and $\bar q(v,\xi)$ denote solutions to \eq{q(v)} and \eq{bax-t}.

To apply the quantization condition \eq{Phi} we have to construct solutions to \eq{bax-t} for small $v$ and $\xi \gg 1$.
This will be done in two steps. First, we solve \eq{bax-t} for large positive $v$ such that $v=O(\xi)$. The solutions are fixed
uniquely by the condition for $q(v,\xi)$ and $\bar q(v,\xi)$ to satisfy \eq{q(v)} for $v\gg \xi$.  At the second step, we
construct solutions to \eq{bax-t} for fixed $v$ and $\xi\to \infty$ and require that the two
sets of solutions can be sewed together  in the transition region $v\ll \xi$. In this way, we obtain the functions $q(v,\xi)$ and $\bar q(v,\xi)$ which satisfy the Baxter equation \eq{bax-t} for arbitrary $v$ and have asymptotic behavior \eq{q(v)}.

\subsection{Shortcut to the solution}

In this subsection, we present a shortcut to finding the exact solution to \eq{Phi}. For this purpose, we concentrate on the states that  start at zero coupling at $\Delta(0) =3,7,11, \dots$ (the remaining states will be discussed in Sect.~\ref{lastqq}). As mentioned above, at strong coupling their scaling dimensions scale as \eq{Delta-as}
with a real valued function $d(\xi)$ depending on the state. We introduce integer positive $N$ to count
different functions $d(\xi)$ in the order in which the corresponding functions $\Delta(\xi)$ approach the plane $\rm Re\, \Delta(\xi)=2$ (see Fig.~\ref{Fig:res}),
 i.e. $N=1$ for $\Delta(0)=3$ state,
$N=2,3$ for $\Delta(0)=7$ states, $N=4,5$ for $\Delta(0)=11$ states and so on.

It is convenient to invert the dependence $d=d(\xi)$ and introduce the function
\begin{align}\label{varphi}
\xi^3 =  d^2 \varphi(d)\,.
\end{align}
We expect from \eq{Delta-as} that $\varphi(d)$ should approach a constant value for $d\to\infty$ and look for its general expression in the form
\begin{align}\label{phi-exp}
\varphi(d) = \varphi_0 + {1\over d^2} \varphi_1 + {1\over d^4} \varphi_2 + \dots\,,
\end{align}
where expansion runs in even powers of $1/d$.

At large $v$ the solutions to the Baxter equation \eq{bax-t} can be constructed as a formal series in $1/v$. The leading term of the expansion is given
by \eq{q(v)}, e.g.
\begin{align}
q(v) = v^{\frac{1+ i\,d }2} \left[1+\frac1{v} q_1 +  \frac1{v^2} q_2 +\dots \right]
\end{align}
and similar for $\bar q(v)$. The coefficients $q_1,q_2,\dots$ can be found by plugging the expansion into \eq{bax-t} and matching the coefficients in front of different
powers of $1/v$. They are given by rational functions in $d$ and have a power-like behavior at large $d$.

Let us now examine the same combination
of the $q$ and $\bar q$ functions that enters into the quantization condition \eq{Phi}
\begin{align}\label{F-form}
F(v)= q(v,\xi) \bar q(v,-\xi)+ \bar q(v,\xi)q(v,-\xi)  = 2v + \frac1{v} F_1 +\frac1{v^3} F_3 + O(1/v^5)\,.
\end{align}
Due to the symmetry of the Baxter equation \eq{bax-t} under $v\to -v$ and $\xi\to -\xi$, the expansion runs in odd powers of $v$. At large $d$ the first few
expansion coefficients are given by
\begin{align}\notag\label{F-coef}
 F_1{}& = -\frac{d^2 }{16
        } \left(8
        \varphi_0-1\right) \left(8
        \varphi_0+1\right)+
        \frac{1}{8} \left(128
        \varphi_0^2-64 \varphi_1
        \varphi_0+1\right) + \dots\,,
 \\   \notag
 F_3{}&=
        \frac{d^4}{3072} \left(8
                \varphi_0-3\right) \left(8
                \varphi_0-1\right) \left(8
                \varphi_0+1\right) \left(8
                \varphi_0+3\right)
\\
{}&
        +\frac{d^2}{384}   \left(-10240
       \varphi_0^4+2048
        \varphi_1\varphi_0^3+224
        \varphi_0^2-160 \varphi_1
        \varphi_0-3\right)+ \dots\,,
\end{align}
where dots denote terms suppressed by powers of $1/d^2$. Notice that $F_1, F_2, \dots$ scale at large $d$ as $F_k\sim d^{k+1}$ and the coefficients in front
of powers of $d$ are invariant under $\varphi_i \to -\varphi_i$.

We would like to emphasize that the relation \eq{F-form} holds at large $v$ whereas in order to solve the quantization condition \eq{Phi} we need to know
the function $F(v)$ for small $v$. One may try to resum the series in  \eq{F-form} and analytically continue it to small $v$. As we show below,
this proves to be a nontrivial task since  the function $F(v)$ has complicated analytical properties along the positive $v-$axis. Instead of following
this route, we use \eq{F-form} and impose the following additional condition: for large but fixed $v$ and $d\to\infty$ the function \eq{F-form} should
scale as
\begin{align}\label{quant-aux}
F(v) = O\left(d^{N-1}\right)\,,
\end{align}
with $N=1,3,5\dots$. As we will see in a moment, this condition fixes unambiguously the coefficients of the expansion \eq{phi-exp} and
yields a prediction for the function $d(\xi)$ in \eq{Delta-as}, which is in a perfect agreement with the numerical results shown in Fig.~\ref{states2}.
However it is not obvious {\it a priori} why the relation \eq{quant-aux} is equivalent to the exact quantization condition \eq{Phi}. We clarify this issue
in section~\ref{sect:as}.

\begin{figure}
\centerline{
        \includegraphics[width = 0.8\textwidth]{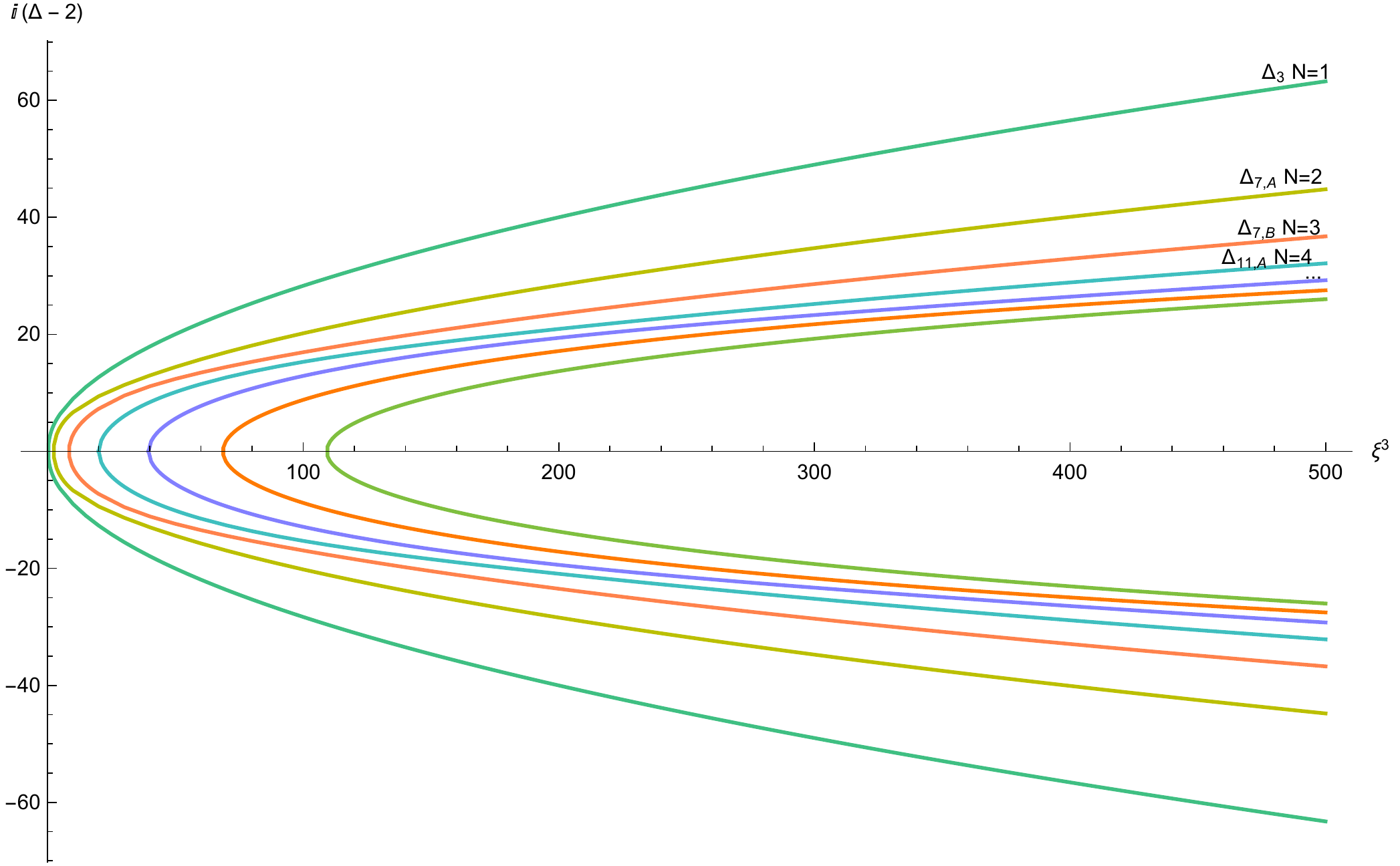} }
        \caption{Scaling of ${|\Delta-2|}$ at large $\xi$ for several states. \la{states2}}
\end{figure}

Let us examine the relation \eq{quant-aux} for a few values of $N$.

\subsubsection{States with odd $N$}

For $N=1$ we find from \eq{F-coef}
that the condition $F_1=O(d^0)$ implies the vanishing of $O(d^2)$ term. In the similar manner, the condition
$F_3=O(d^0)$ translates into the vanishing of $O(d^4)$ and $O(d^2)$ terms. This fixes the values of the coefficients
$\varphi_0=\pm 1/8$ and $\varphi_1=\mp 1/8$. It is quite nontrivial that the same procedure can be applied to higher $F_k$ since
the number of terms to cancel grows with $k$ and the system of equations for the $\varphi-$coefficients become
overdetermined. We checked explicitly up to $F_{23}$ term that this system has two solutions for \eq{phi-exp}
\begin{eqnarray}\nn\la{res3}
\varphi_{N=1}(d)&=&\frac{1}{8}-\frac{1}{8 d^2}-\frac{7}{4
        d^4}-\frac{169}{4
        d^6}-\frac{3511}{2
        d^8}-\frac{209057}{2
        d^{10}}-\frac{33305757}{4
        d^{12}}-\frac{3413828955}{4
        d^{14}}-\frac{438519141555}{4
        d^{16}}\la{frakm3}\\&-&\frac{69161788659565}{4
        d^{18}}
-\frac{13165550516521529}{4
        d^{20}}-\frac{2980425673116579991}{
        4
        d^{22}}+{\cal O}\left(\frac{1}{d^{24}}\right)\,,
\end{eqnarray}
and the second one that differs by the sign. Substituting these expressions into \eq{F-coef} we verify that the  function \eq{F-form} has the
expected asymptotic behavior  \eq{quant-aux}. Moreover, its expansion simplifies significantly and can be easily resummed
\begin{equation}\la{lead}
F_{N=1}(v) =
2 v+\frac{1}{2 v}+\frac{1}{8
        v^3}+\frac{1}{32 v^5}+\frac{1}{128
        v^7}+\frac{1}{512
        v^9}+\dots =
\frac{8 v^3}{4 v^2-1}+{\cal O}\left(\frac{1}{d^2}\right)\,.
\end{equation}
The relations \eq{varphi} and \eq{res3} fix the dependence $d=d(\xi)$ for $N=1$.
As can be seen from Fig.~\ref{states3},  the resulting function
$d_{N=1}(\xi)$ is in agreement with the numerical results for the state with $\Delta(0)=3$.

\begin{figure}
        \centerline{
                \includegraphics[width = 0.7\textwidth]{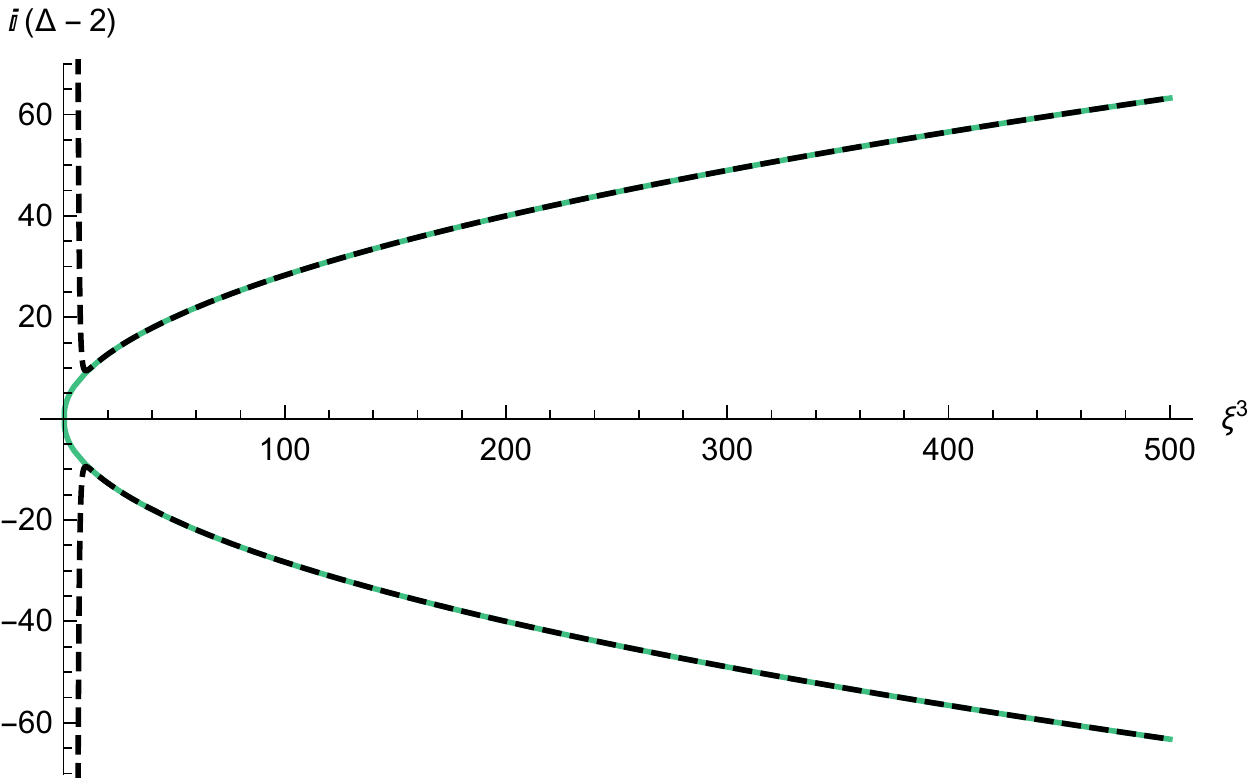} }
        \caption{Comparison of the numerical values for ${\pm|\Delta-2|}$ (solid green line) for the state with $\Delta(0)=3$ with the large $\xi$
        expansion  \eq{varphi} and \eq{res3} (dotted black line). The lines almost coincide for up to very small values of $\xi$. \la{states3}}
\end{figure}

For $N=3$ the relation \eq{quant-aux} does not lead to any condition for $F_1$ in \eq{F-coef} but requires the vanishing of the $O(d^4)$ term in
$F_3$. This gives  a new solution with $\varphi_0=\pm 3/8$. As in the previous case, the condition $F_k=O(d^2)$ for $k=5,7,\dots$ allows us
to determine high $\varphi-$coefficients leading to
\begin{eqnarray}\nn\la{res11}
\varphi_{N=3}(d)&=&\frac{3}{8}-\frac{51}{8 d^2}-\frac{837}{4 d^4}-\frac{81675}{4 d^6}-\frac{5872365}{2
        d^8}-\frac{1075210659}{2 d^{10}}-\frac{470669185863}{4 d^{12}}-\frac{119219450547393}{4
        d^{14}}\\
&-&\frac{34321640048050473}{4 d^{16}}-\frac{11112335132337334359}{4
        d^{18}}+{\cal O}\left(\frac{1}{d^{21}}\right)
\end{eqnarray}
and to the second solution that differs by the sign.
For these values of the coefficients, we get a simple expression for the function \eq{F-form} and \eq{F-coef}
\begin{align}\la{lead1}\notag
F_{N=3}(v) {}&=-
\frac{d^2}{2 v}-\frac{5 d^2}{4 v^3}-\frac{91 d^2}{32 v^5}-\frac{205 d^2}{32 v^7}-\frac{7381
        d^2}{512 v^9}-\frac{33215 d^2}{1024 v^{11}}-\frac{597871 d^2}{8192
        v^{13}}+\dots
\\
{}& =-\frac{8 d^2 v^3}{\left(4 v^2-1\right) \left(4 v^2-9\right)}+{\cal O}\left(d^0\right)\,.
\end{align}
We verified that the function $d_{N=3}(\xi)$ defined by \eq{varphi} and \eq{res11} is in agreement with the numerical value for $|\Delta-2|$ for
one of the two states with $\Delta(0)=7$.

For $N=5$ going along the same lines as before we find from \eq{quant-aux}
\begin{align}
\nn
\varphi_{N=5}(d)&=
\frac{5}{8}-\frac{245}{8 d^2}-\frac{9875}{4 d^4}-\frac{2244125}{4 d^6}-\frac{358818875}{2
        d^8}-\frac{140748665125}{2 d^{10}}-\frac{127555155602625}{4
        d^{12}}
\\ \nn
&- \frac{64666557386856375}{4 d^{14}}-\frac{35994292566223479375}{4
        d^{16}}+{\cal O}\left(\frac{1}{d^{18}}\right)\;.
\\
F_{N=5}(v) {}&= \frac{8 d^4 v^3}{\left(4 v^2-1\right) \left(4 v^2-9\right)\left(4 v^2-25\right)}+{\cal O}\left(d^0\right)\,.
\end{align}
These expressions correctly describe numerical values of $|\Delta-2|$ for
one of the two states with $\Delta(0)=11$. The same pattern persists to higher odd $N$ and $\Delta(0)$.

\subsubsection{States with even $N$}\label{Sect:half}

So far we demonstrated that the relation \eq{quant-aux} describes half of the states with odd $\Delta(0)\ge 5$. The question arises
of how to get the remaining states corresponding to even $N=2,4,\dots$. In an analogy with \eq{F-form} and \eq{quant-aux}, we expect that they should
satisfy the condition
\begin{align}\label{quant-aux1}
F^-(v)=O(d^{N-1})\,.
\end{align}
For $N$ even, large $d$ asymptotics of this function should contain odd powers of $d$ only and, therefore,  $F^-(v)$ should change
the sign under $d\to -d$. Notice that the same transformation exchanges the two solutions to the Baxter equation
$q(v,\xi)$ and  $\bar q(v,\xi)$. The function \eq{F-form} is obviously invariant under $q(v,\xi) \leftrightarrow \bar q(v,\xi)$ and, as a consequence,
its coefficients \eq{F-coef} are even functions of $d$. To get a parity odd function, it is sufficient to flip the sign in \eq{F-form}
\begin{align}\label{F-minus}
F^-(v)= i \left[q(v,\xi) \bar q(v,-\xi) - \bar q(v,\xi)q(v,-\xi) \right]\,.
\end{align}
Its expansion at large $v$ looks as
\begin{align}\nn
F^-(v) &= {4d\varphi_0\over v}
-\frac{d^3 \varphi_0 \left(4
        \varphi_0-1\right) \left(4 \varphi_0+1\right)}{6 v^2}
        \\\nn
        &
+\frac{d^5 \varphi_0 \left(2 \varphi_0-1\right) \left(2 \varphi_0+1\right) \left(4
        \varphi_0-1\right) \left(4 \varphi_0+1\right)}{120 v^4}\\
&- \frac{d^3 \left(2240
        \varphi_0^5-320 \varphi_1 \varphi_0^4-148 \varphi_0^3+60 \varphi_1
        \varphi_0^2+11 \varphi_0-\varphi_1\right)}{120 v^4}+
\dots
\end{align}
where dots denote terms suppressed by powers of $1/v^2$ and $1/d^2$.

We find that for $N=2$ and $N=4$ the conditions $F^-(v) = O(d)$ and $F^-(v) = O(d^3)$, respectively, lead to the system of
equations for the $\varphi-$coefficients whose solutions are
\begin{align} \notag
\nn\varphi_{N=2}(d)&= \frac{1}{4}-\frac{7}{4
        d^2}-\frac{32}{d^4}-\frac{1760}{d^6}-\frac{148000}{d^8}-\frac{16426208}{d^{10}}-\frac{2260059168}{d^
        {12}}-\frac{373366588128}{d^{14}}\\\notag
&- \frac{72726695282208}{d^{16}}-\frac{16512196163543264}{d^{18}}+{\cal O}\left(\frac{1}{d^{21}}\right)\,,
\\\la{resL15}
\nn\varphi_{N=4}(d)&=
\frac{1}{2}-\frac{31}{2
        d^2}-\frac{832}{d^4}-\frac{128704}{d^6}-\frac{28565312}{d^8}-\frac{7898593984}{d^{10}}-\frac{2557811
        948352}{d^{12}}\\
&- \frac{938971375124160}{d^{14}}-\frac{383507398659888960}{d^{16}}
+{\cal O}\left(\frac{1}{d^{19}}\right)\,,
\end{align}
plus the same functions with the signed flipped.
Substituting these expressions into \eq{varphi} we verified that the resulting expressions for $d_{N=2}(\xi)$ and $d_{N=4}(\xi)$
agree with the numerical values of scaling dimensions of the states with $\Delta(0)=7$ and $\Delta(0)=11$.

Evaluating the function \eq{F-minus} for the solutions   \eq{resL15} we obtain
\begin{align}\notag
{}& F^-_{N=2}(v) = {d v^3\over (v + 1) v (v - 1)} + O(1/d^2)\,,
\\
{}& F^-_{N=4}(v) =-{d^3 v^3\over 4(v + 2) (v + 1) v (v - 1) (v - 2)}+O(d) \,.
\end{align}
These expressions have a simple form suggesting a generalization to arbitrary $N$ (see Eq.~\eq{F-poles} below).

\subsubsection{Summary of the Results}

Examining the expressions for the functions $\varphi_N=\xi^3/d^2$ with odd and even $N$ found above we
notice that they are described by the following universal formula
\begin{eqnarray}\label{d-fin}
  \frac{\xi^3}{d^2}&=&\frac{N}{8}-\frac{N \left(2 N^2-1\right)}{8 d^2}-\frac{N^3 \left(3
        N^2+4\right)}{4 d^4}+\frac{-6 N^7-\frac{117 N^5}{4}-7
        N^3}{d^6}\\
\nn&-&\frac{N^3 \left(255 N^6+2664 N^4+3651 N^2+452\right)}{4
        d^8}\\
\nn&-&\frac{N^3 \left(3150 N^8+57315 N^6+195552 N^4+149205
        N^2+12892\right)}{4
        d^{10}}+{\cal O}\left(\frac{1}{d^{11}}\right)\,,
\end{eqnarray}
where only odd powers of $N$ appear. Inverting this series we can find the function $d(\xi)=|\Delta(\xi)-2|$.
We depicted this function for various states on Fig.~\ref{states2}.
It is convenient
however to examine the following combination that enters into the Baxter equation \eq{t-d}
\beq\label{d-def}
{\mathfrak d}\equiv -\frac{(\Delta -3) (\Delta -1)}{4 \xi ^2}=\frac{d^2+1}{4 \xi ^2}\,.
\eeq
We obtain that at strong coupling it is given by
\begin{eqnarray}\la{deltaexp}
{\mathfrak d}&=&
\frac{2 \xi }{N}+\frac{N^2}{2 \xi ^2}+\frac{N^3 \left(3
        N^2+4\right)}{16 \xi ^5}+\frac{N^4 \left(9 N^4+56 N^2+16\right)}{64
        \xi ^8}\\
\nn&+&\frac{N^5 \left(153 N^6+2200 N^4+3728 N^2+512\right)}{1024
        \xi ^{11}}\\
\nn&+&\frac{N^6 \left(195 N^8+5096 N^6+22176 N^4+19584
        N^2+1792\right)}{1024 \xi ^{14}}+O\left(\frac{1}{\xi^{17}
}\right)\,.
\end{eqnarray}
\begin{figure}
\begin{center}
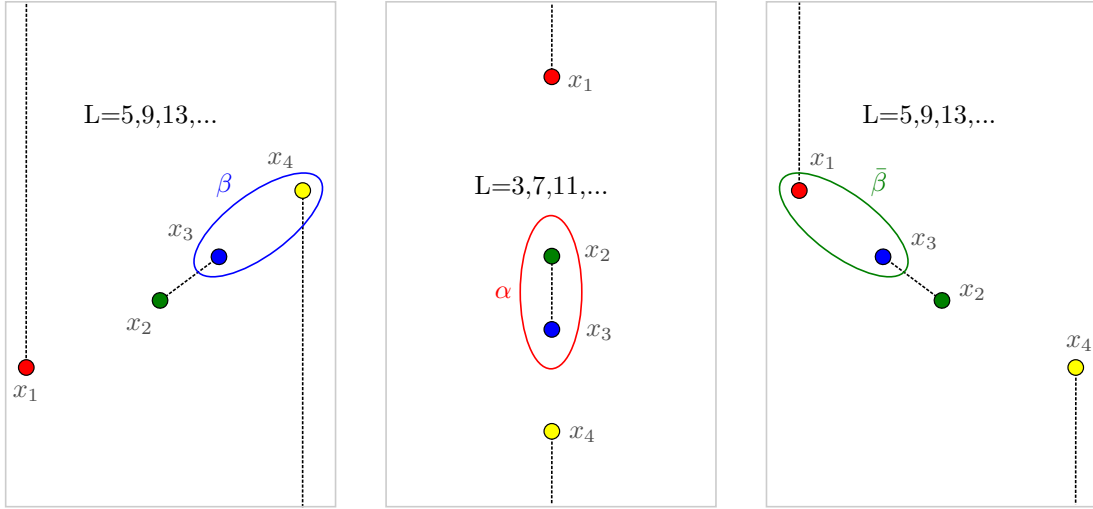
\end{center}

\caption{\la{cont}Integration contours entering the Bohr-Sommerfeld quantization condition for different states. For $L=3+4n$ the integral is taken
around the branch points which collide at large $\mathfrak d$ (middle panel). For $L=5+4n$, for half of the states, the integration goes around the branch points which collide at $\mathfrak d=3 e^{2\pi i/3}$ (left panel) and for another (complex conjugated) half of states the integration goes around the branch points which collide at $\mathfrak d=3 e^{-2\pi i/3}$ (right panel).
 }
\end{figure}%
We expect that this relation holds for all states with $\Delta(0)=3+4n$ for $n=0,1,2,\dots$.  For each given $n>0$ it describes two states
with $N=2n$ and $N=2n+1$. In the next section we will see an independent confirmation for this statement.

Finally, the function $F_N(v)$ is given for general $N$ by
\begin{align}\label{F-poles}
F_N(v) \sim d^{N-1}{v^3\, \Gamma(v-N/2)\over \Gamma(1+v+N/2)} \times \left(1+ O(1/d^2)\right)\,.
\end{align}
We would like to emphasize that this relation was derived at $v\gg 1$ and it cannot hold for small $v$. Indeed, according to its
definition, Eqs.~\eq{F-form} and \eq{F-minus} for odd and even $N$, respectively, the function $F_N(v)$ is built out of
solutions to the Baxter equation which are analytic for positive $v$ and have poles at negative integer $v$. As a consequence,
$F_N(v)$ cannot have poles at positive $v$.

\subsection{Asymptotic regime}\label{sect:as}

In this and in the following subsections we justify the quantization conditions \eq{quant-aux} and \eq{quant-aux1}.
Following the outline described in Sect.~\ref{8.1}, we shall examine the Baxter equation \eq{bax-t} in two regimes
corresponding to different range of the parameters, $1\ll v\ll \xi$ and $v=O(\xi)$. In what follows we refer to them as
to asymptotic and quasiclassical regimes, correspondingly.

The asymptotic regime corresponds to the limit in which transfer matrix $t(v)$ defined in \eq{t-d} is large,
\begin{equation}\label{roots}
t(v)=2 + \frac{ d^2{\mathfrakM}(d) }{v^3}-\frac{d^2+1}{4 v^2}  \equiv {2(v-\lambda_+)(v-\lambda_-)(v-\lambda_0)\over v^3}\,.
\end{equation}
Here in the second relation we introduced the notations for the roots of $t(v)$. At large $d$ we find that two of the roots are large,
 $\lambda_\pm \sim \pm d/(2\sqrt{2})$, and the remaining one $\lambda_0 = 4\varphi(d)$ stays finite in the large $d$ limit.
 As a consequence, the condition $|t(v)|\gg 1$ is verified for
 \begin{align}\label{range}
4\varphi(d) < v \ll d/(2\sqrt{2})\,.
\end{align}
In this regime, we can distinguish two solutions to \eq{bax-t}, $q_{+}(v)$ and $q_{-}(v)$, such that
$q_{+}(v)\gg q_{+}(v+1)$ and $q_{-}(v)\ll q_{-}(v+1)$.  In the leading large $d$ limit they satisfy
\begin{equation}\label{Bax-red}
q_{+}(v)\simeq  \frac{q_{+}(v-1)}{t(v)}\,,\qquad\qquad q_{-}(v+1)\simeq q_{-}(v){t(v)}\,,
\end{equation}
with $v$ satisfying \eq{range}.

Having constructed the solutions $q_{+}(v)$ and $q_{-}(v)$ at large $v$, we use \eq{Bax-red} to
continue them to finite $v$.
Since $t(v)\sim d^2$ for $v=O(d^0)$ the solution $q_{+}(v)$ becomes exponentially large  when approaching
the lower bound in \eq{range}, $q_{+}(v)\sim d^{+\alpha d}$ with some $\alpha\sim 1$. In the similar manner,
$q_{-}(v)$ becomes exponentially small $q_{-}(v)\sim d^{-\alpha d-1}$ in the same limit\footnote{we use the term ``exponentially" in the sense faster-than-any-power grows/decay.}.
To build the solutions satisfying \eq{Bax-red} explicitly
we write them in terms of the well-defined and uniquely fixed $q(v)$ and $\bar q(v)$ in the following way
\begin{align}\notag
{}& q_{+}(v)\equiv q(v)+\bar q(v)\,,
\\
{}& q_{-}(v)\equiv \left(q(v)-\frac{q_0}{q_{+,0}}q_{+}(v)\right)\frac{1}{i}=
\frac{\bar q_0 q(v)-q_0 \bar q(v)}{q_0+\bar q_0}
\frac{1}{i}\,.
\end{align}
where we used notations for $q_0= q(0)$, $\bar q_0= \bar q(0)$ and $q_{+,0}= q_{+}(0)$.
Indeed, generically each of the two independent solutions $q(v)$ and $\bar q(v)$
will contain both $q_+$ and $q_-$. As $q_+$ is defined up to an arbitrary addition of $q_-$ we use this freedom to define $q_+$ as in the first line above. Next, to extract $q_-$ from, say, $q(v)$ we have to project out the growing part $q_+(v)$, i.e. we have to find some coefficient $\alpha$ such that $q(v)-\alpha q_+(v)$ is not exponentially large at $v\sim 1$. This can be done simply by requiring the difference to vanish
at $v=0$\footnote{Alternatively one can set it to zero at any other $v\sim 1$, this will change the definition of $q_-$ by adding to it $q_+$ with an exponentially small coefficient.}.

Inverting the above relations
 can now write $q(v)$ and $\bar q(v)$ in terms of the  solutions to \eq{Bax-red}
\begin{align}\notag\label{qq-new}
{}& q(v)=\frac{q_0}{q_0+\bar q_0}q_{+}(v)+iq_{-}(v)\,,
\\
{}& \bar q(v)=\frac{\bar q_0}{q_0+\bar q_0}q_{+}(v)-iq_{-}(v)\,.
\end{align}
By construction, these expressions are dominated for $v=O(d^0)$ by the first term.  We recall that $q(v)$ and $\bar q(v)$ satisfy Wronskian relation
 $q(v+1)\bar q(v)-\bar q(v+1)q(v)=i\; d$, (see Eq.~\eq{Wr}). Applying \eq{qq-new} and taking into account \eq{Bax-red} we find that it leads to
the following relation for  $q_{-}(v)$ and $q_{+}(v)$
\begin{equation}\la{qqwr}
q_{-}(v)q_{+}(v) = \frac{d}{t(v)-1/t(v+1)}\sim  \frac{d}{t(v)} \;.
\end{equation}
For $v=O(d^0)$ the expression on the right-hand side scales as $O(1/d)$, in agreement with the expected asymptotic behavior
$q_{+}(v)\sim d^{+\alpha d}$ and $q_{-}(v)\sim d^{-\alpha d-1}$.


Let us now examine the function $F(v)$ defined in \eq{F-form}. Replacing the functions $q(v)$ and $\bar q(v)$ by their expressions \eq{qq-new} we
find
\begin{equation}\la{qqqs}
F(v) =
\frac{\left(\dot{q}_0
        \bar{q}_0+q_0
        \dot{\bar{q}}_0\right)  }{\left(\bar{q}_0+q_0\right)
        \left(\dot{\bar{q}}_0+\dot{q}_0\right)} q_{+}(v)
        \dot{q}_{+}(v)
-\frac{i \left(q_0-\bar{q}_0\right)
         }{\bar{q}_0+q_0}\dot{q}_{-}(v)
        q_{+}(v)
        +\frac{i
        \left(\dot{\bar{q}}_0-\dot{q}_0\right)  }{\dot{\bar{q}}_0+\dot
        {q}_0}q_{-}(v)
        \dot{q}_{+}(v)
+2 q_{-}(v) \dot{q}_{-}(v)\;,
\end{equation}
where $\dot{q}_{\pm}(v)$ stands for the functions ${q}_{\pm}(v)$ with $\varphi(d)\to -\varphi(d)$. By construction, this relation holds for $v$ inside the
region \eq{range}. For $v=O(d^0)$ the four terms on the right-hand side of \eq{qqqs} have different scaling behavior at large $d$: the first
term grows exponentially, the last term decreases exponentially {\bf } whereas the two terms in the middle have a power-like
behavior. We notice however that the first term in \eq{qqqs} involves the same combination $\dot{q}_0 \bar{q}_0+q_0\dot{\bar{q}}_0$ that
enters the exact quantization condition \eq{quant}. In other words, for the function $d(\xi)$ satisfying the quantization condition \eq{quant} the
first term in \eq{qqqs} vanishes leading to a power-like scaling of the function $F(v)$ at large $d$. Turning the logic around, we find
that the condition for the function $F(v)$ do not have an exponential growth at large $d$ is equivalent to the quantization condition  \eq{quant}.
As we demonstrated in the previous subsection, the same condition fixes the function $\varphi(d)$.

We can also make an additional consistency test of this observation. After canceling the leading term in \eq{qqqs}
we are left with the two finite terms which we can evaluate explicitly. We can use \eq{Bax-red} to get
\begin{align}\label{rec1}
 {q}_{-}(v)\dot q_{+}(v) =  {q}_{-}(v-1) \dot q_{+}(v-1) { t(v-1) \over \dot t(v)}
= {q}_{-}(v-1)\dot q_{+}(v-1) {v^3 \over (v-1)^3} {v-1-4\varphi\over v+4\varphi}\,.
\end{align}
Here in the second relation we replaced $t(v)$ by its expression \eq{roots} and took into account that $\dot t(v)$ is given by the same expression \eq{roots}
with the sign of all roots flipped. It is then easy to see that the contribution of large roots $\lambda_\pm$ cancels in the ratio of the transfer matrices and
it only depends on the root $\lambda_0=4\varphi(d)$. Subsequenty applying \eq{rec1} we obtain
\begin{align}\label{prod1}
{q}_{-}(v) \dot q_{+}(v)  = c_+  { v^3\,\Gamma(v-4\varphi)\over \Gamma(1+v+4\varphi)}\,,
\end{align}
where the normalization constant is related to the value of the same product at the lower boundary of the region \eq{range},
$c_+ ={q}_{-}(4\varphi)  \dot q_{+}(4\varphi)/(4\varphi)^3$. Repeating the same calculation for $\dot q_{-}(v)
        {q}_{+}(v)$ we get
\begin{align}\label{prod2}
\dot q_{-}(v)
        {q}_{+}(v)=c_-  { v^3\,\Gamma(v+4\varphi)\over \Gamma(1+v-4\varphi)}\,.
\end{align}
The normalization constants $c_+$ and $c_-$ are not independent. Taking the product of the last two
expressions and making use of \eq{qqwr} we obtain
\begin{align}
c_+ c_- ={d^2 v^6 (v+4\varphi)(v-4\varphi)\over t(v) \dot t(v)} \sim {16\over d^2}\,.
\end{align}
To find the dependence of $c_\pm$ on $d$, we can examine the relation \eq{prod1} and \eq{prod2} for $v$ close to the upper bound in \eq{range} and match
them into analogous expressions obtained in the quasiclassical regime $v= O(d)$. One can show that this leads to $c_+/c_-=d^{16\varphi}$ and, as a consequence, for positive $\varphi=O(d^0)$ the contribution of \eq{prod2} to $F(v)$ is suppressed by a power of $1/d$ as compared to that of \eq{prod1}.
We observe that for $\varphi=N/8$ expression on the right-hand side of \eq{prod1}  becomes a rational function  of $v$. Its contribution to
\eq{qqqs} matches precisely the expression \eq{F-poles}! This provides the additional support to the results presented in the previous subsection.

\subsection{Quasiclassical regime}

The Baxter equation \eq{bax-t} has an interesting scaling behavior for $v=O(\xi)$. In this regime, it is convenient to introduce a new variable
$x=v/\xi$ and look for solution to \eq{bax-t} in the form of WKB expansion
\begin{align}\label{q-WKB}
q(v) = \exp\left( \xi \int_{x_0}^x dy\, p(y) dy\right)\,,\qquad v=x \xi\,,
\end{align}
where $1/\xi$ plays the role of Planck constant and the quasimomentum $p(x)$ admits an expansion in powers of $1/\xi$
\begin{equation}\label{p-exp}
p(x)=p_0(x)+\frac{1}{\xi}p_1(x)+\frac{1}{\xi^2}p_2(x)+\dots\,.
\end{equation}
The normalization of $q(v)$ depends on the choice of $x_0$.

Plugging the expansion \eq{p-exp} into the Baxter equation \eq{bax-t} and expanding  both sides in $1/\xi$ we obtain \cite{Korchemsky:1995be,Korchemsky:1996kh}
\begin{align}\notag\label{p's}
{}& e^{p_0(x)}+e^{-p_0(x)} =2 - { \mathfrak d \over x^2} + {1\over x^3}\,,
\\ \notag
{}& p_1(x) = -\frac{1}{2} p_0'(x) \coth (p_0(x))\,,
\\
{}& p_2(x) =\frac{1}{12} p_0''(x) \left(3 \text{csch}^2(p_0(x))+1\right)-\frac{3}{8}
   p_0'(x)^2 \coth (p_0(x)) \text{csch}^2(p_0(x))\,, \quad \dots
\end{align}
Solving the first equation we find that $e^{p_0(x)}$ is an analytic
function on a complex $x-$plane with two cuts running the
between the branch points
at which $e^{p_0(x_i)}=\pm 1$. Introducing the function
\begin{align}\label{curve}
y^2 = (4x^3-\mathfrak d x+1)(-\mathfrak d x+1)\,,
\end{align}
we find that the branch points satisfy $y(x_i)=0$ leading to
\begin{align}\label{disc}
\prod_{i<k} (x_i-x_k)^2 = {\mathfrak d^3- 27\over 16 \,\mathfrak d^6}\,.
\end{align}
The position of the branch points depends on the value of $\mathfrak d$ defined in \eq{d-def}. As follows from the last relation,
two of the branch points coincide for $\mathfrak d\to\infty$ and $\mathfrak d^3=27$.

At large $\mathfrak d$, the branch points are located at $x_1=-(\mathfrak d/4)^{1/2}$, $x_2=1/\mathfrak d$, $x_3=1/\mathfrak d+O(1/\mathfrak d^4)$ and $x_4=(\mathfrak d/4)^{1/2}$. It is convenient to choose the cuts of $p_0(x)$ to run along intervals $(-\infty,x_1]$, $[x_2,x_3]$ and $[x_4,\infty)$ on the real axis, so that the middle cut shrinks into a point for $\mathfrak d\to\infty$. At large positive $x$ we can define two  branches of the quasimomentum $p_{0,\pm}(x)=p_0(x\pm i0)$ corresponding to its value above and below the cut, correspondently,
\begin{align}
p_{0,\pm}(x) = \pm {id \over 2 x\xi } + O(1/x^2)\,,
\end{align}
where we replaced $(-\mathfrak d)^{1/2}= i d/(2\xi)+O(1/d)$. Substituting this expression into \eq{q-WKB} we find that the
corresponding solution to the Baxter equation scales at large $v$ and $d\gg 1$ as $q(v) \sim v^{id/2}$ and $\bar q(v) \sim v^{-id/2}$, in
agreement with \eq{q(v)}.

Applying the above relations we can define the semiclassical solutions to the Baxter equation $q(x\xi)$ and $\bar q(x\xi)$ on the complex $x-$plane with the two cuts. We recall that the quantization condition \eq{Phi} involves another pair of solutions $\dot q(v)=q(v,-\xi)$ and $\dot{\bar q}(v) =\bar q(v,-\xi)$. In the semiclassical approximation, they can be obtained from $q(x\xi)$ and $\bar q(x\xi)$ through the transformation $x\to -x$ and $\xi\to -\xi$. The resulting expressions for $\dot q(x\xi)$ and $\dot{\bar q}(x\xi)$
have analytical properties similar to those of $q(x\xi)$ and $\bar q(x\xi)$ with the only difference that their cuts run along intervals $(-\infty,-x_4]$, $[-x_3,-x_2]$ and $[-x_1,\infty)$ on the real axis. Since $x_1=-x_4=(\mathfrak d/4)^{1/2}$ in the large $d$ limit, the two sets of functions share the same semi-infinite cuts whereas the two `short cuts' are symmetric with respect to the origin.

Let us consider the following ratio
\begin{align}\label{F-aux}
{F(v)\over \bar q(v)  \dot q(v)}=
{ q(v)\over  \bar q(v)} {\dot{\bar q}(v)\over\dot q(v)} + 1 \,.
\end{align}
The exact quantization condition \eq{Phi} requires this function to vanish for $v\to 0$. In the previous subsection we demonstrated
that this condition is equivalent to the requirement for $F(v)$ to have a power-like behavior
\eq{quant-aux} and \eq{quant-aux1} for $v$ satisfying \eq{range}. As we will see in a moment, in the quasiclassical regime, for
$v=O(\xi)$, the same quantization condition follows from the requirement for \eq{F-aux} to be a single-valued function of $v=x\xi$
on a complex $x-$plane with the cuts to be specified below. The advantage of considering the ratio \eq{F-aux} is that it is independent
on the choice of normalization of the solutions of the Baxter equation, or equivalently, on the choice of the point $x_0$ in \eq{q-WKB}
provided that $x_0$ is located away on the cuts.

In the quasiclassical regime, for $v=x\xi$, the expression on the right-hand side of \eq{F-aux} has two short cuts $[-x_3,-x_2]$ and $[x_2,x_3]$.
For the ratio \eq{F-aux} to be a single-valued function on the $x-$plane, it should acquire the same value after going around each of these cuts.
Since ${\dot{\bar q}(v)/\dot q(v)}$ is analytical on $[x_2,x_3]$, the monodromy only comes from $q(v)/ \bar q(v)$. Using \eq{q-WKB} we find
that the above condition translates to the Bohr-Sommerfeld quantization condition
\begin{align}
\exp\left(2\xi \int_{x_2}^{x_3} dx [ p(x+i0)- p (x-i0)]\right) = 1\,,
\end{align}
or equivalently
\begin{align}\label{WKB-quan}
\xi \oint_\alpha dx \, p(x) = i  \pi (N+1)\,.
\end{align}
Here the integration contour encircles the segment $[x_2,x_3]$ and $N$ is an arbitrary integer. For the cut $[-x_3,-x_2]$ we can get
analogous relation by replacing $y\to -y$ and $\xi\to -\xi$. We show in the next subsection that, upon
replacing the quasimomentum in \eq{WKB-quan} by its explicit expression
\eq{p-exp} and  \eq{p's},  the relation \eq{WKB-quan} allows us to
obtain the dependence of $\mathfrak d$ on the coupling $\xi$.

We can use the semiclassical analysis to clarify two issues that were mentioned in the previous subsections. We remind that in order to
reproduce half of the states in section~\ref{Sect:half}, we had to flip the sign between the two terms in the exact quantization condition \eq{Phi} and
impose the condition \eq{F-minus} instead. This can be understood as follows. The relation \eq{F-minus} holds at large $v$ and $d$
and it should be applicable in the quasiclassical regime for $x=O(v/d)$ away from the cuts, that is for  $x_3< x< x_4$. Since the exact
quantization condition holds  at $v=0$ one may wonder whether the first term on the right-hand side of \eq{F-aux} acquires a monodromy as $v$ moves from $v=0$ to large $v$ across the cut $[x_2,x_3]$. Indeed,  $q(v)$ and $\bar q(v)$ are given in this case by the same expression
\eq{q-WKB} in which the integration goes slightly above and below the cut, respectively, so that
the ratio $q(v)/\bar q(v)$ generates the additional
factor
\begin{align}
\exp\left( \xi \int_{x_2}^{x_3} dx [ p(x+i0)- p (x-i0)]\right) = e^{i  \pi (N+1)}  \,,
\end{align}
which flips the sign for even $N$.

\subsection{WKB expansion}

In this subsection we solve the Bohr-Sommerfeld quantization \eq{WKB-quan} and show that the resulting expression for $\mathfrak d(\xi)$
coincides with \eq{deltaexp}. Examining \eq{deltaexp} we find that the function $\mathfrak d(\xi)$ has an interesting scaling behavior
for large $\xi$ and $N$ with their ratio $\mathcal{N}=N/(2\xi)$ fixed
\begin{eqnarray}\la{qqD}
{\mathfrak d}&=&\left(
\frac{1}{\mathcal{N}}
+
2 \mathcal{N}^2+
6\mathcal{N}^5+
36 \mathcal{N}^8+
306 \mathcal{N}^{11}+
3120 \mathcal{N}^{14}+\dots
\right)\\
\nn&+&\frac{1}{\xi^2}\left(
2 \mathcal{N}^3
+56 \mathcal{N}^6
+1100 \mathcal{N}^9
+20384
\mathcal{N}^{12}+\dots\right)+{\cal O}\left(\frac{1}{\xi^4 }\right)\,.
\end{eqnarray}
Notice that the expansion runs in powers of $1/\xi^2$ with the coefficients being nontrivial functions of $\mathcal{N}$.
We shall determine these functions exactly using  \eq{WKB-quan}.

Integrating by parts, we can rewrite  \eq{WKB-quan} in the following equivalent form
\begin{align}\label{BS}
a =-{1\over 2\pi i} \oint_\alpha dx \, x \,p'(x) =  \mathcal N + {1\over 2\xi} \,,
\end{align}
where  $\mathcal{N}=N/(2\xi)$ and the integration goes along the contour encircling the cut $[x_2,x_3]$. Replacing the quasimomentum with its expansion \eq{p-exp} in powers of $1/\xi$
 we obtain analogous expansion for $a$
\begin{align}\label{a-exp}
a = a_0(\mathfrak q) + {a_1(\mathfrak q)\over \xi}+ {a_2(\mathfrak q)\over \xi^2}+ \dots\,.
\end{align}
The first term of the expansion is given by
\begin{align}
a_0 = -{1\over 2\pi i} \oint_\alpha dx \, x \,p_0'(x) = {1\over 2\pi i} \oint_\alpha  {dx (3-2 \mathfrak d x) \over \sqrt{(4x^3-\mathfrak d x+1)(-\mathfrak d x+1)}}\,.
\end{align}
The integral on the right-hand side has a simple interpretation
in terms of a Riemann surface defined by a complex (elliptic) curve \eq{curve}. Namely, $a_0$ is given by the period of the `action' differential
over the $\alpha-$cycle. For $\mathfrak d\to \infty$, the $\alpha-$cycle shrinks into a point and the integrand develops a pole at $x=1/\mathfrak d$.
Expanding the integrand around this point we can easily compute the integral by residues
\begin{align} \notag\label{leadingorder}
a_0 {}& =
\frac{1}{\mathfrak{d}}+\frac{2}{\mathfrak{d}^4}+\frac{18}{\mathfrak{d}^7}
+\frac{240}{\mathfrak{d}^{10}}+{\cal O}\left(\frac{1}{\mathfrak{d}^{13}}\right)
\\
{}&   =
\frac1{\mathfrak{d}}
{}_3F_2\left(\frac{1}{3},\frac{1}{2},\frac{2}{3};1,\frac{3}{2};\frac{2
7}{\mathfrak{d}^3}\right) \,,
\end{align}
where the second relation was obtained from summing the series in $1/\mathfrak{d}$ and it holds for $\mathfrak{d}^3> 27$.

For the second term in \eq{a-exp} we find
\begin{align}
a_1 = {1\over 2\pi i} \oint_\alpha dx \,  p_1(x)  = - {1\over 4\pi i} \oint_\alpha d \log \sinh p_0(x) = \frac12\,,
\end{align}
where the integral takes a universal form and its value does not depend on $\mathfrak{d}$. We observe that the contribution of $a_1$
to the left-hand side of \eq{BS} is given by $1/(2\xi)$ and it matches an analogous term on the right-hand side.

For the third term in  \eq{a-exp},
$a_2=-{1\over 2\pi i} \oint_\alpha dx \, x \,p_2'(x) $, we find
after some algebra a compact representation in terms of hypergeometric functions
\begin{align}\notag\label{a2}
a_2
=\frac{2}{ \mathfrak{d}^5}+\frac{84}{ \mathfrak{d}^8}+\frac{2700}{ \mathfrak{d}^{11}}+\frac{80080}{\mathfrak{d}^{14}}+\frac{2293200}{ \mathfrak{d}^{17}}+\frac{64465632}{ \mathfrak{d}
^{20}}+{\cal O}\left(\frac{1}{ \mathfrak{d}^{23}}\right)
\\
={2\over  \mathfrak{d}^5} \left[\,
_3F_2\left(\frac{3}{2},\frac{5}{3},\frac{7}{3};\frac{5}{2},3;\frac{
27}{\mathfrak{d}^3}\right)+\frac{21 }{\mathfrak{d}^3}
\,
_3F_2\left(\frac{5}{2},\frac{8}{3},\frac{10}{3};\frac{7}{2},4;\frac
{27}{\mathfrak{d}^3}\right)\right]\,.
\end{align}
As before, the second relation holds for $\mathfrak{d}^3> 27$.

It is straightforward to continue this procedure and compute subleading terms in \eq{a-exp}. In this way, we found that all terms
with odd powers of $1/\xi$ vanish
\begin{align}
a_3(\mathfrak{d})=a_5(\mathfrak{d})= \dots =0
\end{align}
In the Appendix \ref{AH} we give expressions for the subleading terms $a_4$, $a_6$ and $a_8$.

\begin{figure}
        \center{\includegraphics[scale=0.6]{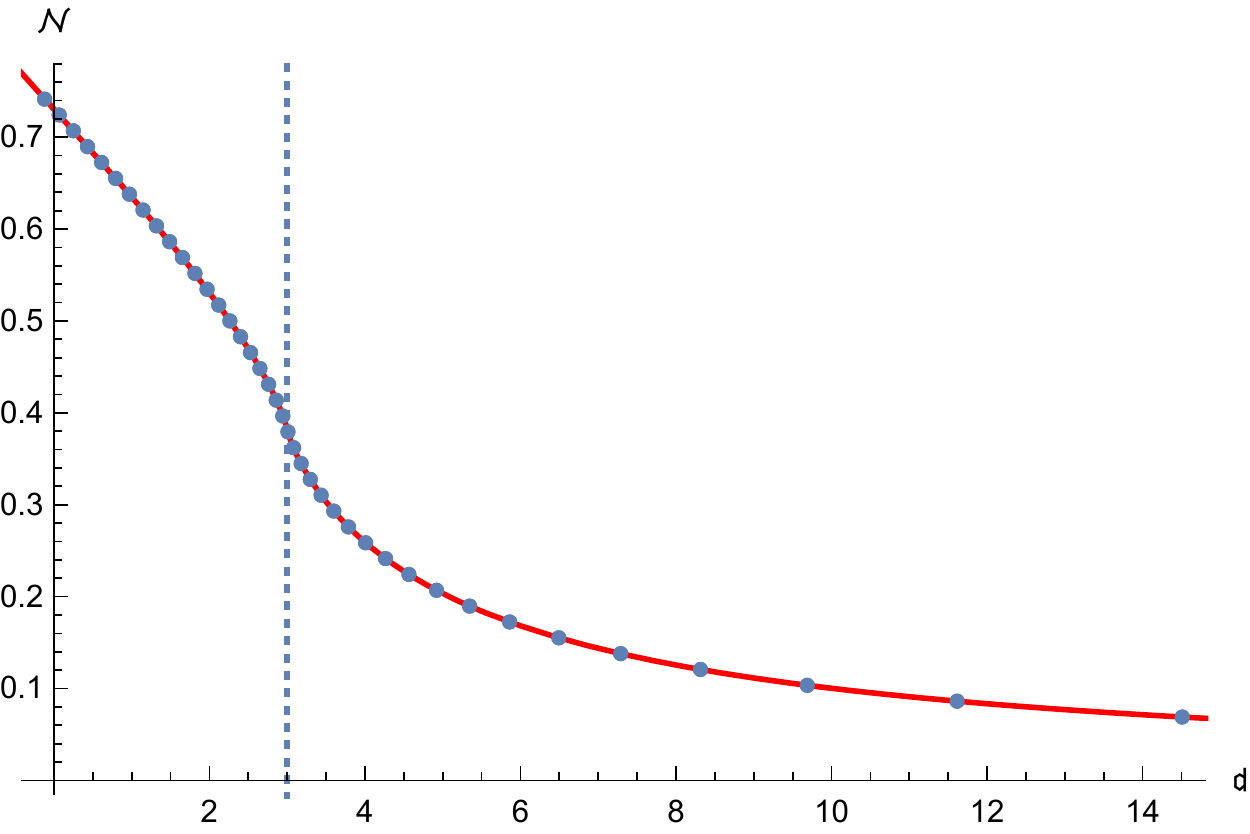}
                \includegraphics[scale=0.6]{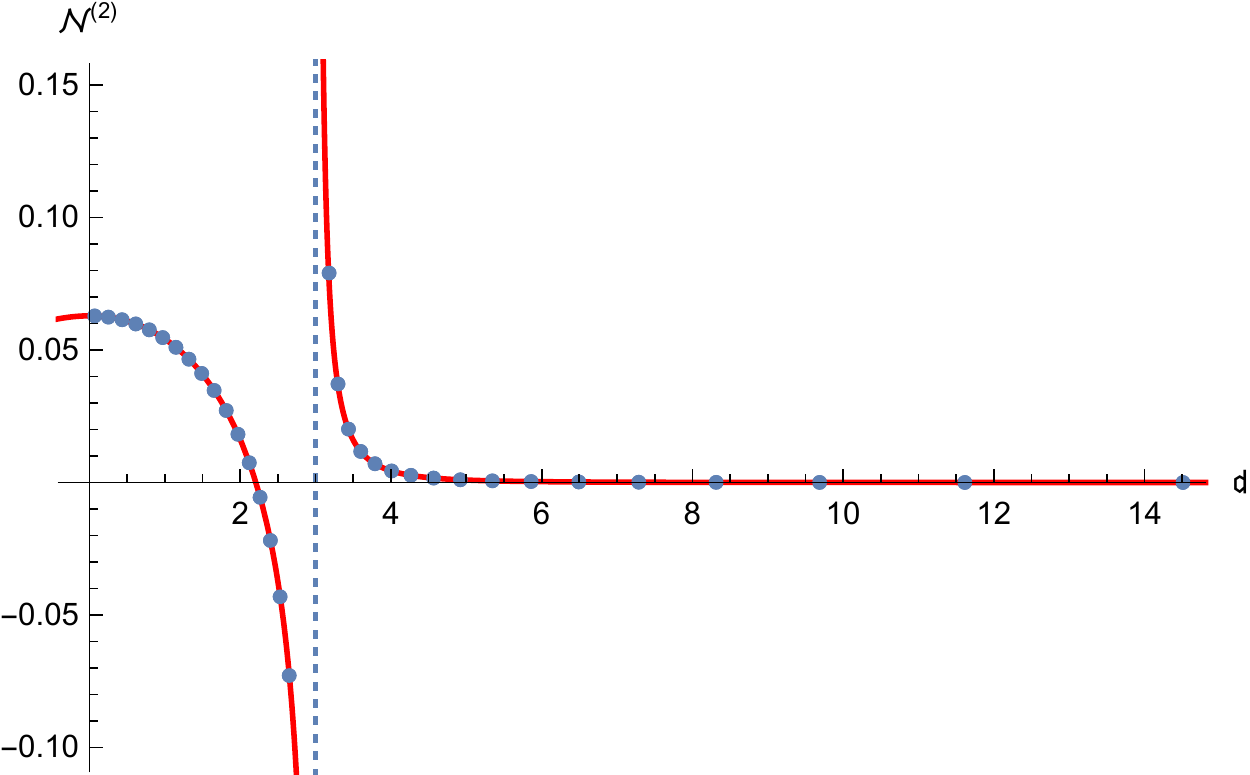}
        \caption{\label{fig:LD} (Left panel)
Comparison of the numerical data for the states with $\Delta(0)=3+4n$ at $\xi=29$ against the WKB expansion \eq{subleading} (with hypergeometric series replaced by their real part for ${\mathfrak d}<3$). (Right panel) Comparison of $O(1/{\xi}^2)$ correction to the WKB expansion \eq{subleading} with the numerical data.
Blue dots denote the numerical data with the leading order correction \eq{leadingorder} subtracted, red line describes $O(1/{\xi}^2)$ correction to \eq{subleading}.
%
        }}
\end{figure}

Finally, we substitute the obtained expressions for $a_0$, $a_1$ and $a_2$ into \eq{BS} and obtain the Bohr-Sommerfeld quantization
condition
\begin{equation}\la{subleading}
{\cal N}=
\frac{\,
_3F_2\left(\frac{1}{3},\frac{1}{2},\frac{2}{3};1,\frac{3}{2};\frac{2
7}{\mathfrak{d}^3}\right)}{\mathfrak{d}}+
\frac{2}{\xi^2}\left(
\frac{
  \,
_3F_2\left(\frac{3}{2},\frac{5}{3},\frac{7}{3};\frac{5}{2},3;\frac{
27}{\mathfrak{d}^3}\right)}{\mathfrak{d}^5}+\frac{21 \,
_3F_2\left(\frac{5}{2},\frac{8}{3},\frac{10}{3};\frac{7}{2},4;\frac
{27}{\mathfrak{d}^3}\right)
}{\mathfrak{d}^8}
\right)
+{\cal O}\left( \frac{1}{\xi^4}\right)\;.
\end{equation}
To find the dependence of $\mathfrak{d}$ on the coupling $\xi$ and $\mathcal{N}=N/(2\xi)$, we have to invert this relation. We verified that
at large $\mathfrak{d}$ this yields \eq{qqD}. We also checked that \eq{subleading}  is in a perfect agreement with the numerical results for
$\mathfrak{d}^3> 27$ (see Fig.~\ref{fig:LD}). This gives a strong support to our conjecture that the Bohr-Sommerfeld quantization condition
\eq{subleading} gives the correct result for the scaling dimensions to all orders in $1/\xi$ expansion.

We would like to emphasize that the hypergeometric series on the right-hand side of \eq{subleading} were obtained by summing up $1/\mathfrak{d}$ expansion.
Curiously, at ${\mathfrak d}=3$ these series develop a logarithmic branch cut. The reason why this singularity appears is that
the two branch points of the curve \eq{curve} collide at this value of ${\mathfrak d}$. Indeed, for ${\mathfrak d}=3$ the branching
points are given by $x_1=-1$, $x_2=1/3$ and $x_3=x_4=1/2$. Going from ${\mathfrak d}\gg 1$ to ${\mathfrak d}=3$ we find that
the $\alpha-$cycle encircling the cut $[x_2,x_3]$ expands whereas the $\beta-$cycle encircling the interval $[x_3,x_4]$ shrinks
into a point. For ${\mathfrak d}<3$ the branch points $x_3$ and $x_4$ develop imaginary part, $x_3=x_4^*$, and move away from the real
axis. The definition of the $\alpha-$cycle becomes ambiguous in this case since there are two possible choices of the cuts, $[x_2,x_3]$ and
$[x_2,x_4]$. The integral \eq{BS} evaluated along the closed contour encircling these cuts takes different complex conjugated values but its real part
is the same for the two cuts. This suggests that for an arbitrary positive ${\mathfrak d}$ the Bohr-Sommerfeld quantization \eq{BS} condition should look as
\begin{align}\label{BS1}
{\rm Re} \left[ -{1\over 2\pi i} \oint_\alpha dx \, x \,p'(x) \right]=  \mathcal N + {1\over 2\xi}\,.
\end{align}
For ${\mathfrak d}>3$ this relation coincides with \eq{BS}.
Evaluation of the integral on the left-hand side of \eq{BS1} leads to \eq{subleading} with hypergeometric series replaced by their real part. We verified
that this relation correctly describes numerical results for ${\mathfrak d}<3$ (see Fig.~\ref{fig:LD}).

\subsection{Strong coupling expansion of $\Delta(0)=5,9,\dots$ states}\la{lastqq}

So far we derived a strong coupling expansion of the states with $\Delta(0)=3,7,11,\dots$. For $\xi\gg 1$ their scaling dimension
$\Delta=2+i d(\xi)$ is given by Eqs.~\eq{d-fin} -- \eq{deltaexp}. We have shown that it can be found by solving the Bohr-Sommerfeld
quantization condition \eq{BS1}  for a particular choice of the cycle $\alpha$ on the Riemann surface \eq{curve}.  This cycle is uniquely
defined by the requirement that the branch points should merge for ${\mathfrak d}\to\infty$.
In this section we argue that the remaining states with $\Delta(0)=5,9,\dots$ also admit an analogous semiclassical description.
The only difference compared to \eq{BS1} is that these states correspond to another choice of the integration contour on the
Riemann surface \eq{curve}.

We remind that the branch points satisfy the relation \eq{disc}. There are four different values of ${\mathfrak d}$ for which
two of these points collide. Two of them, ${\mathfrak d}\to\infty$ and ${\mathfrak d}=3$, we encountered in the previous
subsection. As we will see in a moment, the two remaining ones, ${\mathfrak d}=3\, e^{\pm 2\pi i/3}$, describe the strong coupling
limit of the states $\Delta(0)=5,9,13,\dots$. As follows from the definition \eq{d-def}, the value ${\mathfrak d}=3\, e^{\pm 2\pi i/3}$
corresponds to the following scaling behavior  of $\Delta(\xi)$ at strong coupling
\begin{align}\label{scal}
\Delta(\xi) = 2\sqrt{3}\, \xi\,  e^{\pm i\pi/6} + O(\xi^0)\,.
\end{align}
It should be compared with analogous relation \eq{Delta-as} for the states $\Delta(0)=3,7,11,\dots$. We verified that \eq{scal} correctly describes numerical results for the trajectories
in Fig.~\ref{Fig:res} that start at $\Delta(0)=5,9,\dots$.

To find subleading corrections to \eq{scal} we shall employ the Bohr-Sommerfeld quantization condition analogous to \eq{BS1}.
Following the logic of the previous subsection, the integration contour should be chosen to encircle the pair of branch points
that collide at  ${\mathfrak d}=3\, e^{\pm 2\pi i/3}$. It is easy to see from \eq{curve} that for  this value of ${\mathfrak d}$ the
branch points are aligned along the same ray in a complex plane, $x_1= -e^{\mp 2i\pi/3}$, $x_2=e^{\mp 2i\pi/3}/3$ and $x_{3}=x_4=e^{\mp 2i\pi/3}/2$.
Thus, in the vicinity of  ${\mathfrak d}=3\, e^{\mp 2\pi i/3}$ the integration contour should encircle the segment $[x_3,x_4]$. This
corresponds to the choice of the $\beta-$cycle on the Riemann surface \eq{curve} (see Fig.~\ref{cont}). The resulting
Bohr-Sommerfeld quantization condition reads
\begin{align}\label{BS2}
a_D =-{1\over 2\pi i} \oint_\beta dx \, x \,p'(x) =\mathcal N + {1\over 2\xi}\,,
\end{align}
where $\mathcal{N}=N/(2\xi)$ and the quasimomentum is given by \eq{p-exp} and \eq{p's}.

The relation \eq{BS2} has been previously encountered in the study of semiclassical limit of the $SL(2)$ spin chain \cite{Korchemsky:1996kh}.
Using the results of \cite{Korchemsky:1996kh} we find that in the vicinity of ${\mathfrak d}=3\, e^{\pm 2\pi i/3}$ the first two terms of the
expansion of $a_D$ in powers of $1/\xi$ are given by
\begin{align}\label{ad}
a_D =    {\sqrt{3}\,e^{\pm i\pi/6} \over 2} \left(x-\frac5{12} x^2 +\frac{53}{216}x^3 -\frac{2497}{15552} x^4 + O(x^5)\right) + {1\over 2\xi} + O(1/\xi^2)\,,
\end{align}
where ${\mathfrak d}=3\, e^{\pm 2\pi i/3}(1+x)$ and $x$ is small. Here $O(1/\xi)$ term describes the contribution of $p_1(x)$ correction to
the quasimomentum. As in the previous case, it cancels against the same term on the right-hand side of \eq{BS2}. The series in the first term
in \eq{ad} can be summed up and expressed as discontinuity of a hypergeometric $_3F_2-$series
\begin{eqnarray}\label{ad-exp}
a_D =  \frac{1}{\mathfrak{d}}{\rm Disc}\left[\,
_3F_2 \left(\frac{1}{3},\frac{1}{2},\frac{2}{3};1,\frac{3}{2};\frac{2
7}{\mathfrak{d}^3}\right)\right]+ {1\over 2\xi} + O(1/\xi^2)\,.
\end{eqnarray}
More precisely, the hypergeometric series develops a logarithmic cut that starts at $\mathfrak{d}^3=27$. The discontinuity across this cut
is a rational function of ${\mathfrak d}=3\, e^{\pm 2\pi i/3}(1+x)$ whose expansion at small $x$ matches the first term on the right-hand side of
\eq{ad}. Notice that the same hypergeometric series enters \eq{leadingorder}. This is not accidental of course since the relations \eq{ad-exp}
and \eq{leadingorder} define the period of the same `action' differential over the two cycles on the Riemann surface \eq{curve}.

Solving the Bohr-Sommerfeld quantization condition \eq{BS2} we can obtain the large $\xi$ expansion of ${\mathfrak d}$
\begin{align}\label{d-sub}
{\mathfrak d} = {\mathfrak d}_0(\mathcal N) + {1\over \xi^2} {\mathfrak d}_2(\mathcal N) + \dots
\end{align}
To find the leading term we substitute \eq{ad} into  \eq{BS2} and invert the series
\begin{align}\label{d0}
\mathfrak{d}_0 =   3\, e^{\pm 2i \pi/3}\left[1+\frac{2 }{\sqrt{3} }\bar {\mathcal N}+\frac{5}{9  }\bar {\mathcal N}^2+\frac{22 }{81 \sqrt{3} }\bar {\mathcal N}^3+\frac{43 }{2187  }\bar {\mathcal N}^4 +O(\bar {\mathcal N}^5)\right]\,,
\end{align}
where the notation was introduced for complex $\bar {\mathcal N} = \mathcal N e^{\pm i \pi/6}= Ne^{\pm i \pi/6}/(2\xi)$. In agreement
with our expectations, this relation defines two complex-valued functions $\Delta(\xi)$ (see  \eq{d-def}). We verify that at large $\xi$
these functions have a correct asymptotic behavior \eq{scal}.


To determine the subleading correction to \eq{d-sub}, we exploit the relation between the periods $a$ and $a_D$ mentioned above. It allows us to
find $O(1/\xi^2)$ correction to $a_D$ by taking a discontinuity of \eq{a2}. In this way,  we find
\begin{eqnarray} \label{d2}
\mathfrak{d}_{2}=-\frac{25}{36}-\frac{161}{54
\sqrt{3}}\bar {\mathcal N} -\frac{2459}{1458}\bar {\mathcal N}^2-\frac{3902 }{2187
\sqrt{3}}\bar {\mathcal N}^3 -\frac{14645 }{39366}\bar {\mathcal N}^4 + O(\bar {\mathcal N}^5)\,,
\end{eqnarray}
where $\bar {\mathcal N}$ was defined in \eq{d0}.

The relations \eq{d0} and \eq{d2} were derived at large $\xi$ and $N$ with $N/\xi$ fixed.
Substituting \eq{d0} and \eq{d2} into \eq{d-sub} and collecting terms with the same power of $1/\xi$ we obtain
\begin{eqnarray}\la{l5pred}
{\mathfrak d}=3 \, e^{\pm 2i \pi /3}\left[1 +\frac{ N}{\xi \sqrt{3}}e^{\pm i\pi /6} +\frac{5 \left(3 N^2+5\right)}{108 \xi ^2}e^{\pm i\pi/3}+{ O}\left(\frac{1}{\xi^3 }\right)\right]\;.
\end{eqnarray}
The subleading terms of the expansion up to order $O(1/\xi^8)$ can be found
in Appendix \ref{AH}, see  \eq{moremore}. Using the definition \eq{d-def} we can find an analogous expression for the
scaling dimension
\begin{align}\la{l5pred1}
\Delta = 2\sqrt{3} \, e^{\pm i \pi/6} \xi + N+2 + {(3N^2+17)\over 18 \sqrt{3} \xi}e^{\mp i \pi/6} + O(1/\xi^2)\,,
\end{align}
where expansion runs in powers of $e^{\pm i \pi/6} \xi$.

\begin{figure}
\center{\includegraphics[scale=0.65]{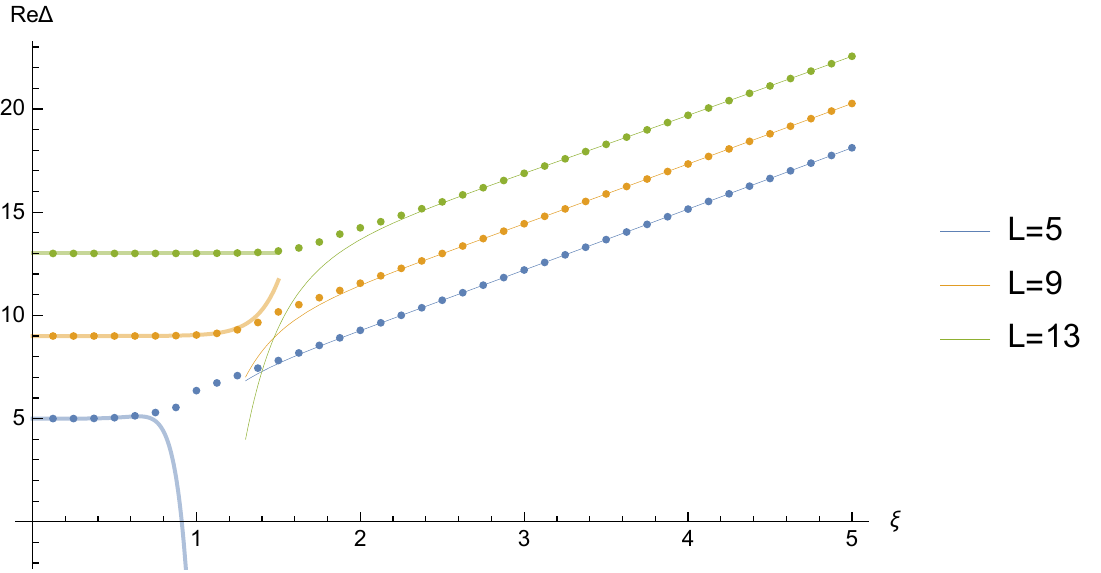}
\includegraphics[scale=0.65]{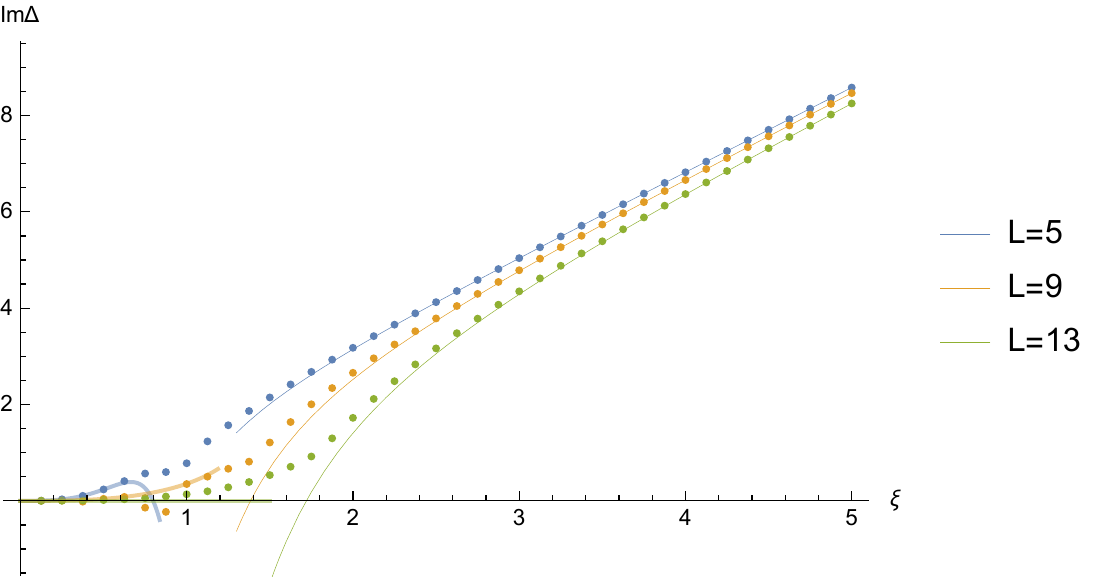}
\caption{\label{fig:LD5} Real and imaginary parts of the dimension $\Delta(\xi)$ as a function of $\xi$ for the states with $L=\Delta(0)=5,9,13$. Dots represent the numerical data, this line is given by our quasi-classical prediction \eq{l5pred} with $N=1,3,5$ correspondingly. The thick line is the weak coupling prediction.}}
\end{figure}

The relations \eq{l5pred} and \eq{l5pred1} are valid at large $\xi$ with $N$ fixed.
We checked them against numerical data on Fig.~\ref{fig:LD5} for different $N$.
We found that the identification of integers $N$ corresponding to different states
is not trivial. Namely, for odd $N$ the relation \eq{l5pred} correctly describes the states with $\Delta(0)=5,9,13,\dots$
whereas for even $N$ it predicts some states which are not present in the spectrum.  To some extend that is expected as these states come in the complex conjugate pairs and in order to have the same average number of states in some interval of the quantum numbers $N$ as for the states with
$\Delta(0)=3,7,11,\dots$ we should miss exactly half of all $N$'s. 

\section{Problems and perspectives}

In this paper, we used the quantum integrability methods to 
compute the exact scaling dimensions of various operators in four-dimensional bi-scalar theory defined by the Lagrangian ~\eqref{bi-scalarL} (also called \(\chi\)FT\(_4\)) in planar limit. This theory has the symmetry \(SU(2,2)\times U(1)^2\) ~--- a subgroup of the original \(PSU(2,2|4)\)  conformal symmetry of the full (untwisted) \({\cal\ N}=4\) SYM theory from which it was obtained in a special double scaling limit combining large imaginary \(\gamma\)-twist and small coupling. Consequently, the operators in this theory are classified by the set of Cartan charges \((\Delta,S,\dot S|J_1,J_2)\).  The basic example of such operators is  \(\tr(\phi_1^J)\)  -- the ``BMN vacuum" operator\footnote{It was  protected BMN vacuum operator in untiwisted \({\cal\ N}=4\) SYM theory but it gets corrections after \(\gamma\)-twisted \({\cal\ N}=4\) SYM    and, consequently, in the bi-scalar \(\chi\)FT\(_4\).} with charges \((\Delta,0,0|J,0)\). The  perturbative  corrections for this operator  are given by so called ``wheel" graphs of Fig.\ref{fig:multiwheel} (a single graph at each non-zero order). Our method makes it possible to compute these conformal graphs at least to \(O({1}/{\epsilon })\) order in dimensional regularization, in an algorithmic way. We computed the \(J=3\) wheel graphs up to  12 loops,  which can be easily pushed further with more computer time or on a more powerful computer. Our numerical computation of anomalous dimension have been done for this operator at virtually unlimited precision for all reasonable  couplings \(\xi\).   Moreover, it is also applicable to the other operators, of the type \eqref{O-def}, with the same charges but a bigger length \(L\) 
corresponding to the insertion of any number of couples of  fields \(\phi_1,\,\phi_1^\dagger\) and \(\phi_2,\,\phi_2^\dagger\) inside   \(\tr(\phi_1^J)\), as well as of the derivatives, where we performed similar analytic and numerical computations. Furthermore, we investigated the strong coupling limit of the anomalous dimensions for this family of operators. In this limit we found that the spectrum is described by a classical algebraic curve and the quantization condition reduces to a modified 
Bohr-Sommerfeld quantization condition with half-integer filling fractions. We managed to develop a systematic strong coupling expansion
to a high order in $1/\xi$.

To compute the spectrum of scaling dimensions \(\Delta\),  we needed to provide two basic ingredients:    i) 
formulate a 4th order finite difference Baxter equation for a quartet of ${\bf Q}-$functions of spectral parameter; ii)  derive quantization condition  fixing the  unique solution with prescribed analytic structure, and at the same time all the constants in Baxter equation and the  spectrum of dimensions \(\Delta(\xi)\) with given charges.

   We managed to  solve the problem i) -- to find the Baxter equation   at any \(J\)  for the operators in the sector \((\Delta,0,0|J,0)\) . We did it in two ways: from AdS\(_5\)/CFT\(_4\) QSC formalism (so far only for the charges \(J=2,3,4)\), where the full Baxter equation is known,  in the double scaling limit leading to the bi-scalar theory; and also from explicit Lax operator of \(SU(2,2)\) conformal spin chain describing the underlying  fishnet graphs of the bi-scalar model~\cite{Gurdogan:2015csr}. We used  the fact that the Baxter equation is a universal object in quantum integrability, independent of the auxiliary representation of Lax operator.
 The resulting Baxter relation, even though cross-checked in many ways, is not derived in a rigorous way in none of two formalisms. Certain very natural assumptions  of symmetry of fundamental transfer-matrices has been used, but not yet formally proven, in derivation from Lax operators. It would be good to check these assumptions algebraically, directly from the from the  spin chain formalism.  The QSC derivation serves as a good cross-check but it also involves certain natural, but unproved assumptions about the scaling behavior of QSC Q-functions in the double scaling limit. One of the ways to verify these assumptions is to use numerics for the full twisted ${\cal N}=4$ SYM and then approach the limit with infinite coupling  numerically.

Concerning the ingredient ii), we  managed to derive the quantization condition from the QSC formalism,  so far  only in the case \(J=3\), where the 4th order Baxter equation nicely factorizes into two 2nd order equations. Its derivation for an arbitrary \(J\) can be  certainly obtained in a similar way but we leave this, likely more involved, calculation to  future publications. In particular for $J=4$ one should reproduce~\cite{Gurdogan:2015csr}  
\begin{equation}
\gamma^{\L=4}_{\mathrm{vac}} =-40\zeta_5\,\xi^8+
\,8\,\bigl[
309\zeta_{11}
+16\zeta_{3,8}
+20\zeta_{5,6}
-4\zeta_{6,5}
+40\zeta_{8,3}
-8\zeta_{3,3,5}
+40(\zeta_{3,5,3}+\zeta_{5,3,3})
-200\, \zeta_{5}^2
\bigr]\,\xi^{16}+\CO(\xi^{24})\;.
\end{equation}

It would be also interesting to derive the quantization condition in a rigorous way, from Lax formalism of the underlying conformal spin chain. It would  involve the computation of  certain spin chain eigenfunctions and matrix elements  relevant to the ``wheel" graphs. Probably, the most adequate formalism for it should be the Sklyanin's separated variables (SOV).   Such a formalism would not only allow to compute in a rigorous way the wheel graphs but also  get hand on a much bigger variety of physical quantities:  more general than in~\cite{Zamolodchikov:1980mb}  fishnet-type graphs at arbitrary loop orders, such as described in~\cite{Caetano:2016ydc}, multi-point correlators and amplitudes~\cite{Chicherin:2017cns} in \(\chi\)FT\(_4\), etc. Recently observed simplifications in the SOV approach to the wave functions~\cite{Gromov:2016itr} could pay an important role here.

Our methods are potentially applicable to the  \(\chi\)FT\(_3\) theory obtained in the double scaling limit of twisted ABJM~\cite{Caetano:2016ydc}, where certain classes of \(\Phi^6\)-type graphs can be computed at any loop order. One can try to apply for it both the Lax \cite{Chicherin:2012yn} and the QSC formalisms \cite{Bombardelli:2017vhk,Cavaglia:2014exa}.  One could also try to approach the same program  for tri-scalar  \(\chi\)FT\(_6\) with  \(\Phi^3\)-type chiral interactions, which seem to define the genuine CFT in planar limit~\cite{Mamroud:2017uyz}.  The 6D twisted SYM "mother"-theory, from which the latter model could be obtained in a double scaling limit, is  unfortunately  not known. 

It would be interesting to include into our formalism more general \(\chi\)FT\(_4\) containing fermions and more of scalars, described in~\cite{Gurdogan:2015csr, Caetano:2016ydc}. This could be done by applying more sophisticated limits to AdS/CFT QSC equations or, alternatively, establishing the Lax formalism for conformal spin chain with fermionic degrees of freedom. The latter would open a way to multi-loop computations of more sophisticated  than \(\Phi^4\) type graphs, containing also the Yukawa couplings. 

It would be also interesting to compute next-to-leading corrections to the double scaling limit of \(\gamma\)-twisted \({\cal N}=4\)  planar SYM which might give a better understanding of the origins of integrability in the full theory. Another, more ambitious direction, would be to generalize the results of the current paper to the full $PSU(2,2|4)$ Heisenberg spin chain, which could led to the prove  of integrability of 
\({\cal N}=4\)  planar SYM from first principles.

\section*{Acknowledgements}

We are thankful to B. Basso, J. Caetano, D.Chicherin, L. Dixon,
D. Grabner,  J. Henn, N.Reshetikhin,  F.Smirnov, M.Staudacher, P.Wiegmann, A.Zamolodchikov   for
discussions. The work of V.K., S.N. and G.S. was supported   by the European Research Council (Programme
``Ideas" ERC-2012-AdG 320769 AdS-CFT-solvable). N.G. and V.K. are grateful to Humboldt
University (Berlin) for the hospitality and financial support of this work in the framework of the ``Kosmos" programme. N.G. wishes to thank STFC for support from Consolidated grant number ST/J002798/1.

\begin{appendices}

\section{Baxter equation from the Lax operator formalism}
\label{subsec:Baxter_from_lax}


We have seen in subsection \ref{subsec:graph_build_transf_mat} that the Hamiltonian $\mathcal{H}_{{\L}}$ can be interpreted
as the transfer matrix of a spin chain of length ${\L}$ with periodic boundary conditions. This operator is invariant
under the action of the conformal $SO\left(1,5\right)$ group, whose
Lie algebra is isomorphic (as a complex Lie algebra) to $\mathfrak{su}\left(2,2\right)$.
In fact, it is well known 
that there exists a whole chain of isomorphisms of complex Lie
algebras
\begin{equation}
\mathfrak{so}\left(2,4\right)\simeq\mathfrak{so}\left(1,5\right)\simeq\mathfrak{so}\left(6\right)\simeq\mathfrak{sl}\left(4\right)\simeq\mathfrak{su}\left(2,2\right)\;,
\end{equation}
suggesting that we can define the $R$-matrix for any of the above Lie algebras. It turns
out that, in order to construct the Baxter $T$-$Q$ relation,
the best choice is to consider $\mathfrak{sl}\left(4\right)$ acting on
the following Hilbert space
\begin{equation}
\mathsf{H}_{\mathfrak{sl}\left(4\right)}=\bigotimes_{j=1}^{{\L}}V\left(\boldsymbol{z}\right)\,, 
\end{equation}
where $V(\boldsymbol{z})=\mathbbm C[\left\{ z_{ij}\right\} _{1\leq j<i<4}]$ is
the vector space of polynomials of arbitrary degree in $z_{ij}$ variables
with complex coefficients. The $R-$matrix for this setting
acts on the space $V\left(\boldsymbol{z}\right)\otimes\pi_{\Lambda}\left[\mathfrak{sl}\left(4\right)\right]$,
where the symbol $\pi_{\Lambda}$ denotes the finite-dimensional representation
with highest-weight $\Lambda$. We denote such 
$R$-matrices  as $R_{\pi,\boldsymbol{\nu}}$, where
$\boldsymbol{\nu}$ is an index that specifies the particular representation
of $\mathfrak{sl}\left(4\right)$ on $V(\boldsymbol{z})$.
The spin chains based on this Hilbert space have been studied for
$\mathfrak{sl}\left(n\right)$ by Derkachov and Manashov in \cite{Derkachov2009,Derkachov2011}. We chose to adhere to their conventions and employ the real parameter $\su=iu$ for the derivation of the Baxter equation.\\

\subsection*{Generic $\mathfrak{sl}\left(4\right)$ Verma modules}
\label{subsubsec:PS}

Here we construct the most general highest-weight representation of
the $\mathfrak{sl}\left(4\right)$ algebra on the vector space $V\left(\boldsymbol{z}\right)$
of polynomials in $6$ complex variables $\boldsymbol{z}=\left\{ z_{ij}\right\} _{1\leq j<i\leq4}$ of arbitrary degree.%
\footnote{These representations
have a deep relationship with the principal series representations
\cite{Gelfand_Naimark_unitary_repr,Zelobenko1973} of the group $\mathrm{SL}\left(4\vert\mathbbm C\right)$
as explained in \cite{Derkachov2009,Derkachov2006b,Chicherin:2012yn}.} This representation is parameterized
by a set of $4$ complex numbers $\boldsymbol{\nu}:=\left(\nu_{1},\nu_{2},\nu_{3},\nu_{4}\right)$,
such that
\begin{equation}
\sum_{k=1}^{4}\nu_{k}=6\,.
\end{equation}
Complex $z_{ij}$ variables can be used to define the lower-triangular matrix
\begin{equation}
Z=\mathbbm I_{4}+\sum_{1\leq j<i\leq4}z_{ij}\hat{e}_{ij}=\begin{pmatrix}1 & 0 & 0 & 0\\
z_{21} & 1 & 0 & 0\\
z_{31} & z_{32} & 1 & 0\\
z_{41} & z_{42} & z_{43} & 1
\end{pmatrix}\,,
\end{equation}
where $\left(\hat{e}_{ij}\right)_{k}^{l}:=\delta_{ik}\delta_{j}^{l}$
are the unit $4\times4$ matrices.
Define further the matrix differential operator
\begin{equation}
\mathcal{D}=\begin{pmatrix}0 & \partial_{21}+z_{32}\partial_{31}+z_{42}\partial_{41} & \partial_{31}+z_{43}\partial_{41} & \partial_{41}\\
0 & 0 & \partial_{32}+z_{43}\partial_{42} & \partial_{42}\\
0 & 0 & 0 & \partial_{43}\\
0 & 0 & 0 & 0
\end{pmatrix}\;.
\end{equation}
The algebra $\mathfrak{sl}(4)$, with generators $e_{ij}$ satisfying the commutation relations
\begin{equation}
\left[e_{ij},e_{kl}\right]=\delta_{jk}e_{il}-\delta_{il}e_{kj}\;,
\end{equation}
can then be represented in $\textrm{Aut}\left[V\left(\boldsymbol{z}\right)\right]$
by defining its generators as
\begin{equation}
E_{ij}^{\boldsymbol{\nu}}\equiv\pi_{\boldsymbol{\nu}}(e_{ij})=-\left(Z\left(\mathcal{D}+\hat{\nu}\right)Z^{-1}\right)_{ji}\;,\label{eq:princ_ser_gen}
\end{equation}
where $\hat \nu={\rm diag}(\nu_1,\nu_2,\nu_3,\nu_4)$. The corresponding
$\mathfrak{sl}(4)$-module
$V_{\boldsymbol{\nu}}$ is irreducible \cite{Bern_Gelf_Gelf_75,Verma1968}
iff $\nu_{ij}:=\nu_{i}-\nu_{j}\notin\mathbb{N}\;,\;\forall i<j$,
otherwise $V_{\boldsymbol{\nu}}$  contains an invariant subspace.

Let us consider a set of complex numbers $\boldsymbol{\sigma}=\{\sigma_{i i+1}\}$ such
that
\begin{equation}
\sigma_{n\, n+1}\in\mathbb{N}\;,\qquad\forall n\geq4-k\;.
\end{equation}
Then it can be shown \cite{Zelobenko1973} that the associated principal
series $\mathfrak{sl}\left(4\right)$-module $V_{\boldsymbol{\sigma}}^{\left(k\right)}$
decomposes as a tensor product:
\begin{equation}
V_{\boldsymbol{\sigma}}^{\left(k\right)}=V_{3-k}\otimes v_{k+1}^{\boldsymbol{\sigma}}\;.
\end{equation}
One of the factors is the infinite-dimensional space $V_{3-k}$ of
polynomials in the variables appearing in the first $3-k$ columns and rows
of $Z$. The other factor $v_{k+1}^{\boldsymbol{\sigma}}$ is a finite-dimensional
$\mathfrak{sl}\left(k+1\right)$ module with highest weight $\Lambda_{\boldsymbol{\sigma}}=\left(\lambda_{k},\cdots,\lambda_{3}\right)$
and $\lambda_{j}=\sigma_{j\, j+1}-1$. Consequently, choosing $k=3$,
we have that the principal series module $V_{\boldsymbol{\sigma}}^{\left(3\right)}$
collapses to a finite dimensional $\mathfrak{sl}\left(4\right)$ module:
\begin{equation}
V_{\boldsymbol{\sigma}}^{\left(3\right)}=v_{4}^{\boldsymbol{\sigma}}=\pi_{\Lambda_{\boldsymbol{\sigma}}}\;,\qquad\Lambda_{\boldsymbol{\sigma}}=\left(\sigma_{12}-1,\sigma_{23}-1,\sigma_{34}-1\right)\;.
\end{equation}
This module is associated to the Young tableau defined by the partition
$\boldsymbol{\ell}=\left\{ \ell_{1},\ell_{2},\ell_{3}\right\} $ with
$\ell_{j}=\sum_{k=j}^{3}\lambda_{k}$.

\subsection*{The general $T-Q$ relation from $\mathfrak{sl}(4)$ invariant $R$-matrix}
\label{sec:Baxter_construction}

Let us consider the following invariant $R$-matrix:
\begin{equation}
\mathcal{R}\,:\quad V_{\boldsymbol{\nu}}\otimes V_{\boldsymbol{\sigma}}\longrightarrow V_{\boldsymbol{\nu}}\otimes V_{\boldsymbol{\sigma}}\;,\qquad\mathcal{R}_{12}\left(\su\right)\mathcal{R}_{13}\left(\su+\mathsf v\right)\mathcal{R}_{23}\left(\mathsf v\right)=\mathcal{R}_{23}\left(\mathsf v\right)\mathcal{R}_{13}\left(\su+\mathsf v\right)\mathcal{R}_{12}\left(\su\right)\;,
\end{equation}
where both $V_{\boldsymbol{\nu}}$ and $V_{\boldsymbol{\sigma}}$
are $\mathfrak{sl}\left(4\right)$ principal series modules, defined above. As was shown in \cite{Derkachov2009,Derkachov2006b},  this operators adimt a factorised form:
\begin{equation}
\mathcal{R}_{12}\left(\su\right)=P_{12}\mathbbm R_{12}^{\left(1\right)}\left(\su-\nu_{1}+\sigma_{1}\right)\cdots\mathbbm R_{12}^{\left(4\right)}\left(\su-\nu_{4}+\sigma_{4}\right)\;,
\end{equation}
where $P_{ij}$ is the permutation operator of spaces $i$ and $j$.
The associated $T$-operator, defined in the usual way as the trace
over the auxiliary space $V_{\boldsymbol{\sigma}}$ of the product
of $J$ $R$-matrices%
\footnote{In order to consistently define the $T$- and $Q$- operators it is
necessary to introduce a regulator in the $R$-matrix. We choose to
avoid this subtlety as it is of no relevance for our goal. Further
informations can be found in \cite{Derkachov2011}.%
}
\begin{equation}
\mathcal{T}_{\boldsymbol{\sigma}}\left(\su\left|\boldsymbol{\nu}\right.\right)=\textrm{tr}_{V_{\boldsymbol{\sigma}}}\left[\mathcal{R}_{10}\left(\su\right)\mathcal{R}_{20}\left(\su\right)\cdots\mathcal{R}_{J0}\left(\su\right)\right]\;,
\end{equation}
will then enjoy a factorised expression
\begin{equation}
\mathcal{T}_{\boldsymbol{\sigma}}\left(\su\left|\boldsymbol{\nu}\right.\right)=\mathcal{Q}_{1}\left(\su+\sigma_{1}\left|\boldsymbol{\nu}\right.\right)\cdots\mathcal{Q}_{4}\left(\su+\sigma_{4}\left|\boldsymbol{\nu}\right.\right)\;,
\end{equation}
where the $Q$-operators are defined as transfer matrices associated
to the factorising operators $\mathbbm R^{\left(j\right)}$:
\begin{equation}
\mathcal{Q}_{j}\left(\su\left|\boldsymbol{\nu}\right.\right)=\textrm{tr}_{V_{\boldsymbol{\sigma}}}\left[\tilde{\mathbbm R}_{10}^{\left(j\right)}\left(\su-\nu_{j}\right)\tilde{\mathbbm R}_{20}^{\left(j\right)}\left(\su-\nu_{j}\right)\cdots\tilde{\mathbbm R}_{J0}^{\left(j\right)}\left(\su-\nu_{j}\right)\right]\;,\qquad\tilde{\mathbbm R}_{k0}^{\left(j\right)}\left(\su\right)=P_{k0}\mathbbm R_{k0}^{\left(j\right)}\left(\su\right)\;.
\end{equation}

This factorisation is a general property, irrespective of
the fact that $V_{\boldsymbol{\sigma}}$ is irreducible or not. If
we were to choose $\boldsymbol{\sigma}$ such that $V_{\boldsymbol{\sigma}}=V_{\boldsymbol{\sigma}}^{\left(k\right)}$,
then the $R$-matrix will assume an upper-triangular form
\begin{equation}
\mathcal{R}=\begin{pmatrix}R & \ast\\
0 & \mathcal{R}'
\end{pmatrix}\;,\qquad R\,:\quad V_{\boldsymbol{\nu}}\otimes v_{k+1}^{\boldsymbol{\sigma}}\longrightarrow V_{\boldsymbol{\nu}}\otimes v_{k+1}^{\boldsymbol{\sigma}}\;,
\end{equation}
which implies
\begin{equation}
\mathcal{T}_{\boldsymbol{\sigma}}\left(\su\left|\boldsymbol{\nu}\right.\right)=T_{\boldsymbol{\sigma}}^{\left(k\right)}\left(\su\left|\boldsymbol{\nu}\right.\right)+\mathcal{T}_{\boldsymbol{\sigma}}'\left(\su\left|\boldsymbol{\nu}\right.\right)\;.
\end{equation}
It is possible to prove, by using Bernstein-Gel'fand-Gel'fand resolution
for finite dimensional $\mathfrak{sl}\left(n\right)$ modules \cite{Derkachov2011,Bern_Gelf_Gelf_75},
that
\begin{equation}
T_{\boldsymbol{\sigma}}^{\left(k\right)}\left(\su\left|\boldsymbol{\nu}\right.\right)=\left[\prod_{j=1}^{3-k}\mathcal{Q}_{j}\left(\su+\sigma_{j}\left|\boldsymbol{\nu}\right.\right)\right]\det_{1\leq i,j\leq k+1}\Big(\mathcal{Q}_{3-k+i}\left(\su+\sigma_{3-k+j}\left|\boldsymbol{\nu}\right.\right)\Big)\;.
\end{equation}
Finally, by choosing $k=3$, we obtain the following nice determinant
expression
\begin{equation}
T_{\boldsymbol{\sigma}}^{\left(3\right)}\left(\su\left|\boldsymbol{\nu}\right.\right)=\det_{1\leq i,j\leq4}\Big(\mathcal{Q}_{i}\left(\su+\sigma_{j}\left|\boldsymbol{\nu}\right.\right)\Big)\;,
\end{equation}
which can be rewritten in terms of Young tableaux indices as
\begin{equation}
T_{\boldsymbol{\ell}}^{\left(3\right)}\left(\su+f_{\boldsymbol{\ell}}\left|\boldsymbol{\nu}\right.\right)=\det_{1\leq i,j\leq4}\Big(\mathcal{Q}_{i}\left(\su+l_{j}\left|\boldsymbol{\nu}\right.\right)\Big)\;,
\end{equation}
where
\begin{equation}
l_{j}=4-j+\ell_{j}\;,\qquad f_{\boldsymbol{\ell}}=\frac{1}{4}\sum_{k=1}^{3}\ell_{k}\;.
\end{equation}
In table \ref{tab:representation_labels} we have collected the labels associated to the fundamental representations\footnote{Note that in our conventions, the defining representation $\mathbf{4}$ is associated to the Young tableau $\boldsymbol{\ell}_1 = \left(1,1,1\right)$ with three boxes, while the conjugate one $\overline{\mathbf{4}}$ to the Young tableau $\boldsymbol{\ell}_1 = \left(1,0,0\right)$ with a single box. This is all a matter of preference, since the $\mathbbm Z_2$ symmetry of the Dynkin diagram makes the two representations isomorphic. However the isomorphism is realised in a complicated fashion at the level of $R$-matrices which results in the fact that the associated transfer matrices are, in general, radically different, as we see later. Our choice was dictated by the request that $t_1$ should be the simplest among the transfer matrices.} $\pi_k$.

\begin{table}

\begin{centering}
\begin{tabular}{|c|c|c|c|}
\hline
Representation & Fundamental weight & Young tableau integers & Principal Series labels\tabularnewline
\hline
\hline
$\boldsymbol{1}$ & $\Lambda_{0}=\left(0,0,0\right)$ & $\boldsymbol{\ell}_{0}=\left(0,0,0\right)$ & $\boldsymbol{\sigma}=\left(3,2,1,0\right)$\tabularnewline
\hline
$\boldsymbol{4}$ & $\Lambda_{1}=\left(0,0,1\right)$ & $\boldsymbol{\ell}_{1}=\left(1,1,1\right)$ & $\boldsymbol{\sigma}=\left(\frac{13}{4},\frac{9}{4},\frac{5}{4},-\frac{3}{4}\right)$\tabularnewline
\hline
$\boldsymbol{6}$ & $\Lambda_{2}=\left(0,1,0\right)$ & $\boldsymbol{\ell}_{2}=\left(1,1,0\right)$ & $\boldsymbol{\sigma}=\left(\frac{7}{2},\frac{5}{2},\frac{1}{2},-\frac{1}{2}\right)$\tabularnewline
\hline
$\overline{\boldsymbol{4}}$ & $\Lambda_{3}=\left(1,0,0\right)$ & $\boldsymbol{\ell}_{3}=\left(1,0,0\right)$ & $\boldsymbol{\sigma}=\left(\frac{15}{4},\frac{7}{4},\frac{3}{4},-\frac{1}{4}\right)$\tabularnewline
\hline
\end{tabular}\protect\caption{Various labels for the fundamental representations of $\mathfrak{sl}\left(4\right)$\label{tab:representation_labels}}

\par\end{centering}

\end{table}

By considering the following identity
\begin{equation}
\det_{1\leq i,j\leq5}\Big(\mathcal{Q}_{i}\left(\su+5-j\left|\boldsymbol{\nu}\right.\right)\Big)=0\;,\qquad\mathcal{Q}_{5}\left(\su\left|\boldsymbol{\nu}\right.\right)\equiv\mathcal{Q}_{k}\left(\su\left|\boldsymbol{\nu}\right.\right)\;,\quad \textrm{for some}\quad k=1,\ldots,4\;,
\end{equation}
and denoting the $T$-operators associated to the $k$-th fundamental
representation as
\begin{equation}
t_{4-k}\left(\su\left|\boldsymbol{\nu}\right.\right)=T_{\mathbbm1_{k}}^{\left(3\right)}\left(\su\left|\boldsymbol{\nu}\right.\right)\;,\qquad\mathbbm1_{k}=\big(\underbrace{1,\ldots,1}_{k},0,\ldots,0\big)\;,
\end{equation}
and understanding $t_{4}\left(\su\right)\equiv t_{0}\left(\su\right)$,
we easily obtain the general form of Baxter $T$-$Q$ relation for a $\pi_{\boldsymbol{\nu}}\left[\mathfrak{sl}(4)\right]$-spin chain \cite{Derkachov2011}
\begin{equation}
\sum_{k=0}^{4}\left(-1\right)^{k}t_{4-k}\left(\left.\su+\frac{k}{4}\right|\boldsymbol{\nu}\right)\mathcal{Q}_{i}\left(\su+4-k\left|\boldsymbol{\nu}\right.\right)=0\;,\qquad\forall i=1,\ldots,4\;.\label{eq:Baxter_EQ_gen_form}
\end{equation}

\subsection*{The fundamental $R$-matrices and the associated Lax operators}

In order to obtain an explicit expression for the transfer matrices
$t_{k}$, we need to compute the traces
\begin{equation}
t_{k}\left(\left.\su\right|\boldsymbol{\nu}\right)=\textrm{tr}_{\pi_{\Lambda_{k}}}\left[R_{10}^{\left(k\right)}\left(\su\right)R_{20}^{\left(k\right)}\left(\su\right)\cdots R_{L0}^{\left(k\right)}\left(\su\right)\right]\;,
\end{equation}
where $R^{\left(k\right)}$ is the $R$ matrix with physical space
$V_{\boldsymbol{\nu}}$ and auxiliary space $\pi_{\Lambda_{k}}$
and $\left(\Lambda_{k}\right)^{i}=\delta_{k}^{i}$ is the $k$-th
fundamental weight.
These $R$-matrices for the fundamental representations are related to the Lax operators $\mathsf{L}_k$ by a normalization:
\begin{equation}
R^{\left(k\right)}\left(\su+\kappa_{k}\right)=X_{k,\boldsymbol{\nu}}\left(\su\right)\mathsf{L}_{k,\boldsymbol{\nu}}\left(\su\right)\;,
\label{eq:R_Lax}
\end{equation}
where $\kappa_{k}$ are some rational numbers and $X_{\pi_{k},\boldsymbol{\nu}}\left(\su\right)$
are functions. The Lax operator for the basic representation is a linear operator on the space $V_{\boldsymbol{\nu}}\otimes\mathbf{4}\left[\mathfrak{sl}(4)\right]$ and is defined as follows:
\begin{equation}
\mathsf{L}_{1,\boldsymbol{\nu}}\left(\su\right) = \su+\hat{e}_{ij}E_{ji}^{\boldsymbol{\nu}}\;,\qquad \left(\hat{e}_{ij}\right)_a^b = \delta_{ia}\delta_j^b\;.
\end{equation}
The higher Lax operators $\mathsf{L}_{2,\boldsymbol{\nu}}$ and $\mathsf{L}_{3,\boldsymbol{\nu}}$ are linear operators on, respectively, $V_{\boldsymbol{\nu}}\otimes\mathbf{6}\left[\mathfrak{sl}(4)\right]$ and $V_{\boldsymbol{\nu}}\otimes\overline{\mathbf{4}}\left[\mathfrak{sl}(4)\right]$. Although they have not a simple form such as $\mathsf{L}_{1,\boldsymbol{\nu}}$, from classical representation theory it is known that
\begin{equation}
\mathbf{6}\left[\mathfrak{sl}(4)\right] = \mathbf{4}\left[\mathfrak{sl}(4)\right] \wedge \mathbf{4}\left[\mathfrak{sl}(4)\right] \;,\qquad \overline{\mathbf{4}}\left[\mathfrak{sl}(4)\right] = \mathbf{4}\left[\mathfrak{sl}(4)\right] \wedge \mathbf{4}\left[\mathfrak{sl}(4)\right] \wedge \mathbf{4}\left[\mathfrak{sl}(4)\right] \;,
\end{equation}
which means that $\mathsf{L}_{2,\boldsymbol{\nu}}$ and $\mathsf{L}_{3,\boldsymbol{\nu}}$ can be defined as follows
\begin{equation}
\mathsf{L}_{2,\boldsymbol{\nu}}\left(\su\right) = \left(\wedge^2\mathsf{L}^2_{1,\boldsymbol{\nu}}\right) \left(\su\right)\;,\qquad \mathsf{L}_{3,\boldsymbol{\nu}}\left(\su\right) = \left(\wedge^3\mathsf{L}^3_{1,\boldsymbol{\nu}}\right) \left(\su\right)\;.
\end{equation}
With the notation $\left(\wedge^k A^k\right)\left(\su\right)$, where $A$ is a linear operator on a space $V$, we denote a linear operator on the space $\bigwedge^k V$, acting as follows
\begin{equation}
\left(\wedge^k A^k\right)\left(\su\right) \phi_1 \wedge \phi_2 \wedge \cdots \wedge \phi_k = \left(A\left(\su-k+1\right)\phi_k\right) \wedge\left(A\left(\su-k+2\right)\phi_{k-1}\right) \wedge \cdots \wedge \left(A\left(\su-1\right)\phi_2\right) \wedge \left(A\left(\su\right)\phi_1\right)\;,
\label{eq:wedgeaction}
\end{equation}
with $\phi_k\in V$. As it will be useful in a short while we also introduce the quantum determinant of the Lax operator $\mathsf{L}_{1,\boldsymbol{\nu}}$:
\begin{equation}
\textrm{qdet}\left[\mathsf{L}_{1,\boldsymbol{\nu}}\left(\su\right)\right] \mathbbm 1 \equiv \mathsf{L}_{4,\boldsymbol{\nu}}\left(\su\right) = \left(\wedge^4 \mathsf{L}^4_{1,\boldsymbol{\nu}}\right)\left(\su\right)\;.
\end{equation}

In order to fix the normalisations $\kappa_k$ and $X_{k,\boldsymbol{\nu}}$ in (\ref{eq:R_Lax}), we apply
the $R$-matrix on the highest-weight state $\Phi_{0}=1\in V_{\boldsymbol{\nu}} \otimes V_{\boldsymbol{\sigma}}$, using the
known expression for the eigenvalues of the general $R$-matrix \cite{Derkachov2009}
\begin{equation}
\mathcal{R}(\su)\cdot\Phi_{0}=p\left(\su\right)\prod_{1\leq i<j\leq4}\frac{\Gamma\left(\su-\nu_{i}+\sigma_{j}+1\right)}{\Gamma\left(-\su+\nu_{j}-\sigma_{i}+1\right)}\,\Phi_{0}\;,
\end{equation}
with $p\left(\su\right)$ being a periodic function whose exact expression
is irrelevant for our needs. Plugging in the values of the labels $\boldsymbol{\sigma}$ corresponding to the fundamental representations (see table \ref{tab:representation_labels}), we obtain
\begin{align}
\frac{r_{0}(\su+1)}{r_{0}(\su)}&=\delta(\su+1)\delta(\su+2)\delta(\su+3)\;,\qquad\qquad\quad\;\frac{r_{1}\left(\su+\frac{3}{4}\right)}{r_{0}(\su)}=\delta(\su+2)\delta(\su+3)(\su-\nu_{4}+1)\;,\nonumber \\
\\
\frac{r_{2}(\su+\frac{1}{2})}{r_{0}(\su)}&=-\delta(\su+3)(\su+\nu_{3}+2)(\su+\nu_{4}+2)\;,\qquad\frac{r_{3}\left(\su+\frac{1}{4}\right)}{r_{0}(\su)}=-(\su+\nu_{2}+3)(\su+\nu_{3}+3)(\su+\nu_{4}+3)\;, \nonumber
\end{align}
where we denoted as $r_{k}\left(\su\right)$ the eigenvalue of $R^{\left(k\right)}\left(\su\right)$
on the highest weight state $\Phi_{0}$ and we introduced the notation
\begin{equation}
\delta\left(\su\right) = \prod_{j=1}^4 \left(\su-\nu_j\right)\;.
\end{equation}

Now we will compare these eigenvalues with those obtained by acting with the Lax operators $\mathsf{L}_{k,\boldsymbol{\nu}}$ on the highest weight states
\begin{equation}
\Phi_{1}=\begin{pmatrix}0\\
0\\
0\\
1
\end{pmatrix}\;,\qquad\Phi_{2}=\begin{pmatrix}0\\
0\\
0\\
1
\end{pmatrix}\wedge\begin{pmatrix}0\\
0\\
1\\
0
\end{pmatrix}\;,\qquad\Phi_{3}=\begin{pmatrix}0\\
0\\
0\\
1
\end{pmatrix}\wedge\begin{pmatrix}0\\
0\\
1\\
0
\end{pmatrix}\wedge\begin{pmatrix}0\\
1\\
0\\
0
\end{pmatrix}\;,\qquad \Phi_k\in V_{\boldsymbol{\nu}}\otimes\left(\bigwedge^k\mathbf{4}\left[\mathfrak{sl}(4)\right]\right)\;.
\end{equation}
These are easily computed by using the explicit representation of the operators $E_{ji}$ (\ref{eq:princ_ser_gen}) and the definition (\ref{eq:wedgeaction}):
\begin{align}
 \mathsf{L}_{1,\boldsymbol{\nu}}(\su)\cdot\Phi_{1}&=(\su-\nu_{4})\Phi_{1}\;,\nonumber \\
 \mathsf{L}_{2,\boldsymbol{\nu}}(\su)\cdot\Phi_{2}&=-\left(\su-\nu_{3}\right)\left(\su-\nu_{4}\right)\Phi_{2}\;, \\
 \mathsf{L}_{3,\boldsymbol{\nu}}(\su)\cdot\Phi_{3}&=-\left(\su-\nu_{2}\right)\left(\su-\nu_{3}\right)\left(\su-\nu_{4}\right)\Phi_{3}\;.\nonumber
\end{align}
We also easily obtain
\begin{equation}
\textrm{qdet}\left[\mathsf{L}_{1,\boldsymbol{\nu}}(\su)\right]= \delta\left(\su\right)\;.
\end{equation}
By direct comparison we then make the following identifications
\begin{align}
 &R^{\left(0\right)}\left(\su+1\right)=r_{0}\left(\su\right)\delta(\su+1)\delta(\su+2)\delta(\su+3)\;, \quad R^{\left(1\right)}\left(\su+\frac{3}{4}\right)=r_{0}\left(\su\right)\delta(\su+2)\delta(\su+3)\mathsf{L}_{1,\boldsymbol{\nu}}\left(\su+1\right)\;,\nonumber\\ \label{eq:relation_R_Lax}
 \\
 &R^{\left(2\right)}\left(\su+\frac{1}{2}\right)=r_{0}\left(\su\right)\delta(\su+3)\mathsf{L}_{2,\boldsymbol{\nu}}\left(\su+2\right)\;,\qquad R^{\left(3\right)}\left(\su+\frac{1}{4}\right)=r_{0}\left(\su\right)\mathsf{L}_{3,\boldsymbol{\nu}}\left(\su+3\right)\;.\nonumber
\end{align}
Note that we might have identified Lax operators with fundamental $R$-matrices also as
\begin{align}
 & R^{\left(2\right)}\left(\su+\frac{1}{2}\right)=r_{0}\left(\su\right)\delta(\su+3)\delta\left(\su+2\right)\left[\mathsf{L}_{2,\boldsymbol{\nu}}^{t}\left(\su+2\right)\right]^{-1}\;, \nonumber \\
 \\
 & R^{\left(3\right)}\left(\su+\frac{1}{4}\right)=r_{0}\left(\su\right)\delta\left(\su+3\right)\left[\mathsf{L}_{1,\boldsymbol{\nu}}^{t}\left(\su+3\right)\right]^{-1}\;,\nonumber
\end{align}
giving us the relations
\begin{equation}
\mathsf{L}_{1,\boldsymbol{\nu}}^{t}\left(\su\right)\mathsf{L}_{3,\boldsymbol{\nu}}\left(\su\right)=\delta\left(\su\right)\;,\qquad\mathsf{L}_{2,\boldsymbol{\nu}}^{t}\left(\su\right)\mathsf{L}_{2,\boldsymbol{\nu}}\left(\su\right)=\delta\left(\su\right)\;,
\end{equation}
which can be rewritten as
\begin{align}
&\mathsf{L}_{1,\boldsymbol{\nu}}^{t}\left(\su\right)\left(\wedge^3 \mathsf{L}^3_{1,\boldsymbol{\nu}}\right)\left(\su\right) = \left(\wedge^4 \mathsf{L}^4_{1,\boldsymbol{\nu}}\right)\left(\su\right) = \textrm{qdet}\mathsf{L}_{1,\boldsymbol{\nu}}\left(\su\right)\mathbbm1\;,\nonumber \\
\\
&\left(\wedge^2 \mathsf{L}^2_{1,\boldsymbol{\nu}}\right)^t\left(\su\right) \left(\wedge^2 \mathsf{L}^2_{1,\boldsymbol{\nu}}\right)\left(\su\right) = \left(\wedge^4 \mathsf{L}^4_{1,\boldsymbol{\nu}}\right)\left(\su\right) = \textrm{qdet}\mathsf{L}_{1,\boldsymbol{\nu}}\left(\su\right)\mathbbm1\;.\nonumber
\end{align}
These relations are the generalisations to the quantum case of the equalities
\begin{equation}
A^{t}\left(\wedge^3 A^3\right)=\det A\,\mathbbm1 \;, \qquad \left(\wedge^2 A^2\right)^t \left(\wedge^2 A^2\right)= \det A\,\mathbbm1\;,
\end{equation}
valid for any $4\times4$ matrix \cite{Wini_10}.

\subsection*{The monodromy matrix as a Manin matrix and Talalaev's formula}

From now on we will drop the explicit dependence on the labels $\boldsymbol{\nu}$.

Thanks to the identifications (\ref{eq:relation_R_Lax}) and defining the Monodromy matrix $\mathsf{M}\left(\su\right)$ of the spin chain as follows
\begin{equation}
\mathsf{M}\left(\su\right) = \frac{1}{[\delta\left(\su\right)]^J}\,\overset{1}{\mathsf{L}}_1\left(\su\right)\overset{2}{\mathsf{L}}_1\left(\su\right)\cdots \overset{J}{\mathsf{L}}_1\left(\su\right)\;,
\end{equation}
we are able to write the general Baxter equation (\ref{eq:Baxter_EQ_gen_form}) in the following form
\begin{align}
\mathcal Q_i\left(\su-1\right)  +  &\textrm{tr}\left[\left(\wedge^2\mathsf{M}^2\right)\left(\su+1\right)\right]\mathcal Q_i\left(\su+1\right) + \textrm{tr}\left[\left(\wedge^4\mathsf{M}^4\right)\left(\su+3\right)\right]\mathcal Q_i\left(\su+3\right)=\nonumber \\
=&\textrm{tr}\left[\mathsf{M}\left(\su\right)\right]\mathcal Q_i\left(\su\right) + \textrm{tr}\left[\left(\wedge^3\mathsf{M}^3\right)\left(\su+2\right)\right]\mathcal Q_i\left(\su+2\right) \;.
\label{eq:BAXTER}
\end{align}
Now we notice that the following relations hold
\begin{equation}
\left[\mathsf{M}_{ij}\left(\su\right)e^{\partial_\su},\mathsf{M}_{kl}\left(\su\right)e^{\partial_\su}\right] = \left[\mathsf{M}_{kj}\left(\su\right)e^{\partial_\su},\mathsf{M}_{il}\left(\su\right)e^{\partial_\su}\right]\;,\qquad \forall i,j,k,l=1,\ldots ,4\;.
\end{equation}
These are the defining relations for a Manin matrix \cite{Manin1988}, a particular class of matrices with not necessarily commuting entries. They are a natural extension of the matrices over commutative rings, in the sense that most of the standard theorem of linear algebra continue to hold true for them \cite{Chervov2009}. Of particular interest for us is the fact that the spectral determinant of $\mathsf{M}\left(\su\right)e^{\partial_\su}$ has the following expansion
\begin{equation}
\textrm{cdet}\left[t \mathbbm 1-\mathsf{M}\left(\su\right)e^{\partial_\su}\right] = \sum_{k=0}^4 t^{4-k}(-1)^k\textrm{tr}\left[\left(\wedge^k \mathsf{M}^k\right)\left(\su+k-1\right)\right] e^{k\partial_\su}\;,
\end{equation}
where ``cdet" stands for the column ordered determinant
\begin{equation}
\textrm{cdet}\left[A\right] = \sum_{\sigma\in\mathfrak S_4} (-1)^{\vert\sigma\vert}A_{\sigma(1)1}A_{\sigma(2)2}A_{\sigma(3)3}A_{\sigma(4)4}\;.
\end{equation}
As a consequence, the Baxter $T-Q$ equation (\ref{eq:BAXTER}) takes the following suggestive expression
\begin{equation}
\textrm{cdet}\left[\mathbbm 1-\mathsf{M}\left(\su\right)e^{\partial_\su}\right]\mathcal Q_i\left(\su-1\right) = 0\;,
\label{eq:baxter_Talalaev}
\end{equation}
which is the quantum version of a classical spectral curve equation. The above expression has first appeared in the works of Talalaev \cite{Talalaev2004} and we thus refer to it as ``Talalaev's formula".

\subsection*{The explicit computation of the transfer matrices}

More important on the practical level is the fact that the higher traces
\begin{equation}
\mathsf t_k\left(\su\right) = \textrm{tr}\left[\left(\wedge^k \mathsf{M}^k\right)\left(\su+k-1\right)\right]\;,
\end{equation}
satisfy the Newton's identities \cite{Chervov2008}
\begin{equation}
k \mathsf t_k\left(\su\right) e^{k\partial_\su}= \sum_{j=1}^k \left(-1\right)^{j-1}\mathsf t_{k-j}\left(\su\right)e^{(k-j)\partial_\su}\textrm{tr}\left[\left(\mathsf{M}\left(\su\right)e^{\partial_\su}\right)^j\right]\;,\qquad \mathsf{t}_0\left(\su\right) = 1\;.
\end{equation}
Explicitly we have
\begin{equation}
\mathsf t_1\left(\su\right) = \textrm{tr}\left[\mathsf{M}\left(\su\right)\right]\;,\qquad \mathsf t_2\left(\su\right) = \frac{1}{2} \textrm{tr}\left[\mathsf{M}\left(\su\right)\right] \textrm{tr}\left[\mathsf{M}\left(\su+1\right)\right] - \frac{1}{2} \textrm{tr}\left[\mathsf{M}\left(\su\right)\mathsf{M}\left(\su+1\right)\right]
\end{equation}
\begin{align}
\mathsf t_3\left(\su\right) = &\frac{1}{6}\textrm{tr}\left[\mathsf{M}\left(\su\right)\right]\textrm{tr}\left[\mathsf{M}\left(\su+1\right)\right]\textrm{tr}\left[\mathsf{M}\left(\su+2\right)\right] +\frac{1}{3} \textrm{tr}\left[\mathsf{M}\left(\su\right)\mathsf{M}\left(\su+1\right)\mathsf{M}\left(\su+2\right)\right] \\
- &\frac{1}{3}\textrm{tr}\left[\mathsf{M}\left(\su\right)\right]\textrm{tr}\left[\mathsf{M}\left(\su+1\right)\mathsf{M}\left(\su+2\right)\right] - \frac{1}{6}\textrm{tr}\left[\mathsf{M}\left(\su\right)\mathsf{M}\left(\su+1\right)\right]\textrm{tr}\left[\mathsf{M}\left(\su+2\right)\right] \;. \nonumber
\end{align}
Thanks to these relations we are now in the position to derive the expressions for the coefficients of Baxter $T-Q$ equation (\ref{eq:BAXTER}). To this end, we introduce the following objects
\begin{equation}
\mathscr E^{(j)}_{l k} = \sum_{1\leq a_1 <\cdots < a_j\leq J} \sum_{i_1,\ldots ,i_{j-1} =1}^4 \overset{a_1}{E}_{i_1 k} \overset{a_2}{E}_{i_2 i_1} \cdots \overset{a_j}{E}_{l i_{j-1}}\;,
\end{equation}
and
\begin{equation}
\mathfrak q^{[j_1,\ldots ,j_m]} = \textrm{tr}\left[\mathscr E^{(j_1)}\cdots\mathscr E^{(j_m)}\right]\;,\qquad \mathfrak q_k\equiv \mathfrak q^{[k]}\;.
\label{eq:charges}
\end{equation}
One easily see that
\begin{equation}
\mathsf{M}\left(\su\right) =\frac{1}{\delta\left(\su\right)^J}\left( \su^J\mathbbm 1 +\sum_{k=1}^J \mathscr E^{(k)} \su^{J-k}\right)\;,
\end{equation}
and, since\footnote{This is because $\mathfrak q_1 = \sum_{a=1}^J \sum_{i=1}^4 \overset{a}{E}_{ii}$ and the generators of $\mathfrak{sl}(4)$ satisfy the property $\sum_{i=1}^4 E_{ii}=0$.} $\mathfrak q_1 = 0$,
\begin{equation}
\delta\left(\su\right)^J\mathsf t_1\left(\su\right) =4\su^J + \sum_{k=2}^J \mathfrak q_k \su^{J-k}\;.
\label{eq:t_1}
\end{equation}
With some basic algebra we obtain the following expressions
\begin{align}
\delta\left(\su\right)^J\delta\left(\su+1\right)^J\mathsf t_2\left(\su\right) = &6\su^J\left(\su+1\right)^J + \frac{3}{2}\sum_{j=2}^J\mathfrak{q}_j\left[\su^J\left(\su+1\right)^{J-j}+\left(\su+1\right)^J\su^{J-j}\right] +\nonumber \\
+&\frac{1}{2}\sum_{j,k=1}^J \left[\mathfrak q_j\mathfrak q_k-\mathfrak q^{[j,k]}\right]\su^{J-j}\left(\su+1\right)^{J-k}\;,
\label{eq:t_2}
\end{align}
\begin{align}
\delta\left(\su\right)^J & \delta\left(\su+1\right)^J\delta\left(\su+2\right)^J\mathsf t_3\left(\su\right) = 4\su^J\left(\su+1\right)^J \left(\su+2\right)^J +\nonumber \\
+&\sum_{j=2}^J\mathfrak q_j\left[\su^J\left(\su+1\right)^J\left(\su+2\right)^{J-j}+\su^J\left(\su+1\right)^{J-j}\left(\su+2\right)^{J}+\su^{J-j}\left(\su+1\right)^J\left(\su+2\right)^{J}\right] + \nonumber \\
+&\frac{1}{6}\sum_{j,k=2}^J\mathfrak q_j\mathfrak q_k \left[3\su^{J}\left(\su+1\right)^{J-j}\left(\su+2\right)^{J-k}+2\su^{J-j}\left(\su+1\right)^{J-k}\left(\su+2\right)^{J}+\su^{J-j}\left(\su+1\right)^{J}\left(\su+2\right)^{J-k}\right]+\nonumber \\
+&\frac{1}{3}\sum_{j,k=1}^J\mathfrak q^{[j,k]}\left[\su^{J}\left(\su+1\right)^{J-j}\left(\su+2\right)^{J-k}-\su^{J-j}\left(\su+1\right)^{J-k}\left(\su+2\right)^{J}-3\su^{J-j}\left(\su+1\right)^{J}\left(\su+2\right)^{J-k}\right] \nonumber \\
+&\frac{1}{6}\sum_{j,k,l=1}^J\left[\mathfrak q_j\mathfrak q_k\mathfrak q_l-\mathfrak q^{[j,k]}\mathfrak q_l-2\mathfrak q^{[k,l]}\mathfrak q_j+2\mathfrak q^{[j,k,l]}\right] \su^{J-j}\left(\su+1\right)^{J-k}\left(\su+2\right)^{J-l}\;.
\label{eq:t_3}
\end{align}

The quantities $\mathfrak q^{[i_1,\ldots ,i_m]}$ (\ref{eq:charges}) appearing in the expressions of the $\mathsf{t}_k$ are local conserved charges. It can be shown with straightforward computation that they all commute amongst themselves
\begin{equation}
\left[\mathfrak q^{[i_1,\ldots ,i_m]},\mathfrak q^{[j_1,\ldots j_n]}\right]=0 \;, \qquad \forall i_1,\ldots ,i_m,j_1,\ldots j_n = 1,\ldots J\;.
\end{equation}
Their eigenvalue is, in general, dependent on the particular state of the spin chain Hilbert space. However, the following charge
\begin{equation}
\mathbbm C^{(J)}_m = \frac{1}{m}\mathfrak q^{\overbrace{[1,\ldots ,1]}^m}\;,
\end{equation}
is global, in the sense that it only depends on the labels of the representation $V_{\boldsymbol{\theta}}$ of the full spin chain\footnote{The Hilbert space $$\mathsf{H}_{\mathfrak{sl}(4)} = \bigotimes_{j=1}^J V_{\boldsymbol{\nu}}\left(\boldsymbol{z}\right)\;,$$ of the spin chain decomposes as a direct integral of representations $V_{\boldsymbol{\theta}}$ \cite{Williams1973}: $$\mathsf{H}_{\mathfrak{sl}(4)}  = \int^{\oplus} V_{\boldsymbol{\theta}} d\boldsymbol{\theta}\;.$$ The eigenvalues of the charges $\mathbbm C^{(J)}_m$ only depend on which of these summands the state of the spin chain belongs to, not on the specific state considered.}: it is thus a Casimir operator and we call it ``total Casimir". On the other hand, not all of the charges $\mathfrak q^{[i_1,\ldots ,i_m]}$ are independent. There exist relations amongst them which also involve the usual Casimir operator
\begin{equation}
\mathbbm C_m = \frac{1}{m} \sum_{i_1,\ldots i_{m}=1}^4 \overset{a}{E}_{i_2, i_1} \overset{a}{E}_{i_3, i_2}\cdots \overset{a}{E}_{i_1, i_m}\;,\qquad \forall a=1,\ldots J\;,
\end{equation}
whose eigenvalue only depends on the labels of the representation $V_{\boldsymbol{\nu}}$ of the single spins in the chain. One can obtain the explicit expression for the eigenvalues of the Casimirs by acting with them on the vacuum; the results for $m=2,3,4$ are
\begin{align}\nonumber
\mathbbm C_2 &= \frac{1}{2} \sum_{j=1}^{4} \nu_j^2-7\;,\qquad \mathbbm C_3 = \frac{1}{3}\sum_{j=1}^{4} \nu_j^3-\frac{5}{3}\mathbbm C_2 -12 \;,\qquad \mathbbm C_4 = \frac{1}{4}\sum_{j=1}^{4} \nu_j^4 - \frac{3}{2}\mathbbm C_3-\frac{9}{2}\mathbbm C_2-\frac{49}{2}\;,\nonumber \\  
\label{eq:Casimirs}
\mathbbm C^{(L)}_2 &= \frac{1}{2} \sum_{j=1}^{4} \theta_j^2-7\;,\qquad \mathbbm C^{(L)}_3 = \frac{1}{3}\sum_{j=1}^{4} \theta_j^3-\frac{5}{3}\mathbbm C^{(L)}_2 -12 \;,\qquad \mathbbm C^{(L)}_4 = \frac{1}{4}\sum_{j=1}^{4} \theta_j^4 - \frac{3}{2}\mathbbm C^{(L)}_3-\frac{9}{2}\mathbbm C^{(L)}_2-\frac{49}{2}  \;,
\end{align}
The relations mentioned above are deformations of the classical Newton's identities amongst symmetric polynomials and power sums; the first few are as follows
\begin{align}
\mathbbm C^{(J)}_2 - J \mathbbm C_2 &= \mathfrak q_2\;, \nonumber \\
\mathbbm C^{(J)}_3 - J\mathbbm C_3 &-\frac{4}{3}\left(\mathbbm C^{(J)}_2 - J \mathbbm C_2\right) = \mathfrak q^{[1,2]} - \mathfrak q_3\;, \label{eq:def_newt_id}\\
\mathbbm C^{(J)}_4 - J \mathbbm C_4 &-3\left(\mathbbm C^{(J)}_3 - J\mathbbm C_3\right) +2 \mathfrak q^{[1,2]} = \mathfrak q^{[1,1,2]}-\frac{1}{2}\mathfrak q^{[2,2]}-\mathfrak q^{[1,3]}+\mathfrak q_4\;. \nonumber
\end{align}

The following functions
\begin{align}
\tau_1\left(\su\right)=&\,\delta\left(\su\right)^J\mathsf{t}_1\left(\su\right) \;,\qquad \tau_2\left(\su\right)=\delta\left(\su-\frac{1}{2}\right)^J\delta\left(\su+\frac{1}{2}\right)^J\mathsf t_2\left(\su-\frac{1}{2}\right)\;, \\
\qquad \tau_3\left(\su\right)=&\,\delta\left(\su-1\right)^J \delta\left(\su\right)^J\delta\left(\su+1\right)^J\mathsf t_3\left(\su-1\right)\;, \nonumber
\end{align}
possess an interesting symmetry property. In fact let us explicitly denote their dependence on the local conserved charges:
\begin{equation}
\tau_1\left(\su\right)=\tau_1\left(\su\left\vert\mathfrak q^{[\boldsymbol{\alpha}_m]}\right.\right)\;,\qquad \tau_2\left(\su\right)=\tau_2\left(\su\left\vert\mathfrak q^{[\boldsymbol{\alpha}_m]}\right.\right)\;,\qquad \tau_3\left(\su\right)=\tau_3\left(\su\left\vert\mathfrak q^{[\boldsymbol{\alpha}_m]}\right.\right)\;,
\end{equation}
where $\boldsymbol{\alpha}_m=\left(\alpha_1,\alpha_2,\cdots ,\alpha_m\right)$ is a multi-index and we will denote $\vert\boldsymbol{\alpha}_m\vert = \sum_{k=1}^m \alpha_k$. Then it is easily checked that
\begin{align}
\tau_1\left(-\su\left\vert(-1)^{\vert\boldsymbol{\alpha}_m\vert}\mathfrak q^{[\boldsymbol{\alpha}_m]}\right.\right) =&\, (-1)^J\tau_1\left(\su\left\vert\mathfrak q^{[\boldsymbol{\alpha}_m]}\right.\right)\;,\quad \tau_2\left(-\su\left\vert(-1)^{\vert\boldsymbol{\alpha}_m\vert}\mathfrak q^{[\boldsymbol{\alpha}_m]}\right.\right) = \tau_2\left(\su\left\vert\mathfrak q^{[\boldsymbol{\alpha}_m]}\right.\right)\;,\label{eq:symmetry_of_taus}\\
\tau_3\left(-\su\left\vert(-1)^{\vert\boldsymbol{\alpha}_m\vert}\mathfrak q^{[\boldsymbol{\alpha}_m]}\right.\right) =&\, (-1)^J\tau_3\left(\su\left\vert\mathfrak q^{[\boldsymbol{\alpha}_m]}\right.\right)\;.\nonumber
\end{align}
With the help of relations (\ref{eq:def_newt_id}) we find that the first few terms in the $\su\rightarrow \infty$ expansion of the functions $\tau_k$ are
\begin{align}
\tau_1\left(\su\right)=&\,4 \su^J+\left(\mathbbm C^{(J)}_2-J\mathbbm C_2\right) \su^{J-2} + \mathfrak q_3 \su^{J-3} + \mathfrak q_4 \su^{J-4} \cdots \;, \nonumber \\
\tau_2\left(\su\right)=&\,6 \su^{2J} + \left(2\mathbbm C^{(J)}_2 - 3J\mathbbm C_2 -\frac{3}{2}J\right)\su^{2J-2}+\left(2\mathfrak q_3 - \left(\mathbbm C^{(J)}_3-J\mathbbm C_3\right)+\frac{4}{3}\left(\mathbbm C^{(J)}_2-J\mathbbm C_2\right)\right) \su^{2J-3} + \nonumber \\
+ & \upsilon_2^{(4)}\su^{2J-4} +\cdots \;, \label{eq:taus} \\
\tau_3\left(\su\right)=&\,4 \su^{3J} + \left(\mathbbm C^{(J)}_2-3J\mathbbm C_2-4J\right)\su^{3J-2} +\left(\mathfrak q_3 - \left(\mathbbm C^{(J)}_3-2J\mathbbm C_3\right)+\frac{4}{3}\left(\mathbbm C^{(J)}_2-2J\mathbbm C_2\right)\right)\su^{3J-3} +\nonumber \\
+ & \upsilon_3^{(4)} \su^{3J-4} + \cdots \nonumber \;,
\end{align}
with
\begin{align}
\upsilon_2^{(4)} = &2 \mathfrak q_4 -\mathfrak q^{[1,1,2]} + 2 \mathfrak q_3 +\mathbbm C^{(J)}_4 - J\mathbbm C_4 - \left(\mathbbm C^{(J)}_3 - J\mathbbm C_3\right) +\frac{1}{2}\left(\mathbbm C^{(J)}_2 - J\mathbbm C_2\right)^2- \nonumber \\
- &\frac{3J+4}{6}\left(\mathbbm C^{(J)}_2 - J\mathbbm C_2\right)+J\frac{J-1}{4}\mathbbm C_2 +3J \frac{J-1}{16}\;,
\end{align}
and
\begin{align}
\upsilon_3^{(4)} = &\upsilon_2^{(4)} - \mathfrak q_4 +\mathbbm C^{(J)}_4 - J \mathbbm C_4-3\left(\mathbbm C^{(J)}_3 - J\mathbbm C_3\right) -\frac{1}{2}\left(\mathbbm C^{(J)}_2-J \mathbbm C_2\right)^2-J\mathbbm C_2\left(\mathbbm C^{(J)}_2-J \mathbbm C_2\right)-\nonumber\\
-&\frac{J-4}{2}\mathbbm C^{(J)}_2+3J\frac{3J-5}{4}\mathbbm C_2+29J\frac{J-1}{16}\;.
\end{align}

\subsection*{The gauged form of the Baxter equation}\label{subsubsec:Baxter_for_BMN}

There is one final manipulation we wish to apply to the Baxter equation (\ref{eq:BAXTER}) before specifying the representation $\pi_{\boldsymbol{\nu}}$. First of all, let us write it as follows
\begin{align}
\delta\left(\su-\frac{1}{2}\right)^J&\delta\left(\su+\frac{1}{2}\right)^J\delta\left(\su+\frac{3}{2}\right)^J\mathcal Q_i\left(\su-\frac{3}{2}\right)  +  \delta\left(\su+\frac{3}{2}\right)^J\tau_2\left(\su\right)\mathcal Q_i\left(\su+\frac{1}{2}\right) +\mathcal Q_i\left(\su+\frac{5}{2}\right)=\nonumber \\
=&\delta\left(\su+\frac{1}{2}\right)^J\delta\left(\su+\frac{3}{2}\right)^J\tau_1\left(\su-\frac{1}{2}\right)\mathcal Q_i\left(\su-\frac{1}{2}\right) + \tau_3\left(\su+\frac{1}{2}\right)\mathcal Q_i\left(\su+\frac{3}{2}\right) \;.
\label{eq:BAXTER2}
\end{align}
Considering the asymptotic limit $\su\rightarrow\infty$ of its coefficients, we see from (\ref{eq:taus}) that they all behave as powers, although of different order:
\begin{equation}
\su^{12J} \mathcal Q_i\left(\su-\frac{3}{2}\right)+6 \su^{6J} \mathcal Q_i\left(\su+\frac{1}{2}\right) + \mathcal Q_i\left(\su+\frac{5}{2}\right) - 4 \su^{9 J} \mathcal Q_i\left(\su-\frac{1}{2}\right) - 4 \su^{3J} \mathcal Q_i\left(\su+\frac{3}{2}\right) \sim 0\;.
\end{equation}
This is in contradiction with the expected asymptotic behaviour of the functions $q$ (\ref{eq:q_asympt}), which lead us to search for a function $\mathsf{f}$ such that $q\left(\su\right) = \mathsf{f}\left(\su\right)\mathcal Q\left(\su\right)$ and the coefficients of the equation for $q$ all have the same leading power of $\su$ for $\su\rightarrow\infty$. The relation is easily found to be
\begin{equation}
\mathcal{Q}_{i}\left(\su\right)=q_i\left(\su-\frac{1}{2}\right)\prod_{j=1}^{3}\Gamma\left(\su-\nu_{j}+1\right)^J\;,
\end{equation}
and the gauged functions $q_i\left(\su\right)$ satisfy the following Baxter equation
%
%
\begin{equation}
A_{\boldsymbol{\nu}}\left(\su+1\right)q\left(\su+2\right) - B_{\boldsymbol{\nu}}\left(\su+\frac{1}{2}\right)q\left(\su+1\right)+C_{\boldsymbol{\nu}}\left(\su\right)q\left(\su\right)-D_{\boldsymbol{\nu}}\left(\su-\frac{1}{2}\right)q\left(\su-1\right)+E_{\boldsymbol{\nu}}\left(\su-1\right)q\left(\su-2\right)=0\;,
\label{eq:gauged_baxter}
\end{equation}
where the coefficients are
\begin{align}
&A_{\boldsymbol{\nu}}\left(\su\right) = \prod_{j=1}^{3}\left(\su-\nu_{j}+\frac {3}{2}\right)^{{\L}}\;,\qquad B_{\boldsymbol{\nu}}\left(\su\right) = \tau_3\left(\su\right)\;,\qquad C_{\boldsymbol{\nu}}\left(\su\right) = \left(\su-\nu_{4}+\frac{3}{2}\right)^{{\L}}\tau_{2}\left(\su\right)\,,\nonumber \\
&D_{\boldsymbol{\nu}}\left(\su\right) = \prod_{j=1}^{2}\left(\su-\nu_{4}+j\right)^{{\L}}\tau_{1}\left(\su\right)\;,\qquad E_{\boldsymbol{\nu}}\left(\su\right) = \prod_{j=1}^{3}\left(\su-\nu_{4}+j-\frac{1}{2}\right)^{{\L}}\,.
\end{align}

\subsection*{The free scalar field representation}\label{subsubsec:free_scalar_rep}

In order to apply the Baxter equation (\ref{eq:gauged_baxter}) to the conformal spin chain that we encountered in Sect.~\ref{sec:conf_spin_chain_picture}, we have to replace
the parameters $\boldsymbol{\nu}$ with their values for a free scalar field with conformal charges $(1,0,0)$
\footnote{For the scalar operators (\ref{O-def}) carrying the charges $(\Delta,0,0)$, the corresponding parameters $\boldsymbol{\nu}$ are given by
$
\boldsymbol{\nu}_{(\Delta,0,0)}=\left(3-\frac{\Delta}{2},2-\frac{\Delta}{2},1+\frac{\Delta}{2},\frac{\Delta}{2}\right)\,.
$}
\begin{align}\label{rhoBMN}
\boldsymbol{\nu}_{(1,0,0)}=\left(\frac{5}{2},\frac{3}{2},\frac{3}{2},\frac{1}{2}\right)\,.
\end{align}
For this particular choice, one readily checks that
\begin{equation}
A_{\boldsymbol{\nu}_{(1,0,0)}} \left(\su\right) =  \left(\su-1\right)^{{\L}} \su^{2{\L}}\;, \qquad E_{\boldsymbol{\nu}_{(1,0,0)}} \left(\su\right) =\su^{{\L}} \left(\su+1\right)^{{\L}} \left(\su+2\right)^{{\L}}\;.
\end{equation}
Substituting (\ref{rhoBMN}) into (\ref{eq:gauged_baxter}) and dividing both sides of the equation by $\su^{\L}\left(\su+1\right)^{\L}$
we arrive at
\begin{align}
&\left(\su+1\right)^{{\L}}q_{i}\left(\su+2\right)+\left(\su-1\right)^{\L} q_{i}\left(\su-2\right)  +\frac{\tau_2\left(\su\right)}{\su^{\L}} q_{i}\left(\su\right) = \nonumber \\
& = \frac{\tau_3\left(\su+\frac{1}{2}\right)}{\su^{\L}\left(\su+1\right)^{\L}}q_{i}\left(\su+1\right)+ \tau_1\left(\su-\frac{1}{2}\right)q_{i}\left(\su-1\right)\,,
\label{eq:gauged_baxter_BMN_symmetric}
\end{align}
%
where, using the results (\ref{eq:taus}), we have
\begin{align}\notag
\tau_1\left(\su\right){}&  =4\su^{{\L}}+\frac{\alpha+3{\L}-4}{2}\su^{{\L}-2}+\mathfrak q_3\su^{{\L}-3}+\mathfrak q_4 \su^{{\L}-4}+\sum_{k=5}^{{\L}}\mathfrak q_k \su^{{\L}-k}\;,
\\\notag
 \tau_2\left(\su\right){}&=6\su^{2{\L}}+\left(3{\L}+\alpha-4\right)\su^{2{\L}-2}+2\mathfrak q_3\su^{2{\L}-3}+\upsilon_2^{(4)}\su^{2{\L}-4}+\sum_{k=5}^{2{\L}}\upsilon_2^{(k)}\su^{2{\L}-k}\;,
\\\label{eq:tau3_allL}
 \tau_3\left(\su\right) {}&=4\su^{3{\L}}+\frac{\alpha+{\L}-4}{2}\su^{3{\L}-2}+\mathfrak q_3 \su^{3{\L}-3}+
 \\ \nonumber
{}& \qquad+\left(\upsilon_2^{(4)}-\mathfrak q_4-\frac{(\alpha-4)^2+4{\L}(\alpha-5)+7{\L}^2}{16}\right)\su^{3{\L}-4}+\sum_{k=5}^{3{\L}}\upsilon_3^{(k)}\su^{3{\L}-k}\;.
\end{align}
Here $\alpha = \left(\Delta-2\right)^2$ and $\mathfrak q_k$ are integrals of motion defined in (\ref{eq:charges}). The coefficients $\upsilon_2^{(k)}$ and $\upsilon_3^{(k)}$ are some complicated combinations the of local charges $\mathfrak q^{[\boldsymbol{\alpha}]}$. Their explicit expressions can be obtained from (\ref{eq:t_1}), (\ref{eq:t_2}) and (\ref{eq:t_3}).


Up to now, we have used solely the representation theory which allowed us to fix $9$ coefficients amongst the $6{\L}$ ones. We need further constraints. As it has been shown in Sect.~\ref{sec:BaxterfromQSC}, considering the equation (\ref{eq:gauged_baxter_BMN_symmetric}) as a double scaling limit of the Baxter equation in $\gamma$-twisted $\mathcal N=4$ SYM   (see Eq.~(\ref{Baxter})) means we must be able to bring it to the  symmetric form \eq{eq:symmetricform}.
Imposing these symmetries on the functions (\ref{eq:tau3_allL}), will allow us to fix some of the unknown coefficients
of the functions $\tau_{k}$. First of all, we see that, in order to have the coefficients in front of $q(\su+1)$ and $q(\su-1)$ in
(\ref{eq:gauged_baxter_BMN_symmetric})
 to be, respectively, $B(\su+\frac12)$ and $B(\su-\frac12)$ for some function $B(u)$, the transfer matrices have to satisfy
 the following relation
\begin{equation}
\tau_1\left(\su\right) = \frac{\tau_3\left(\su\right)}{\left(\su^2-\frac{1}{4}\right)^{\L}}\;.
\label{eq:symmetry_1_3}
\end{equation}
Since $\tau_1(\su)$ is, by definition, a polynomial of order ${\L}$, this relation implies that $\tau_3(\su)$ should have
${\L}$-th order zero at $\su=\pm\frac{1}{2}$. Thus we can recast part of the relation (\ref{eq:symmetry_1_3}) into the following $2{\L}$ requirements
\begin{equation}
\frac{\partial^k}{\partial \su^k} \tau_3(\su)\Bigg\vert_{\su=\pm\frac{1}{2}} = 0\;,\qquad k=0,1,\cdots,{\L}-1\;.
\label{eq:polynomiality_1}
\end{equation}
In addition to this, we have to actually impose the term by term equality of $ {\tau_3(\su)}/{\left(\su^2-\frac{1}{4}\right)^{\L}}$ with $\tau_1(\su)$. Summing it all up, we will be left with ${\L}+[{\L}/2]-3$ unfixed coefficients\footnote{The reason why they are ${\L}+[{\L}/2]-3$ and not ${\L}+[{\L}/2]-4$, as one would expect by simple counting, is that in $\tau_3$, the coefficient $\upsilon_3^{(4)}$ of $\su^{3{\L}-4}$ is not independent from $\upsilon_1^{(4)}$ and $\upsilon_2^{(4)}$.}, before we impose further restrictions on the set of solutions.

Note that the above argument based on the requested symmetry of the Baxter equation agrees with the following observation. Remember the symmetry (\ref{eq:symmetry_of_taus}) of the functions $\tau_k(u)$:
\begin{equation}
\tau_k\left(-\su\left\vert(-1)^{\vert\boldsymbol{\alpha}_m\vert}\mathfrak q^{[\boldsymbol{\alpha}_m]}\right.\right) =\, (-1)^{kJ}\tau_k\left(\su\left\vert\mathfrak q^{[\boldsymbol{\alpha}_m]}\right.\right)\;.
\end{equation}
Then, substituting (\ref{eq:symmetry_1_3}) into the Baxter equation (\ref{eq:gauged_baxter_BMN_symmetric}), we find that
this equation remains invariant under $u\to -u$ and $ q^{[\boldsymbol{\alpha}]}\to(-1)^{\vert\boldsymbol{\alpha}\vert}\mathfrak q^{[\boldsymbol{\alpha}]}$. This means that for any
solution $q(u,\Delta,q^{[\boldsymbol{\alpha}]})$ to  (\ref{eq:gauged_baxter_BMN_symmetric}) there should exist another solution
$q(-u,\Delta, (-1)^{\vert\boldsymbol{\alpha}\vert}\mathfrak q^{[\boldsymbol{\alpha}]})$ describing the state with the same scaling dimension $\Delta$. Thus, the states with
nonzero $q^{[\boldsymbol{\alpha}]}\;\textrm{s.t.}\;\vert\boldsymbol{\alpha}\vert\in2\mathbbm Z+1$ have a two-fold degeneracy with respect to $\Delta$.
To avoid the degeneracy, we have to require  that all charges with odd indices should vanish  \cite{Korchemsky:1995be}
\begin{align}\label{zero-q}
q^{[\boldsymbol{\alpha}]}=0\;,\quad\forall \boldsymbol{\alpha}\;\textrm{s.t.}\;\vert\boldsymbol{\alpha}\vert\in2\mathbbm Z+1\;.
\end{align}
From this requirement it directly follows that the transfer matrices have then a definite parity in this case, $\tau_1\left(\su\right)=(-1)^{\L} \tau_1\left(-\su\right)$, $\tau_2\left(\su\right)=\tau_2\left(-\su\right)$ and $\tau_3\left(\su\right)=(-1)^{\L}\tau_3\left(-\su\right)$, and are given by
\begin{align} \notag
\tau_1(\su) {}&=4\su^{{\L}}+\frac{3{\L}+\alpha-4}{2}\su^{{\L}-2}+\sum_{k=2}^{[{\L}/2]}\mathfrak q_{2k}\su^{{\L}-2k}\,,
\\
\tau_2(\su) {}&=6\su^{2{\L}}+\left(3{\L}+\alpha-4\right)\su^{2{\L}-2}+\sum_{k=2}^{\L}\upsilon_2^{(2k)}\su^{2({\L}-k)}\,.
\\
\tau_3\left(\su\right) {}&=4\su^{3{\L}}+\frac{\alpha+{\L}-4}{2}\su^{3{\L}-2}+\left(\upsilon_2^{(4)}-\mathfrak q_4-\frac{(\alpha-4)^2+4{\L}(\alpha-5)+7{\L}^2}{16}\right)\su^{3{\L}-4}+ \nonumber
 \\ \nonumber
{}& \qquad+\sum_{k=3}^{[3{\L}/2]}\upsilon_3^{(2k)}\su^{3{\L}-2k}\;.
\label{eq:tau_2_symm}
\end{align}

Finally, asking the symmetry (\ref{eq:symmetry_1_3}), the Baxter equation takes the form \eqref{eq:symmetricform} and \eqref{BC} advocated in section~\ref{sec:BaxterBootstrap}.
For ${\L}=2$, ${\L}=3$ and ${\L}=4$ the expressions \eq{eq:tau_2_symm} contain, respectively, $0$, $1$ and $3$ unfixed coefficients.

 As it has been showed in Sect.~\ref{sec:quantization},
the rest of coefficients have to satisfy the additional quantisation conditions. The relation  (\ref{zero-q}) imposes a tight selection rule on the states of the spin chain.
It is these states that play a distinguished role in our analysis as they allow us to find the scaling dimensions of the operators (\ref{O-def}) for an
arbitrary coupling.

Let us examine relations (\ref{eq:tau3_allL}) and (\ref{eq:tau_2_symm}) for few lowest values of the length ${\L}$ of the spin chain.

\subsubsection*{Baxter equation for ${\L}=2$}

In this case, the general expressions for the transfer matrices (\ref{eq:tau3_allL}) are
\begin{align}\notag
{}& \tau_1\left(\su\right)=4\su^{2}+\frac{\alpha+2}{2}\;,\qquad \tau_2\left(\su\right)=6\su^{4}+\left(\alpha+2\right)\su^{2}+\upsilon_2^{(4)}\;,
\\
{}& \tau_3\left(\su\right)=4\su^{6}+\frac{\alpha-2}{2}\su^{4}+\left(\upsilon_2^{(4)}-\frac{\alpha^2+4}{16}\right)\su^{2}+\upsilon_3^{(5)}\su+\upsilon_3^{(6)}\;.
\end{align}
Imposing the conditions (\ref{eq:polynomiality_1})  we obtain  that all coefficients are fixed
\begin{equation}
\upsilon_2^{(4)} = \alpha \frac{\alpha-4}{16}\;,\qquad \upsilon_3^{(5)} = 0\;,\qquad \upsilon_3^{(6)} = \frac{\alpha+2}{32}\;,
\end{equation}
where $\alpha = \left(\Delta-2\right)^2$. The resulting Baxter equation (\ref{eq:gauged_baxter_BMN_symmetric}) takes then the expected form \eq{L2BaxterLax}. 

\subsubsection*{Baxter equation for ${\L}=3$}

Imposing the symmetry (\ref{zero-q}) we find for the transfer matrices (\ref{eq:tau3_allL})
\begin{align}
{}& \tau_1\left(\su\right)=4\su^{3}+\frac{\alpha+5}{2}\su\;,\qquad \tau_2\left(\su\right)=6\su^{6}+\left(\alpha+5\right)\su^{4}+ \upsilon_2^{(4)} \su^2+\upsilon_2^{(6)}\;,
\\ \notag
{}& \tau_3\left(u\right)=4\su^{9}+\frac{\alpha-1}{2}\su^{7}+\left(\upsilon_2^{(4)}-\frac{\alpha\left(\alpha+4\right)+19}{16}\right)\su^{5}
 +\sum_{k=5}^{9}\upsilon_3^{(k)}\su^{9-k}
\;.
\end{align}
Requiring the transfer matrices to satisfy (\ref{eq:symmetry_1_3}) we can fix $6$ coefficients
\begin{align}
&\upsilon_3^{(5)} =  \upsilon_3^{(7)} =\upsilon_3^{(9)} =  0\;,\qquad  \upsilon_2^{(4)} = \frac{(\alpha-1)^2}{16}\;,\qquad \upsilon_3^{(6)} = \frac{3\alpha+13}{32}\;, \qquad \upsilon_3^{(8)} = -\frac{\alpha+5}{128} \;.
\end{align}
The resulting expressions for $\tau_k$ are
\begin{align}\notag
{}& \tau_1\left(\su\right)=\su\left(4\su^{2}+\frac{\alpha+5}{2}\right)\;,\qquad
\\\notag
{}&
\tau_2\left(\su\right)=6\su^{6}+\left(\alpha+5\right)\su^{4} +\frac{(\alpha-1)^2}{16} \su^2-m^2\;,
\\
{}& \tau_3\left(\su\right)=\left(\su^2-\frac{1}{4}\right)^3\su\left(4\su^{2}+\frac{\alpha+5}{2}\right)\;,
\end{align}
where we defined $m^2=-\upsilon_2^{(6)}$. Substituting these relations into (\ref{eq:gauged_baxter_BMN_symmetric}) we arrive at \eq{L3BaxterLax}.

\subsubsection*{Baxter equation for ${\L}=4$}

As in the previous case, we start from the following $\tau_k$ functions
\begin{align}\notag
{}& \tau_1\left(\su\right)=4\su^{4}+\frac{\alpha+8}{2}\su^2+\mathfrak q_4\;,\qquad
\tau_2\left(\su\right)=6\su^{8}+\left(\alpha+8\right)\su^{6}+ \upsilon_2^{(4)} \su^4+\upsilon_2^{(6)}\su^2+\upsilon_2^{(8)}\;,
\\
{}& \tau_3\left(\su\right)=4\su^{12}+\frac{\alpha}{2}\su^{10}+\left(\upsilon_2^{(4)}-\mathfrak q_4-\frac{\alpha\left(\alpha+8\right)+48}{16}\right)\su^{9}+\sum_{k=5}^{12}\upsilon_3^{(k)}\su^{12-k}\;,
\end{align}
The constraints (\ref{eq:polynomiality_1}) fix $8$ coefficients as
\begin{align}
\upsilon_3^{(5)}&= \upsilon_3^{(7)}= \upsilon_3^{(9)}= \upsilon_3^{(11)} =  0\;,\qquad \upsilon_3^{(6)} = \mathfrak q_4-\upsilon_2^{(4)}+\frac{\alpha(\alpha+3)+28}{16}\;, \nonumber\\
\qquad \upsilon_3^{(8)} &= \frac{3}{8}\left(\upsilon_2^{(4)}-\mathfrak q_4\right)-\frac{\alpha(3\alpha+4)+54}{128}\;, \qquad  \upsilon_3^{(10)}=\frac{1}{16}\left(\mathfrak q_4 - \upsilon_2^{(4)}\right)+\frac{\alpha(2\alpha+1)+24}{512}\;,\nonumber \\
 \upsilon_3^{(12)} &= \frac{1}{256}\left(\upsilon_2^{(4)}-\mathfrak q_4\right)-\frac{\alpha^2+8}{4096}\;,
\end{align}
and the transfer matrcies $\tau_k$ simplify as follows
\begin{align}\notag
{}& \tau_1\left(\su\right)=4\su^{4}+\frac{\alpha+8}{2}\su^2+\mathfrak q_4\;,\qquad
\\\notag
{}&
\tau_2\left(\su\right)=6\su^{8}+\left(\alpha+8\right)\su^{6}+ \upsilon_2^{(4)} \su^4+\upsilon_2^{(6)}\su^2+\upsilon_2^{(8)}\;,
\\
{}& \tau_3\left(\su\right)=\left(\su^2-\frac{1}{4}\right)\left[4\su^{4}+\frac{\alpha+8}{2}\su^{2}+\upsilon_2^{(4)}-\mathfrak q_4-\frac{\alpha^2+8}{16}\right]\;.
\end{align}
In order to satisfy (\ref{eq:symmetry_1_3}), we  have  to impose one more condition
\begin{equation}
\upsilon_2^{(4)} = 2\mathfrak q_4+\frac{\alpha^2+8}{16}\;.
\end{equation}
We are thus left with $3$ coefficients and  the Baxter equation (\ref{eq:gauged_baxter_BMN_symmetric})  matches  \eq{L4BaxterLax} upon
redefinition of the parameters
\begin{equation}
\mathfrak q_4 = b\;,\qquad \upsilon_2^{(6)}=-c_1\;,\qquad \upsilon_2^{(8)}=c_2\;.
\end{equation}

\section{QSC supplementary relations}
\subsection*{Coefficients of 4th order Baxter equation}
\label{sec:Coefficients}
The determinants $D_i$ used the equation \eqref{Baxter} are defined as follows:
\beq
D_0=\det\left(\begin{matrix}
\bP^{1[+2]} & \bP^{2[+2]} & \bP^{3[+2]} & \bP^{4[+2]} \\
\bP^{1} & \bP^{2} & \bP^{3} & \bP^{4} \\
\bP^{1[-2]} & \bP^{2[-2]} & \bP^{3[-2]} & \bP^{4[-2]}\\
\bP^{1[-4]} & \bP^{2[-4]} & \bP^{3[-4]} & \bP^{4[-4]}
\end{matrix}\right)
,\qquad
D_1=\det\left(\begin{matrix}
\bP^{1[+4]} & \bP^{2[+4]} & \bP^{3[+4]} & \bP^{4[+4]} \\
\bP^{1} & \bP^{2} & \bP^{3} & \bP^{4} \\
\bP^{1[-2]} & \bP^{2[-2]} & \bP^{3[-2]} & \bP^{4[-2]}\\
\bP^{1[-4]} & \bP^{2[-4]} & \bP^{3[-4]} & \bP^{4[-4]}
\end{matrix}\right),
\eeq

\beq
D_2=\det\left(\begin{matrix}
\bP^{1[+4]} & \bP^{2[+4]} & \bP^{3[+4]} & \bP^{4[+4]} \\
\bP^{1[+2]} & \bP^{2[+2]} & \bP^{3[+2]} & \bP^{4[+2]} \\
\bP^{1[-2]} & \bP^{2[-2]} & \bP^{3[-2]} & \bP^{4[-2]}\\
\bP^{1[-4]} & \bP^{2[-4]} & \bP^{3[-4]} & \bP^{4[-4]}
\end{matrix}\right)
,\qquad
\bar D_1=\det\left(\begin{matrix}
\bP^{1[-4]} & \bP^{2[-4]} & \bP^{3[-4]} & \bP^{4[-4]} \\
\bP^{1} & \bP^{2} & \bP^{3} & \bP^{4} \\
\bP^{1[+2]} & \bP^{2[+2]} & \bP^{3[+2]} & \bP^{4[+2]}\\
\bP^{1[+4]} & \bP^{2[+4]} & \bP^{3[+4]} & \bP^{4[+4]}
\end{matrix}\right)
\eeq

\subsection*{Formulas for $\Delta$ through ansatz coefficients}
\label{sec:Formulas}
This appendix contains expressions for $\Delta$ through the coefficients of the anatz \eqref{fg_ansatz}.

For $L=2$
\beqa\nonumber
&&(\Delta-2)^2=-\left[{\left(\kappa
   -\hat{\kappa }\right)^2 \hat{\kappa } \left(\hat{\kappa }+1\right) \left(\kappa
   \hat{\kappa }-1\right)^2}\right]^{-1}\left[
        -2 g^8 \left(\kappa -\hat{\kappa }\right)^2 \left(\hat{\kappa }-1\right)^2
   \left(\hat{\kappa }+1\right) \left(\kappa  \hat{\kappa }-1\right)^2 c_{4,1}^2
\right.+\\ \nonumber
&&\left.
   +2 g^6
   \left(\kappa -\hat{\kappa }\right)^2 \left(\hat{\kappa }-1\right)^2 \left(\hat{\kappa
   }+1\right) \left(\kappa  \hat{\kappa }-1\right)^2 c_{4,2}
\right.+\\ \nonumber
&&\left.
   -2 i g^4 \left(\kappa
   -\hat{\kappa }\right) \left(\hat{\kappa }-1\right) \hat{\kappa } \left(\kappa
   \hat{\kappa }-1\right) \left(\left(\hat{\kappa }^2+1\right) \kappa ^2-4 \hat{\kappa }
   \kappa +\hat{\kappa }^2+1\right) c_{4,1}
\right.+\\ \nonumber
&&\left.
   -2 i \left(\kappa ^2-1\right) \left(\kappa
   -\hat{\kappa }\right) \left(\hat{\kappa }-1\right)^2 \hat{\kappa } \left(\hat{\kappa
   }+1\right) \left(\kappa  \hat{\kappa }-1\right) c_{3,-1}
\right.+\\ \nonumber
&&\left.
   +2 \left(\kappa -\hat{\kappa
   }\right)^2 \left(\hat{\kappa }-1\right)^2 \left(\hat{\kappa }+1\right) \left(\kappa
   \hat{\kappa }-1\right)^2 c_{2,0}
\right.+\\ \nonumber
&&\left.
   -2 \left(\hat{\kappa }+1\right) \left(\hat{\kappa }
   \left(-\left(\hat{\kappa }^3+\hat{\kappa }\right) \kappa ^4-\left(\hat{\kappa
   }-3\right) \left(3 \hat{\kappa }-1\right) \left(\hat{\kappa }^2+1\right) \kappa ^3
\right.\right.\right.+\\ \nonumber
&&\left.\left.\left.
   -2
   \left(\hat{\kappa } \left(\hat{\kappa } \left(\left(\hat{\kappa }-7\right) \hat{\kappa
   }+18\right)-7\right)+1\right) \kappa ^2
\right.\right.\right.+\\ \nonumber
&&\left.\left.\left.
   -\left(\hat{\kappa }-3\right) \left(3
   \hat{\kappa }-1\right) \left(\hat{\kappa }^2+1\right) \kappa -\hat{\kappa }
   \left(\hat{\kappa }^2+1\right)\right)-2 g^2 \left(\kappa -\hat{\kappa }\right)^2
   \left(\hat{\kappa }-1\right)^2 \left(\kappa  \hat{\kappa }-1\right)^2\right)
   \right]
   \eeqa

For $L=3$
 \beqa
 \nonumber
 &&(\Delta-2)^2=\left[\left(\kappa -\hat{\kappa }\right)^{2} \hat{\kappa } \left(\hat{\kappa }+1\right)
   \left(\kappa  \hat{\kappa }-1\right)^2 \right]^{-1}  \left[-2 g^{12} \left(\kappa -\hat{\kappa }\right)^2 \left(\hat{\kappa }-1\right)^2
   \left(\hat{\kappa }+1\right) \left(\kappa  \hat{\kappa }-1\right)^2 c_{4,1}^2
\right.+\\ \nonumber
&&\left.
   +2 g^8
   \left(\kappa -\hat{\kappa }\right)^2 \left(\hat{\kappa }-1\right)^2 \left(\hat{\kappa
   }+1\right) \left(\kappa  \hat{\kappa }-1\right)^2 c_{4,2}
\right.+\\ \nonumber
&&\left.
   -4 i g^6 \left(\kappa
   -\hat{\kappa }\right) \left(\hat{\kappa }-1\right) \hat{\kappa } \left(\kappa
   \hat{\kappa }-1\right) \left(\left(\hat{\kappa }^2+\hat{\kappa }+1\right) \kappa
   ^2-\left(\hat{\kappa } \left(\hat{\kappa }+4\right)+1\right) \kappa +\hat{\kappa
   }^2+\hat{\kappa }+1\right) c_{4,1}
\right.+\\ \nonumber
&&\left.
   -2 i \left(\kappa ^2-1\right) \left(\kappa
   -\hat{\kappa }\right) \left(\hat{\kappa }-1\right)^2 \hat{\kappa } \left(\hat{\kappa
   }+1\right) \left(\kappa  \hat{\kappa }-1\right) c_{3,-2}
\right.+\\ \nonumber
&&\left.
   +2 \left(\kappa -\hat{\kappa
   }\right)^2 \left(\hat{\kappa }-1\right)^2 \left(\hat{\kappa }+1\right) \left(\kappa
   \hat{\kappa }-1\right)^2 c_{2,-1}
\right.+\\ \nonumber
&&\left.
   +\left(\hat{\kappa }+1\right) \left(6 g^2
   \left(\kappa -\hat{\kappa }\right)^2 \left(\hat{\kappa }-1\right)^2 \left(\kappa
   \hat{\kappa }-1\right)^2+\hat{\kappa } \left(\hat{\kappa } \left(6 \hat{\kappa
   }^2+\hat{\kappa }+6\right) \kappa ^4+2 \left(6 \hat{\kappa }^4-19 \hat{\kappa }^3-19
   \hat{\kappa }+6\right) \kappa ^3
\right.\right.\right.+\\ \nonumber
&&\left.\left.\left.
   +\left(\hat{\kappa } \left(\hat{\kappa }
   \left(\left(\hat{\kappa }-36\right) \hat{\kappa }+148\right)-36\right)+1\right) \kappa
   ^2+2 \left(6 \hat{\kappa }^4-19 \hat{\kappa }^3-19 \hat{\kappa }+6\right) \kappa
   +\hat{\kappa } \left(6 \hat{\kappa }^2+\hat{\kappa }+6\right)\right)\right)\right]
 \eeqa

\subsection*{QSC equations on  $\Omega_i^j$}\label{app:Omega}
We follow the derivation in \cite{Gromov:2016rrp} replacing complex conjugation with the reflection $u\to -u$.
We assume that the state is parity invariant which implies that at the level of the $\bP_a$ functions we have
\begin{eqnarray}
\bP_a(-u)={\lambda_a}^b\bP_b(u)\;\;,\;\;
\bP^a(-u)={\lambda^b}_a\bP^b(u)
\end{eqnarray}
for some constant coefficients ${\lambda_a}^b$ and ${\lambda^b}_a$, which obey
${\lambda_a}^b{\lambda^a}_c=\delta_{c}^b$. Let us show that
\begin{eqnarray}
{\Omega_i}^j(u)=-Q_{a|i}(-u+i/2){\lambda^{a}}_b Q^{b|j}(u-i/2)\;.
\end{eqnarray}
Indeed
\begin{eqnarray}
{\Omega_i}^j(u)\bQ_j(u)=-Q_{a|i}(-u+i/2){\lambda^{a}}_b {\bP}^b(u)=-Q_{a|i}(-u+i/2){\bP}^a(-u)=\bQ_i(-u)\;.
\end{eqnarray}
using identities like $\bP_a\bP^a=0$ and $\bQ_i=-Q_{a|i}(u\pm i/2)\bP^a$
and $\bQ^i=+Q^{a|i}(u\pm i/2)\bP_a$ it is easy to check that ${\Omega_i}^j(u+i)={\Omega_i}^j(u)$.

Furthermore, we can easily find the discontinuity of ${\Omega_i}^j(u)$:
\begin{equation}
{{\tilde\Omega}_i}^{\;j}
(u)-{\Omega_i}^j(u)=-\tilde\bQ_i(-u)\tilde{\bQ}^j(u)+
\bQ_i(-u){\bQ}^j(u)\;.
\end{equation}

\section{Details of the derivation of the Baxter equation from QSC}
\la{sec:appQSC}
Here we give details of the derivation of the Baxter equation from QSC approach for $J=2,3,4$.
\subsection{Left-right Symmetry}
Most problems solved using QSC method before possessed so-called left-right symmetry, which in particular means that the indices can be raiser or lowered with a constant matrix:
\beq\label{LRsym}
\bQ^i=\chi^{ij}\bQ_j\,,\qquad  \bP^a=\chi^{ab}\bP_b\,,\qquad
\chi= {\small  \left( \begin{matrix}
                0&0&0&-1 \\
                0&0&1&0 \\
                0&-1&0&0 \\
                1&0&0&0
        \end{matrix}\right)}
\eeq
States of twisted \sym$\;$ are not left-right symmetric for general twists, however for the particular case of operator $\tr \phi_1^J$
and twists \eqref{twists} indices can still be raised by $\chi$ if one also exchanges $\kappa$ with $\hat \kappa$:
\beq
\bQ^i=\chi^{ij}\bQ_j\big|_{\kappa\leftrightarrow\hat\kappa}\,,\qquad \bP^a=\chi^{ab}\bP_b\big|_{\kappa\leftrightarrow\hat\kappa}\;.
\label{LRsymmetry}\eeq

\subsection{Baxter equation from double scaling limit of QSC}

In this section we will derive a finite-difference equation for $\bQ_i(u)$ in the double scaling limit. Our starting point will be the Baxter equation \eqref{Baxter}. The computation is similar to that in \cite{Gromov:2016rrp}:
we need to construct an ansatz for $\bP_a$, expand it in the double scaling limit and plug into \eqref{Baxter}. The ansatz will necessarily contain many unknown coefficients which we will fix by solving \eqref{Baxter} expanded at large $u$. In the process we will also get a relation between $\Delta$ and the coefficients of the ansatz for $\bP_a$.

Let us start with an ansatz for $\bP_a$: along the lines of \cite{Kazakov:2015efa}, we pull out of $\bP_a$ the exponential and power-like prefactors, leaving the part which scales like 1 at infinity:
\beqa\label{Pa1}\nn
\label{Pa2}\nn
&&\bp_a=\left\{A_1 f_1(u),A_2 f_1(-u),A_3 g_1(u),A_4 g_1(-u)\right\}\\[2mm]
&&\bp^a=\left\{ f_2(u), f_2(-u), g_2(u), g_2(-u)\right\}.
\eeqa
Here we choose $A^a=1$ and $A_a$ according to \eqref{AA}.

Remember that $\bP_a$ and $\bP^a$ have only one cut ~-- Zhukovsky cut on the real axis. This cut can be resolved by considering $\bP_a$ as a function of Zhukovsky variable $x(u)$. In other words, we can represent
$f_i,g_i$ as series in $x(u)$:
\beqa\label{fg_ansatz}\nn
&&f_1=1+g^{2\L} \sum\limits_{n=1}^{\infty}\frac{g^{2n-2}c_{1,n}}{(g x)^n}\\\nn
&&g_1=(g x)^{-\L}\left(u^\L+  \sum_{k=0}^{\L-1} c_{2,-k}u^k+ \sum\limits_{n=1}^{\infty}\frac{g^{2n}c_{2,n}}{(g x)^n}\right)\\\nn
&&f_2=(g x)^{-\L}\left(u^\L+  \sum_{k=0}^{\L-1} c_{3,-k}u^k+ \sum\limits_{n=1}^{\infty}\frac{g^{2n}c_{3,n}}{(g x)^n}\right)\\
&&g_2=1+g^{2\L} \sum\limits_{n=1}^{\infty}\frac{g^{2n-2}c_{4,n}}{(g x)^n}
\eeqa
This ansatz was constructed so as to satisfy the condition that $\bP_a$ stays finite as $g\rightarrow0$ and $\tilde\bP_a$ grows as ${\cal O}(g^{-\L})$. One can see that this is the case if we assume $c_{a,k}\sim 1$. Indeed, the operation tilde (monodromy around a branch point of the Zhukovsky cut) transforms $x(u)$ to $1/x(u)$. Since we want to keep $u=g(x+1/x)$ finite in the weak coupling regime, when $x\sim 1/g\to\infty$,  each power of $x$ should be compensated by at least one power of $g$.
Plugging the ansatz \eqref{fg_ansatz} into  \eqref{Pa2} we get
\beqa \nn
&&\bP_1\sim(g x)^{-\L/2}(1+\dots)\,, \\\nn
&&\bP_2\sim(g x)^{-\L/2}(1+\dots)\,, \\\nn
&&\bP_3\sim(g x)^{-\L/2}(\text{Poly}(u)+\dots)\,,\\
&&\bP_4\sim(g x)^{-\L/2}(\text{Poly}(-u)+\dots)\,,
\eeqa
where $\text{Poly}(u)$ is some polynomial in $u$. The scaling condition for \(\bP_a\) ~--- the finiteness in the weak coupling limit \(g\to 0\) ~--- is obviously satisfied. Now consider $\tilde \bP_a$:
\beqa\nn
&&\tilde\bP_1\sim g^{-\L}(g x)^{\L/2}\left(1+g^{2\L-2}\sum\limits_{n=1}^{\infty} c_{1,n}(g x)^n\right), \\\nn
&&\tilde\bP_2\sim g^{-\L}(g x)^{\L/2}\left(1+g^{2\L-2}\sum\limits_{n=1}^{\infty} c_{1,n}(-g x)^n\right), \\\nn
&&\tilde\bP_3\sim g^{-\L}(g x)^{-\L/2}\left( \text{Poly}(u) + \sum\limits_{n=1}^{\infty} c_{2,n}(g x)^n\right), \\
&&\tilde\bP_4\sim g^{-\L}(g x)^{-\L/2}\left( \text{Poly}(-u) + \sum\limits_{n=1}^{\infty} c_{2,n}(-g x)^n\right).
\eeqa
The scaling condition for $\tilde \bP_a$ is satisifed as well. To summarize: The scaling of the terms in $f_1,g_2$ is constrained by the scaling  of $\tilde \bP$. The scaling of the polynomial terms in $g_1,f_2$ is constrained by finiteness of $\bP$ as $g\rightarrow0$ and the scaling of the singular terms in $g_1,f_2$ ~--- by $\tilde \bP$.
We also want to have a consistency with the weak coupling result\cite{Kazakov:2015efa}, that is why the potentially singular parts in $f_1,g_2,\tilde g_2,\tilde f_1$ have a \(g^{2L}\) prefactor for the infinite sums.

In order to proceed we need to constrain the coefficients $c_{a,k}$ of the ansatz above as much as we can. We also want to be able to express $\Delta$ through $c_{a,n}$. To this end, we plug $\bP_a$ given by the ansatz into \eqref{Baxter} and expand this equation at $u\rightarrow\infty$ assuming that the asymptotics of $\bQ_i$ are given by \eqref{asymptoticsQi}. This yields a system of equations for coefficients $c_{a,n}$ and $\Delta$. The structure of this system of equations depends on $L$. Typically we have to expand the Baxter equation to the fourth subleading order in $1/u$ to get a non-trivial relation for $\Delta$ and a closed system for $c_{a,n}$. We performed the calculations for $\L=2,3,4$.

For $\L=2$ and $\L=3$ the resulting relations for $\Delta$ have a form
\beqa\nn
&&(\Delta-2)^2=f_2(c_{2,0},c_{3,-1},c_{4,1},c_{4,2},\kappa,\hat \kappa)\,, \\
&&(\Delta-2)^2=f_3(c_{2,-1},c_{3,-2},c_{4,1},c_{4,2},\kappa,\hat \kappa)\,,
\eeqa
where $f_2,f_3$ are rational functions of their arguments given in appendix \ref{sec:Formulas}.

Now we need to take the double scaling limit of the relations just derived and of the equation \eqref{Baxter}.
Let us define (see \eq{kap})
\beqa
&&s=\sqrt{\kappa \hat \kappa} \,, \qquad r=(\kappa/\hat\kappa)^\L\,, \qquad \xi= g s\,.
\eeqa
The double scaling limit consists of taking $g\rightarrow0$ and $s\rightarrow\infty$ when keeping $\xi$ constant. The parameter $r$ is not entering the Lagrangian \eqref{bi-scalarL}, so we expect it to drop out from final result, although it may be present in the intermediate computations.

Each coefficient $c_{a,k}$ is a regular function of $g,\kappa,\hat \kappa$ and so it needs to be expanded in the powers of $g$:
\beq
c_{a,k}=\sum\limits_{m=0}^{\infty}c_{a,k,m}(\xi)(g/\xi)^{2m}
\eeq
Then we plug these expanded coefficients back into the ansatz \eqref{fg_ansatz}. The resulting expanded $\bp_a$ for $\L=3$ look as follows:
\beqa
\label{scaled_p} \nn
&&\bp_1(u)=1+\frac{4i-2i r-r c_{3,-1,2}}{r u }(g/\xi)^4+\\\nn
&&+\frac{4 i \xi^\L -2ir \xi^\L-r \xi^\L c_{3,-1,2}+r c_{1,2,0} \xi^{3\L} u+r c_{1,1,1}\xi^\L u^2}{r u^3}(g/\xi)^6+\dots\\\nn
&&\bp_2(u)=\bp_1(-u)\\ \nn
&&\bp_3(u)=1-\frac{2i r}{(r-1)u}-\frac{r(r+1)}{(r-1)^2 u^2}+\\\nn
&&+\left(\frac{c_{2,-1,1}}{u}+\frac{\sqrt{r}(r(\Gamma_0-4)-4-\Gamma_0)}{2(r-1)u^2}-\frac{4 i r \sqrt{\xi}}{(r-1)u^3}-\frac{2 r \sqrt{\xi}(r+1)}{(r-1)^2 u^4}\right)(g/\xi)^2+\dots\\
&&\bp_4(u)=\bp_3(-u)
\eeqa
where
\beq
(\Delta-2)^2=\Gamma_0+O(g^2)
\eeq
and
\beq
A_1=\frac{r}{r-1}-\frac{4\sqrt{r}}{(r-1)}(g/\xi)^2+\frac{r+7}{(r-1)}(g/\xi)^4+\dots
\eeq

Now we are ready to find the equation for $\bQ_i$ in the double-scaling limit. Plug the expanded $\bP_a$ back into \eqref{Baxter}. To simplify the equation we define
\beq\label{qQ}
q_i(u)=\bQ_i(u)u^{-\L/2}.
\eeq

Since the details of computations vary depending on $\L$, we performed computations for $\L=2,3,4$ and present the results below. In all cases the computations are performed with finite $r$ and we confirm that the final equation becomes non-singular as $r\rightarrow0$\footnote{We assume that the coefficients $c_{a,k,n}$ stay finite as $r\rightarrow 0$, which is confirmed by comparing with the perturbative solution}. Notice that this was not the case for the intermediate quantities like \eqref{scaled_p}. This allows us to set $r$ to 0, which makes more coefficients disappear. Notice that $r\rightarrow 0$ limit is exactly the parameter scaling leading to the Lagrangian \eqref{bi-scalarL}, thus the fact that this limit is non-singular in the final equation is a good sign.

\section{Details of the quasi-classical expansion}\la{AH}

Here we give more orders for the WKB expansion discussed in section~\ref{lastqq}. It is convenient to define the functions via their large $\mathfrak{d}$ expansion, instead of hypergeometric functions
\begin{eqnarray}
{\cal N}^{(a)}=\sum_{n=0}^\infty c^{(a)}_n {\mathfrak d}^{-3n}\,,
\end{eqnarray}
where
\begin{eqnarray}\nn
c_n^{(0)}&=&\frac{3 \Gamma (3 n)}{\mathfrak{d} (2 n+1) \Gamma
(n) \Gamma (n+1)^2}\,, \\\nn
c_n^{(2)}&=&
\frac{3^{3 n+\frac{9}{2}} (n+1) \Gamma
\left(n+\frac{5}{3}\right) \Gamma
\left(n+\frac{7}{3}\right)}{4 \pi
\mathfrak{d}^5 (2 n+3) \Gamma (n+1) \Gamma
(n+3)}\,, \\\nn
c_n^{(4)}&=&
\frac{\left(7 n^3+39 n^2+60 n+24\right) \Gamma
(3 n+6)}{40 \mathfrak{d}^6 (2 n+3) \Gamma
(n+1) \Gamma (n+2) \Gamma (n+4)}\,,\\\nn
c_n^{(6)}&=&\frac{\left(279 n^6+3991 n^5+22511 n^4+63257
n^3+92122 n^2+64512 n+16128\right) \Gamma (3
n+6)}{1680 \mathfrak{d}^7 (2 n+3) \Gamma (n+1)
\Gamma (n+2) \Gamma (n+5)}\,,\\
c_n^{(8)}&=&
\frac{3^{3 n+\frac{11}{2}} p_n \Gamma
\left(n+\frac{8}{3}\right) \Gamma
\left(n+\frac{10}{3}\right)}{44800 \pi
\mathfrak{d}^8 (2 n+3) \Gamma (n+1) \Gamma
(n+6)}\,,
\end{eqnarray}
where
\begin{eqnarray}
p_n&=&3429 n^8+78737
n^7+762569 n^6+4040795 n^5+12701786
n^4+24004388 n^3\\
\nn&+&26281256 n^2+14927040
n+3225600\;.
\end{eqnarray}
One can use the above expression to find the $\log$ singularity at ${\mathfrak d}=3$. This singularity is controlled by the large $n$ asymptotics the above coefficients.
The coefficient in front of the $\log(\mathfrak d-3)$, computed in this way, and
multiplied by $2\pi i$ gives the expansion around $\mathfrak d-3$
for the $\beta-$cycle integral. It produces the following expansion
for $\mathfrak d$ for fixed $N$ and large $\xi$ as explained in the
section \ref{lastqq}
%
\begin{align}
\nn\mathfrak{d}  =  e^{\pm 2i \pi/3} & \bigg[
3  +\frac{\sqrt{3} N}{\bar\xi }
+\frac{5 \left(3
N^2+5\right)}{36 \bar\xi ^2}
+\frac{  \left(11 N^2+161\right)
N}{108 \sqrt{3} \bar\xi ^3}
+\frac{ \left(129
N^4+14754 N^2+16781\right)}{34992 \bar\xi
^4}\\
&- \nn
\frac{\left(57
N^4-11706 N^2-89599\right) N}{52488 \sqrt{3}
\bar\xi ^5}
-\frac{ \left(831 N^6-131805
N^4-5307099 N^2-4448263\right)}{5668704 \bar\xi
^6}
  \\
\nn&+ \frac{\sqrt{3}
\left(1017 N^6+280749 N^4+99326923
N^2+501673551\right) N}{306110016 \bar\xi
^7}\\ \la{moremore}
 &+ \frac{3 \left(633 N^6-36036 N^4+27925062
N^2+527935324\right)
N^2+1122439675}{344373768 \bar\xi
^8}\bigg]\;,
\end{align}
where $\bar\xi= \xi e^{\mp i\pi/6}$.

\end{appendices}

\bibliography{biblioL3}
\bibliographystyle{JHEP}

\end{document}